\documentclass[usegraphicx,usenatbib]{mn2e}

\usepackage{psfig,dynamics}

\def\nreso{$N_{\rm reso}$}
\def\nneigh{$\bar N_{\rm neigh}$}

\title{Numerical requirements for simulations of self gravitating
and non-self gravitating disks} 

\author[Andrew F. Nelson]{Andrew F. Nelson\thanks{
                   Email address:andy.nelson@lanl.gov}
                   \thanks{UKAFF Fellow}
                   \thanks{Current Address:
                        Los Alamos National Laboratory, X-2 MS T087,
                        Los Alamos NM, 87545, USA} \\
        School of Mathematics, University of Edinburgh,
           Edinburgh Scotland EH9 3JZ, United Kingdom }

\begin{document}

\maketitle

\begin{abstract}

We define three requirements for accurate simulations that attempt to
model circumstellar disks and the formation of collapsed objects (e.g.
planets) within them. First, we define a resolution requirement based
on the wavelength for neutral stability of self gravitating waves in
the disk, where a Jeans analysis does not apply. For particle based or
grid based simulations, this criterion takes the form, respectively,
of a minimum number of particles per critical (`Toomre') mass or
maximum value of a `Toomre number', $T= \delta x/\lambda_T$, where the
wavelength, $\lambda_T$, is the wavelength for neutral stability for
waves in disks. The requirements are analogues of the conditions for
cloud collapse simulations as discussed in \citet{BB97} and
\citet{Truelove97}, where the required minimum resolution was shown to
be twice the number of neighbors per Jeans mass or 4-5 times the local
Jeans wavelength, $\lambda_J$, for particle or grid simulations,
respectively.

We apply our criterion to particle simulations of disk evolution and
find that in order to prevent numerically induced fragmentation of the
disk, the Toomre mass must be resolved by a minimum of six times the
average number of neighbor particles used. We investigate the origin
of the apparent discrepancy between the number of particles required
by the cloud and disk fragmentation criteria and find that it is due
largely to ambiguities in the definition of the Jeans mass, as used by
different authors. We reconcile the various definitions, and when an
identical definition of the Jeans mass is used, the condition that
$J\la 1/4$ in the Truelove condition is equivalent to requiring about
10-12 times the average number of neighbor particles per Jeans mass in
an SPH simulation, reducing the difference between simulations of
disks and clouds to about two. While the numbers of particles per
critical mass are similar for both the Jeans and Toomre formalisms,
the Toomre requirement is more restrictive than the Jeans requirement
when the local value of the Toomre stability parameter $Q$ falls below
about one half.

Second, we require that particle based simulations with self gravity
use a variable gravitational softening, in order to avoid inducing
fragmentation by an inappropriate choice of softening length. We show
that using a fixed gravitational softening length for all particles
can lead either to artificial suppression or enhancement of structure
(including fragmentation) in a given disk, or both in different
locations of the same disk, depending on the value chosen for the
softening length. Unphysical behavior can occur whether or not the
system is properly resolved by the new Toomre criterion. 

Third, we require that three dimensional SPH simulations resolve the
vertical structure with at least $\sim4$ particle smoothing lengths
per scale height at the disk midplane, a value which implies a
substantially larger number per vertical column because the disk
itself extends over many scale heights. We suggest that a similar
criterion applies to grid based simulations. We demonstrate that
failure to meet this criterion leads to underestimates in the midplane
density of up to 30--50\% at resolutions common in the literature. As
a direct consequence, gas pressures will be dramatically
underestimated and simulations of self gravitating systems may
artificially and erroneously inflate the likelihood of fragmentation.
We outline an additional condition on the vertical resolution in
simulations that include radiative transfer in order to ensure a
correct description of the cooling, specifically that the temperature
structure near the disk photosphere must be well resolved. As an
example, we demonstrate that for an isentropic vertical structure, the
criterion translates to resolution comparable to $H/20$ near the disk
photosphere, to avoid serious errors in transfer rates of thermal
energy in and out of the disk.

Finally, we discuss results in the literature that purport to form
collapsed objects and conclude that many are likely to have violated
one or more of our criteria, and have therefore made incorrect
conclusions regarding the likelihood for fragmentation and planet
formation.

\end{abstract}


\section{Introduction}\label{intro}

The formation of collapsed objects, both in the context of the
collapse of molecular cloud cores and in the later context of the
clumping of material in a circumstellar disk is very difficult to
model numerically because of the huge range of spatial scales
involved. Accurate simulation of the collapse usually requires that
all of these scales be well resolved, if the result is not to be
contaminated by numerically induced fragmentation.

For three dimensional (3D) simulations, \citet{Truelove97} defined a
minimum resolution condition for the numerical validity of a
simulation that models the collapse of a molecular cloud core using a
grid based hydrodynamic code. Contemporary work by \citet[][hereafter
BB97]{BB97} has examined necessary resolution conditions for the
collapse of a similar system in the context of Smoothed Particle
Hydrodynamics (SPH) simulations, and also discussed the influences
that choices of the form of gravitational softening may have on the
results. Both of these works define minimum resolution criteria in the
context of a Jeans collapse of a gas cloud resolved in 3D, but while
\citet{Truelove97} observe that fragmentation is enhanced by a failure
of the criterion, BB97 only observe enhanced fragmentation if the
gravitational softening used in their particle based simulations is
smaller than the hydrodynamic smoothing. Otherwise, they observe that
fragmentation can be unphysically delayed in a system where it is
known to occur. Later work of \citet{HGW06} explores the problem of
Jeans collapse in isolation. They find that SPH simulations of systems
with initial conditions like that of the original linearized analysis
do not fragment artificially, even when under resolved. 

A number of recent works in the field of planet and brown dwarf
formation \citep{DynI,DynII,Pick98,Pick00a,Pick03,Boss98,Boss00,
Boss02,Boss04,Mayer02,Mayer04,RABB,ArmHan99,Luf04} have discussed the
formation of planets and brown dwarfs via gravitational fragmentation
in disks. In disks systems like those discussed, the Jeans formalism
used to develop the \citealt{Truelove97} and BB97 criteria is not
valid both because the disk scale height is usually small compared to
the Jeans wavelength and because of the existence of shear, which
plays a role as important as self gravity for the growing structures.
As yet however, no analogous resolution criterion exists for disks,
which may be modeled in either fully in two dimensions (2D),
effectively integrating over the vertical coordinate, or in 3D, for
which no vertical integration is assumed, but for which the Jeans
wavelength based criterion still may not apply.

In addition to requirements that simulations resolve the wavelengths
of the relevant instabilities sufficiently, simulations must also
satisfy a number of other criteria if they are to be believed.
Approximations made outside the realm of the physical model, perhaps
used to model the behavior of unresolved or poorly resolved
phenomena, must not drive the results of the simulations themselves.
In the context of simulations of disks, particularly those using
particle based hydrodynamic methods such as SPH, an important
consideration will be the implementation of gravitational softening.
At a more fundamental level, the algorithm used to solve the 
equations used to model physical system must do so accurately, without
becoming unstable or generating large errors through some other sort
of deficiency. A quick glance through the literature \citep[see e.g.
textbooks of][]{hockeast88,fletcher97,leveque02} will demonstrate to
even a casual observer that the study of numerical methods in relation
to their stability is one in which extensive studies on many topics
have been performed. Verification of methods on physically relevant
test problems however, is comparatively more widespread but may often
suffer from insufficient detail or generality. While many studies
discuss the fidelity of the numerical solution on some simple or
contrived test problem, often no test problems sufficiently similar to
the physical system under study can even be devised.  

In section \ref{sec:numerical-factors}, we first extend the previous
3D work of \citet{Truelove97} and BB97 to self gravitating thin disk
systems, then outline alternatives for gravitational softening in
particle simulations, and the possible consequences each choice may
have on results. In section \ref{sec:testing}, we define a test
problem with which we determine appropriate values for resolution for
particle simulations. We continue in section \ref{sec:softening} with
a discussion of the gravitational softening assumed in particle
simulations, and the specific numerical issues encountered in a study
of disks. Next, in section \ref{sec:disks-3d}, we discuss the
application of the criterion to simulations done in 3D, and define a
test problem to validate the accuracy of numerical codes attempting to
model disks in 3D. Using this test problem, we demonstrate that
failure to resolve the vertical structure of disks will lead to large
errors in the numerical solution for the evolution of the entire
physical system. Finally, in section \ref{sec:remarks}, we summarize
our results and discuss them in the context of the models presented in
the literature.

\section{Numerical factors affecting the results of disk 
simulations}\label{sec:numerical-factors}

A numerical simulation of any physical system may suffer from
inaccuracies from several different origins. For example, an incorrect
physical model or incorrect initial conditions will produce results
irrelevant to the system being modeled. On the side of numerics, a
shortcoming in the numerical method may erroneously trigger some
physical process to become active in the evolution, where in reality,
no such physical process is important. A shortcoming in the numerical
method may also trigger effects of purely artificial origin. In this
section, we first describe the conditions under which we may expect a
numerically induced, but physically based instability leading to
fragmentation to be present in simulations of disks. Secondly, we
discuss alternative treatments of the hydrodynamics and gravity in
particle simulations on the spatial scales of the smoothing and
softening lengths in particle simulations. We describe conditions
under which we may expect their implementations to influence a
simulation, possibly also leading to artificially induced
fragmentation. Finally, we discuss the strengths and weaknesses of 
2D and 3D treatments of a system, the meaning of physical quantities
realized in a 2D approximation and consequences that may arise when
the assumptions underlying one or the other treatment break down.

\subsection{Resolution criteria in cloud and disk fragmentation 
simulations}\label{sec:resol}

A condition on the minimum resolution to ensure the collapse of a
cloud is of physical rather than numerical origin was defined by
\citet{Truelove97}, using the ratio of the local grid resolution,
$\delta x$, and local Jeans wavelength, $\lambda_J$, in the fluid:
\begin{equation}\label{eq:Jeans-cond}
J = {{\delta x}\over {\lambda_J}} 
\end{equation}
where $\delta x$ is the size of a grid cell and $\lambda_J$ is
the local Jeans wavelength:
\begin{equation}\label{eq:Jeans-wavel}
\lambda_J = \left({{\pi c^2_s}\over{G\rho}}\right)^{1/2},
\end{equation}
and $c_s$ is the sound speed, $\rho$ is the volume density and $G$ is
the gravitational constant. To obtain a form more useful in particle
based numerical methods (e.g. SPH) where the resolution element is a
unit of mass rather than of length, BB97 used the Jeans mass
as defined from energy considerations \citep{Toh82} for a homogeneous
sphere, to define an analogous criterion for the maximum resolvable
density. Generalizing their result to a gas with $d$ internal degrees 
of freedom we find:
\begin{equation}\label{eq:Mjeans}
M_J^{Energy} =  \left[{{3}\over{4\pi}}
                 \left({{5d}\over{6\gamma}}\right)^3\right]^{1/2}
                       {{c_s^3}\over{(G^3\rho)^{1/2}}} 
\end{equation}
where the superscript `Energy' is included to distinguish this
definition of $M_J$ from two others defined in the next section,  
$\gamma$ is the ratio of specific heats and $d$ is the number of
degrees of freedom, equal to 3 for a monotonic ideal gas. This
equation yields a maximum resolvable density\footnote{Note that this
equation in our previous conference proceeding \citet{N03} was
erroneously stated and should be disregarded.}:
\begin{equation}\label{eq:max-voldens}
\rho_{\rm max} = \left[{{3}\over{4\pi}}
                     \left({{5d}\over{6\gamma G}}\right)^3\right]
                         {{c_s^6}\over{(m_p N_{\rm reso})^2}},
\end{equation}
where we equate the Jeans mass $M_J$ to a sum of $N_{\rm reso}$ SPH
particle masses, $m_p$, that are required to resolve it. 

An exactly analogous stability condition can be made for rotationally
supported (i.e. disk) systems using the local Toomre wavelength,
$\lambda_T$,
\begin{equation}\label{eq:Toom-cond}
T = {{\delta x}\over {\lambda_T}}, 
\end{equation}
where $\lambda_T$ is the wavelength which defines neutral
stability in disks.  We can derive $\lambda_T$ from the dispersion
relation for waves in disks, whose solution \citep[see
e.g.,][]{LinLau79} has four branches:
\begin{equation}\label{eq:k-disks}
k= \pm k_0\left( 1 \pm
          \sqrt{ 1 - Q^2\left(1 - \nu^2\right) }  \right), 
\end{equation}
corresponding to leading and trailing, short and long wavelength spiral
density waves. The variables $k_0$, $Q$ and $\nu$ are defined by
\begin{eqnarray}\label{eq:k0}
k_0 = {{ \pi G \Sigma}\over { c_s^2 }},  \ \ \ \ \ 
Q = {{ \kappa c_s}\over {\pi G \Sigma}} \ \ \ \ {\rm and}  \ \ \ \
\nu = {{ \left( \omega - m\Omega\right)}\over {\kappa}}
\end{eqnarray}
respectively.  Physically, these variables are the wave number, the
well known Toomre $Q$ parameter and the Doppler shifted pattern
frequency of wave of symmetry $m$ (i.e. with $m$ spiral arms),
normalized to the local epicyclic frequency, $\kappa$, in the disk.
They depend on the disk's surface density $\Sigma$, its orbit
frequency, $\omega$, as well as the pattern frequency, $\Omega$ and
symmetry, $m$, (i.e. the number of spiral arms) of the spiral waves.
Neutral stability is defined by the condition that the term under the
square root is zero, for which the wave number is $|k|=k_0$. The
critical wavelength corresponding to this wave number is:
\begin{equation}\label{eq:Toom-wavel} 
\lambda_T =   {{2c_s^2 }\over{G\Sigma}}.
\end{equation}

\citet{GalDyn} derive an expression for the longest wavelength
that will be unstable in a given disk as
\begin{equation}\label{eq:disk-critwavel}
\lambda_c = {{4 \pi^2 G\Sigma }\over{ \kappa^2 } },
\end{equation}
however this wavelength is of limited relevance to the question of
numerical resolution because it is neither the most unstable nor the
shortest unstable wavelength in the disk. As noted by them,
when the disk first becomes unstable (i.e. $Q=1$), $\lambda_c$ is a
factor of two longer than $\lambda_T$. Due to our interest in
determining a critical resolution requirement, which requires the
shorter length scale be used, we use the wavelength definition of eq.
\ref{eq:Toom-wavel} rather than that of eq. \ref{eq:disk-critwavel} in
defining our criterion.

As in the 3D case, a form useful for particle based simulations can be
obtained, this time by defining a Toomre mass. Unlike the 3D case
however, due to the more complicated geometry of the mass
distribution, no easily derivable form of the Toomre mass may be
determined from energy considerations analogous to that in eq.
\ref{eq:Mjeans}. We therefore use a circular volume element to define 
\begin{equation}\label{eq:Mtoomre}
M_T = \pi\Sigma \left({{\lambda_T}\over{2}}\right)^2  =
                    {{ \pi c_s^4}\over{G^2 \Sigma}}.
\end{equation}
A maximum resolvable surface density follows directly as:
\begin{equation}\label{eq:max-surfdens}
\Sigma_{\rm max} = \left({{\pi}\over{G^2}}\right)
                    {{c_s^4}\over{m_p N_{\rm reso}}},
\end{equation}
where the symbols have the same meanings as in equation
\ref{eq:max-voldens} above. Determination of appropriate values for
$T$ and \nreso\  for disk simulations will be the subject of section
\ref{sec:testing}.

It will frequently be useful to determine the resolution required for
the same overall morphology but with varying stability, as required
for a parameter study varying a disk's Toomre $Q$ value. Given the
same overall morphology, $Q$ becomes a function of sound speed only,
which is also the only term in equation \ref{eq:max-surfdens} that will
vary. A useful re-parameterization of this expression to illustrate
the sensitivity of the resolution to the disk stability directly will
therefore be to replace the sound speed with $Q$ through its
definition:
\begin{equation}\label{eq:max-surfdens-Q}
\Sigma_{\rm max} = {{\pi^5 G^2 \Sigma^4}\over{\kappa^4}}
                    {{Q^4}\over{m_p N_{\rm reso}}}.
\end{equation}
The maximum resolvable surface density is thus proportional to the
fourth power of $Q$. In order to obtain identical effective
resolution, as quantified by identical values of $\Sigma_{\rm max}$,
a simulation with $Q=1$ must therefore be resolved with $Q^4\approx 5$
more particles than a simulation with $Q=1.5$.

\subsection{Consistent application of the criteria across simulations
of different types}\label{sec:consistent}

The criteria in equations \ref{eq:Jeans-wavel} and
\ref{eq:max-voldens} can easily be shown to be equivalent up to
constant factors by casting the former as a Jeans mass. The mass
inside a sphere of radius $\lambda_J/2$ is: 
\begin{equation}\label{eq:sphere-wavel-Mjeans}
M_J^{Sphere} = {{4\pi}\over{3}}\rho\left({{\lambda_J}\over{2}}\right)^3 
              =  {{4\pi^{5/2}}\over{24}} {{c_s^3}\over{(G^3\rho)^{1/2}}}.
\end{equation}
so that comparison of equations \ref{eq:Mjeans} (determined from a
similar spherical volume assumption) and equation
\ref{eq:sphere-wavel-Mjeans} yields the desired identification. Yet
another definition of the Jeans mass is given in \citet{KMK04} using a
cubic rather than spherical volume element as:
\begin{equation}\label{eq:cube-wavel-Mjeans}
M_J^{Cube} = \rho \lambda_J^3 = \pi^{3/2} {{c_s^3}\over{(G^3\rho)^{1/2}}}
\end{equation}

In each of these definitions, the constant factors differ.
Specifically, the three forms $M_J^{Energy}:M_J^{Sphere}:M_J^{Cube}$,
yield values of the Jeans mass in a ratio of $0.89:2.92:5.57$ relative
to each other, assuming a monatomic ideal gas. Analogously for the 2D
case, using a square area element, $\lambda_T^2$ rather than circular,
would result in a definition of Toomre mass or density that is a
factor $4/\pi$ larger than stated in equations \ref{eq:Mtoomre} and
\ref{eq:max-surfdens}, respectively. 

Although these constant factors may indeed be regarded as
insignificant in a global sense since the basic functional dependence
is the same, their quantification is important because it allows us to
interpret the results of simulations obtained using various versions
of the criterion. For example, \citet{Truelove97} found empirically
that $J\la 1/4$ was sufficient to suppress numerical instabilities in
a test problem, which implies a mass per grid zone of at most
$m_{zone}\sim J^3M_J\approx M_J/64$. For particle simulations, BB97
showed that numerical instability could be suppressed in a similar
test problem by resolving the local Jeans mass with particles less
massive than $m_p\sim M_J/N_{\rm reso}\approx M_J/100$. Although
ostensibly very similar, these criteria differ by a hidden factor of
$\sim5.57/0.89\approx6.2$ from each other due to the differing
proportionality factors, with the particle based criterion being more
conservative (requiring more particles to obtain equivalent resolution
for a given mass distribution). 

In other words, the $\sim100$ particle ($\approx\times2$\nneigh)
condition of BB97 translates to a $\sim300$ particle
($\approx\times6$\nneigh) condition condition if we use the Jeans mass
defined by equation \ref{eq:sphere-wavel-Mjeans} or a $\sim600$
particle ($\approx\times12$\nneigh) condition if we use the Jeans mass
definition of equation \ref{eq:cube-wavel-Mjeans}.  In general, for
equivalent resolution of the physical parameters of the flow, a
particle simulation will require roughly ten times as many fluid
elements as a grid code like the one discussed in \citet{Truelove97}.
With the same number of fluid elements (particles or grid cells) a
grid simulation will be able to resolve higher mass densities before
becoming numerically unstable than a particle simulation. 

\subsection{The relative sizes of the Jeans and Toomre wavelengths}
\label{sec:sizes}

The critical wavelength for disks from equation \ref{eq:Toom-wavel} is
linearly dependent on the temperature to mass ratio through the sound
speed and mass density, while the Jeans wavelength in equation
\ref{eq:Jeans-wavel} is dependent only on its square root. Thus, the
Toomre criterion will be more strict on smaller spatial scales than
the Jeans criterion, but less strict on larger scales.

We can determine the relative sizes of the two wavelengths, and the
crossover point at which both are equal, from their ratio:  
\begin{equation}\label{eq:wavel-ratio}
{{\lambda_T}\over{\lambda_J}} = 2 
     \left({{ \rho c_s^2}\over { \pi G\Sigma^2}}\right)^{1/2}.
\end{equation}
If we assume that the disk is near Keplerian so that
$\Omega\approx\kappa$, that the volume density and the surface density
are related by $\rho = \Sigma/2fH$, where $f\approx1$ is a coefficient
specifying the exact proportionality between surface and volume
densities, and that the local disk scale height is $H=c_s/\Omega$ (but
see section \ref{sec:bad-approx} below), then we can combine equation
\ref{eq:Jeans-wavel} and \ref{eq:Toom-wavel} to produce 
\begin{equation}\label{eq:TJ-relative} 
\lambda_T \approx \sqrt{ {2 Q}\over{f} } \lambda_J. 
\end{equation}
Thus if the local value of $Q$ falls below about one half, as it may
in regions beginning to fragment, the Toomre instability wavelength
will be a more restrictive criterion for numerical simulations than
the Jeans criterion used by \citet{Truelove97}. 

On first inspection, equation \ref{eq:TJ-relative} and the wavelength
proportionalities that went into it would seem to be backwards. When
the collapse is well underway ($Q<<1$), rotation ceases to matter so
that the collapse should proceed according to a Jeans prescription. In
other words, we would expect that the Jeans wavelength should be
smaller than the Toomre wavelength. Indeed, this would be the case for
a collapsing region that is much smaller than the disk scale height,
but this equation shows that for the initial stages of collapse
relevant to the analytic wavelength derivations, it is not the
presence or absence of rotation that is relevant, but rather the
dimensionality of the problem that plays the most important role. 

\subsection{The application and applicability of the Jeans and Toomre
criteria to numerical simulations}\label{sec:applicability}

The mathematical analysis leading to the Jeans resolution criterion is
fully three dimensional, while that leading to the Toomre criterion is
limited to two dimensions: the equations are integrated over the $z$
coordinate. The difference is important because the Jeans analysis may
not be valid in a disk simulation (even those evolved using fully 3D
models) because it assumes a homogeneous medium which is infinite in
all three spatial coordinates. By definition, a disk structure will
violate this assumption since the matter will always be condensed into
a midplane above and below which relatively little matter lies. The
violation will be important if the Jeans wavelength is comparable to
or larger than the disk scale height, determined for the local
conditions in a given disk. In this case, even the initial conditions
of a system may not satisfy the underlying assumptions of the
analysis. Taking the specific example shown in figure
\ref{fig:Boss-crit}b of section \ref{sec:bad-approx} below, we find
that Jeans wavelength is long compared to the disk scale height
everywhere, so that the Jeans analysis is indeed inapplicable. 

On the other hand, the analysis leading to the Toomre wavelength is
valid in the limit of a thin, rotating system for which a `surface
density' is a meaningful concept, whether or not a given simulation is
actually performed in two or three dimensions. By construction of the
analysis itself, disks fall into this category, so at least we may
construct initial conditions that satisfy the underlying model. As in
the case of the Jeans wavelength, by examining figure
\ref{fig:Boss-crit}b, we see that the Toomre wavelength is also long
in comparison to the disk scale height. In this case, the large ratio
means only that the analytic assumptions become more valid, not less.
Violation of the model assumptions may still be important with
wavelengths comparable to a disk scale height scale because of the
neglect of vertical motion and structure in the analysis. 

Violation of the assumptions in the analytic derivations may also
occur at times after simulations have evolved for some time because
the analyses assume that variations of all quantities from their
initial values are small, even if the initial conditions satisfy all
requirements of the linearized analyses. For example, a density
perturbation may be only slightly enhanced from the local background
in a fragmenting molecular cloud, while the background potential is
characterized by a comparatively steep overall gradient due to some
large structure nearby. A second example may be that fluid velocities
are large while other perturbations are small because the initial
state was seeded with some spectrum of turbulent velocity
perturbations. Finally, numerical simulations may not solve the
equations of hydrodynamics accurately, or may do so with only poor
fidelity for some problems. 

Unfortunately, we will find that in the most relevant range of
parameter space for disk simulations, both the Jeans and Toomre
wavelengths reach values comparable to the disk scale height and local
perturbations reach amplitudes beyond those for which linearized
analyses are strictly valid. As with the original 3D analysis of
collapsing clouds, we therefore resort to numerical simulations to
determine the required `safety factor' (i.e. the values of \nreso,
`$T$' and `$J$' above for disk or cloud collapse analyses) for which
we can be reasonably confident that the basic features of that
analysis are not violated, even though it may be applied in a region
where its assumptions may be called into question. 

Although there is only one criterion for disk evolution and one for
cloud collapse, there are effectively five implementations of them as
applied to disk systems that could be used under different
circumstances. First the equations \ref{eq:Jeans-cond} or
\ref{eq:Toom-cond} might be used directly: a 2D simulation could use
the directly available surface density, or a 3D simulation could use
the directly available volume density. The criteria could also be used
indirectly using the disk scale height to convert between volume
density and surface density. Finally, a surface density could be
obtained from a 3D simulation by directly integrating over the $z$
coordinate, for use in the Toomre criterion. We will show in section
\ref{sec:bad-approx} that using the approximate indirect forms (i.e.
making the volume or surface density conversion using the disk the
scale height) can yield seriously discrepant values of the stability
wavelength, perhaps leading to erroneous conclusions regarding the
veracity of a given simulation.

\subsection{Hydrodynamic smoothing, gravitational softening, and
our implementations of them}\label{sec:our-softsmooth}

In all numerical simulations, the modeler would like to resolve the
largest range of spatial scales possible, so that both the smooth and
highly inhomogeneous regions are accurately modeled. In this regard, a
limit on the resolution will be the scale on which the gravitational
and hydrodynamic forces can be resolved. For grid based simulations,
resolution of both the hydrodynamic features in the flow and
gravitational forces will be related to the local dimensions of the
grid \citep[see e.g.][]{FMA,Pick03}.

For particle simulations, the limits will be related to two length
scales, one for gravity and one for hydrodynamics, each defining the
spatial extent of the particles in different ways. In order to make
clear that they are distinct from each other, we will make a 
distinction between the terms used to refer to each. Specifically we 
will refer to gravitational `softening' lengths and hydrodynamic 
`smoothing' lengths for particles, to describe each effect. 

\subsubsection{Smoothing}\label{sec:our-smoothing}

Fluid quantities in a particle based simulation using SPH are
reconstructed from the positions of the particles at a given time.
Contributions of particles are weighted according to a smoothing
function (the `kernel') and all contributions are summed to define
each of the hydrodynamic quantities at the current location of each
particle \citep{Mon92}. Quantities such as density are calculated
using the kernel directly, while forces due to pressure are calculated
using its derivative. In each case our implementation follows the
discussion in \citet{Benz90} and, in the work presented here, we use
the now standard spline kernel of \citet{MonLat85} given by:
\begin{equation}\label{eq:SPHkern}
W(r_{ij},h_{ij}) = {\sigma \over h_{ij}^{\nu}}
   \cases{ 1 - {{3}\over{2}}v^2 + {{3}\over{4}}v^3 & if $ 0 < v < 1$ ;\cr
           {{1}\over{4}}(2-v)^3                    & if $ 1 < v < 2$ ;\cr 
           0                                       & otherwise.      ;\cr }
\end{equation}
Here, $\nu$ is the number of dimensions, $v=r_{ij}/h_{ij}$, $r_{ij}=
|\mathbf{r_i} - \mathbf{r_j}|$, $h_{ij}= (h_i + h_j)/2$ and $\sigma$
is the normalization with values of $10/(7\pi)$ and $1/\pi$ in two or
three dimensions respectively. 

We will also study the effect of an important modification of that
derivative described by \citet{TC92}, which takes the form:
\begin{equation}\label{eq:mod-kerderiv}
{{d W(r_{ij},h_{ij})}\over{dr_{ij}}} = {\sigma \over h_{ij}^{\nu+1} } 
   \cases{ -1                      & if $ 0   < v < 2/3$ ;\cr
           -3v + {{9}\over{4}}v^2  & if $ 2/3 < v < 1$   ;\cr
           -{{3}\over{4}}(2-v)^2   & if $ 1   < v < 2$   ;\cr  
            \phantom{-}0           & otherwise           ;\cr }
\end{equation}
and acts to reduce unphysical particle clumping
\citep{Steinmetz96,Thacker00}, by providing a net repulsive pressure
force even at zero separation. The modification affects the region
$v<2/3$ where in its unmodified form, the derivative decreases
monotonically to zero as $v$ itself approaches zero. Price (2005,
personal communication) notes that this modification may cause changes
in the effective sound speed determined from linear analyses, because
normalization conditions on the kernel's first and second derivatives
are not satisfied. The normalization conditions will be exact only in
the limits where the smoothing lengths approach zero, neighbor counts
approach infinity and the distribution of neighbors over the kernel
volume allows an accurate correspondence between a summed and an
integral form of the equations. Since none of these three conditions
are realized in practice, it is difficult to evaluate what
consequences may result in actual simulations. In section
\ref{sec:2dtest-evo} we investigate the differences between using the
standard and modified kernel derivatives on the outcomes of our
simulations. Detailed analyses of the numerical stability and errors
resulting from this modified formulation are beyond the scope of this
work and we defer such discussions to the future. 

\subsubsection{Softening}\label{sec:our-softening}

An important subtlety of particle simulations is to employ a
gravitational softening length that removes the singularity in the
force obtained when two particles in a simulation approach each other
too closely, effectively modifying Newton's law of gravitation. Such a
modification is equivalent to defining a mass density distribution for
the particles, so that they include an assumption of some spatial
extent, rather than that they are point like objects. Modifying the
gravitational force is acceptable in hydrodynamic and collisionless
$N$-body simulations because particles do not in fact represent single
particles in the underlying system, but only a statistical
representation of the local distribution of gas or particles. The
question we face is how to implement an appropriate gravitational
softening length in our simulations which, on one hand, prevents
unphysically large forces from developing between particles, and on
the other, is small enough to allow small scale features to develop in
the flow. 

Two alternatives for implementing softening that are common throughout
the literature are first to assume each particle has the mass
distribution of a Plummer sphere, so that the force law is modified to
the form of a Plummer force law \citep{Romeo94,Romeo97,Athan00}, or to
use a spline based kernel as discussed by \citet{Benz90}, who
interprets the smoothing kernel used to realize the hydrodynamic
quantities as a mass distribution. Two variants are common in the
latter case, first to use a softening that varies with the local
conditions (usually chosen to be identical to the SPH smoothing length
of each particle) or second, to fix the softening to a single constant
value at the beginning of the simulation. Examples of variable kernel
softening are found in codes used by \citet{Benz90,SM93}, BB97 and our
own previous work, while the TreeSPH, GADGET and GASOLINE codes of
\citet{hk89,springel2001} and \citet{wadsley2003}, respectively, use
fixed softening. 

We implement the variable kernel softening variant using same kernel
used for the SPH quantities given by equation \ref{eq:SPHkern}. For
gravitational softening, the masses used to compute the acceleration is
modified from its Newtonian form according to the prescription that a
source particle's mass is reduced from its true value by a factor
proportional to the volume enclosed by a sphere whose radius is the
separation between the particles:
\begin{equation}\label{eq:mass-kernsoft} 
\hat m = m_p \cases{ 2\pi \int_0^{r_{ij}} W  u   du  & if $n_d=2$;  \cr
                     4\pi \int_0^{r_{ij}} W  u^2 du  & if $n_d=3$  \cr} 
\end{equation}
where $m_p$ is the mass of the particle from which a force
contribution is to be calculated and $n_d$ is the number of spatial
dimensions. The sink particle, on which forces are calculated, is
assumed to be a point mass, so that for two particles of mass, $m_1$
and $m_2$, the force exerted by particle 1 on particle 2 is 
\begin{equation}\label{eq:forc-kernsoft}
{\mathbf F} = {{-G \hat m_1 m_2 \hat {\mathbf r}_{12}}\over{r_{12}^2}},
\end{equation}
where ${\mathbf r}_{12}={\mathbf r_1} - {\mathbf r_2}$ is the
separation between the particles one and two. Except for a change of
sign, this definition is manifestly invariant to exchanging the
identities of the two particles, and therefore conserves momentum
exactly. In 3D, and as the separation between two particles decreases,
the gravitational force between them will also decrease, ultimately to
zero at $|r|=0$, because $\hat m$ depends on $r^3$.

In 2D, $\hat m$ is proportional only to $r^2$ and the force instead
approaches a {\it non}-zero constant value as the separation decreases
to zero, rather than zero. The physical reason for the discrepancy is
that while equation \ref{eq:mass-kernsoft} implies a 2D structure for
the mass distribution, equation \ref{eq:forc-kernsoft} retains a 3D
structure for the forces they cause. On the scale of the interparticle
separation, the scale height of the disk not be negligible, so that in
fact it is truly three dimensional. We are aware of only one treatment
that attempts to account for the conflict between the 2D and 3D
behaviors \citep{Koller}, by assuming a vertical structure and
numerically integrating the contributions to the forces on a point
mass over the $z$ coordinate. Thereafter Koller applies the derived
correction factor to the forces, tabulated as a function of distance.
We expect Koller's treatment will not be generally suitable for
particle simulations because it requires different tables for
different vertical structures and because, even with the modification,
he finds that a Plummer-like softening term is still required, though
it can be made much smaller in magnitude. 

Instead, we propose an alternative form of softening, still based on
the 2D kernel softening discussed above, with the modification that
the effective mass $\hat m$ is multiplied by an additional
modification factor so that 
\begin{equation}\label{eq:mhat-2d}
\hat m' =  \cases{ (3v-{{4}\over{9}}v^2) \hat m     & if $v<2/3$;  \cr
                    \hat m                          & if $v>2/3$.  \cr} 
\end{equation}
The choice of this exact form of softening is arbitrary, but is
motivated by three desirable conditions on the force within and
outside the softened region. We require that the force decreases
linearly to zero at zero separation, yielding a truly collisionless
form. Second, at separations $>2h$, we require the force returns to
its correctly normalized, perfectly Newtonian form and, finally, the
derivative of the force is continuous, so that the force varies
smoothly at all separations. Coincidentally, for $v<2/3$, this form
duplicates the algebraic form of the derivative of the standard
kernel, a fact that will prove advantageous for obtaining ratios
between gravitational and pressure forces near unity in section
\ref{sec:softsmooth-limits}.

Throughout this paper we implement gravitational softening using
equation \ref{eq:mass-kernsoft} with the variable smoothing length,
$h$, as its length scale. We will perform separate series' of
simulations employing either the original mass distribution defined by
equation \ref{eq:mass-kernsoft} and incorporating the kernel in
equation \ref{eq:SPHkern}, or the modified mass distribution of
equation \ref{eq:mhat-2d}.

\subsection{The relative merits of fixed and variable
softening}\label{sec:comparison}

An important advantage of variable gravitational softening is that it
allows a modeler to soften the gravitational forces on the same length
scale used for generating the hydrodynamic quantities. The variation
is required because the hydrodynamic quantities are generated using an
approximately (or exactly) fixed number of `neighbors' but the local
particle density is not constant. In order to retain the approximately
fixed neighbor count, the smoothing must be correspondingly smaller in
high density regions than in low density regions. In order to retain
the equality over the duration of a simulation, the softening must
also be allowed to vary according to local conditions. 

The advantage of softening with the same length scale as smoothing is
that large imbalances in the gravitational and hydrodynamic forces
between pairs of particles cannot develop due to particles being in
range of one or the other of the lengths, but not both. The
disadvantage is that energy may no longer be conserved because no
account is made of the change in the internal mass distribution of
particles as their smoothing length changes. Contributions to
gravitational potential energy dependent on such changes will
therefore not be evaluated correctly.

In the case of fixed softening, an advantage is that it conserves
energy, with disadvantages including the fact that the smallest
resolvable length scale is both fixed at the beginning of the
simulation and is the same everywhere. A collapsing or expanding body
may quickly reach a size where the flow is dominated by the softening
in one region, while in another the flow may become unphysically
point-like.

Several previous works have discussed the consequences of the
alternatives in simulations of different types, but no clear consensus
has emerged from the discussion. For example, BB97 note that the Jeans
wavelength, cast in the form of a Jeans mass, must be well resolved by
both the gravitational softening length and by the SPH smoothing
length, used respectively to ensure numerical stability in the code
and to produce hydrodynamic quantities from the particle distribution.
For simulations in which the Jeans wavelength is much larger than
either the smoothing or softening lengths (i.e., that the region is
stable against gravitational collapse), BB97 claim that little
difference in behavior should be expected. However, for a marginally
stable mass distribution (e.g. one Jeans mass distributed over a
volume of radius one Jeans length), fragmentation could be
artificially suppressed or enhanced by changing the ratio of the
gravitational softening to SPH smoothing length, because of the large
force imbalances that develop. To avoid such artificial results, they
recommend a variable gravitational softening set to the same length
scale as the hydrodynamic smoothing. In the alternative, when fixed
softening must be used, they recommend a softening length no smaller
than the value that would cause the resolution condition of equation
\ref{eq:max-voldens} to be violated.

Other work \citep{Thacker00} fixes the softening length, but allows
the smoothing to vary, down to a limit of either $\epsilon/2$ or
$\epsilon/20$. They show that a simulation including cooling will
produce much more fragmentation in the latter case, and conclude that
smoothing lengths must be restricted to be greater than the softening
length to avoid artificial fragmentation. Although implementing fixed
softening rather than variable, their conclusion appears consistent
with that of BB97, but takes no account of artificially suppressed
fragmentation. On the other hand, \citet{WCN04} examine fixed
softening models where the smoothing is allowed to vary with and
without the constraint that it may not decrease below the softening
length. They conclude that the smoothing does not need to be equal to
softening and that smoothing must not be constrained because
hydrodynamic shocks cannot be properly resolved on size scales smaller
than the softening length. 

\subsection{The limits and meaning of resolution in the context of
softening and smoothing}\label{sec:softsmooth-limits}

For SPH, where hydrodynamic quantities are derived from interpolations
between pairs of particles, a necessary mathematical condition for the
interpolation kernel is that it be continuous and have a continuous
first derivative \citep[see e.g.][for additional details]{Mon92}.
Physically, the requirement is equivalent to the statement that
contributions to hydrodynamic quantities like mass density and mutual
pressure forces undergo no discontinuous jumps as pairs of particles
approach each other. An unfortunate consequence of the two continuity
requirements is that for kernels like equation \ref{eq:SPHkern},
particles that approach each other experience pressure forces that
first increase with decreasing separation, then decrease to zero as
the separation between them decreases to zero.

The consequence is unfortunate first, because, to the extent that we
can regard the approach of two particles (rather than a large sample
of particles) as representing a compression, the fact that the
pressure force decreases to zero as they approach coincidence means
that our intuitive expectation that pressure forces continue to
increase during a compression is violated. Secondly, and as noted
above, it may lead to unphysical particle clumping. As described by
\citet{Herant94} for non-self gravitating simulations, the reason is
that during the natural course of a simulation, particles that find
themselves closer than a critical separation distance ($v=2/3$ for the
kernel in equation \ref{eq:SPHkern}) where the pressure force is at
its maximum, experience relatively smaller mutual pressure forces
pushing them apart, and relatively similar external forces perturbing
their motion. Because of these small forces, particles continue to
travel together with an end state in which pairs of particles actually
coincide, defining what Herant calls a `pairing instability'.

The astute reader will not fail to note that discussing a `pairing
instability' in the context of a SPH simulation would seem to be
rather irrelevant. SPH particles are of course not assumed to
represent actual physical particles at all, but rather some
(statistical) realization of an underlying physical system. Therefore
any such discussion would appear to be meaningless. In our defense, we
point out that our discussion is of a failure mode for the method
where the statistical assumptions break down, and therefore will
require a careful analysis.

As in the case for smoothing, softened mutual gravitational forces
first increase in magnitude as pairs of particles approach each other,
then decrease to zero as they approach still further. In this case
however, the decrease is not due to any constraint on the kernel, but
rather on the combined assumptions that the softening represents a
mass distribution and that, at some separation, the mass enclosed by a
sphere whose radius equals that separation is less than the total
mass. Then, by invoking Gauss's law, only the fraction of mass inside
the sphere contributes to the force. In 2D simulations, where the
geometric arguments leading to the Gauss's law result do not hold
\citep[see e.g. figure 3 of][]{DynI}, it is necessary to relax the
strict identification of the kernel with a mass distribution. For the
purpose of avoiding infinite force contributions at coincidence,
softening according to this procedure is effective, though as we show
below, problems remain due to the fact that the force still does not
decrease to zero at coincidence. 

In both cases, the reason for the decrease in force magnitude at small
distances is to avoid numerical pathologies: for gravity, the
development of unphysical point-like force contributions, for
hydrodynamics, unphysical discontinuities in the forces and other
hydrodynamic quantities. The important point to note regarding both
softening and smoothing is that on such scales, the magnitudes of the
forces are dependent on assumptions made outside the realm of the
physical model. In other words, the forces computed there are under
resolved and any developing phenomena sensitive to effects in that
region can be due only to an external assumption rather than to any
physical process.

In order to ensure physically valid simulations, it will therefore be
important to ensure that effects originating on unresolved scales do
not drive the results. Moreover, because they act on similar scales
but with effects of opposing sign, it will be important to consider
the resolution limits of both softening and smoothing lengths
together. Setting a softening length much smaller than the smoothing
length is equivalent to the statement that at some distances pressure
forces are under resolved (i.e. that an error is made in evaluating
them), but gravitational forces are not under resolved (that no error
is made in evaluating them). The same statement is true in reverse
when the smoothing length is smaller than the softening. 

Regardless of whether or not one or the other force actually can be
resolved better than the other, the numerical assumptions always limit
the effective, physically correct force resolution to the larger of
the two scales. As pointed out by BB97, the consequences of differing
values of softening and smoothing lengths are net force imbalances of
up to a factor seven at small separations, even when the softening and
smoothing are only different by a factor of two. Much larger force
imbalances develop when the ratios become more extreme, and BB97
attribute the unphysical outcomes of several simulations presented in
the literature to this source.

\begin{figure}
\rotatebox{0}{\includegraphics[width=85mm]{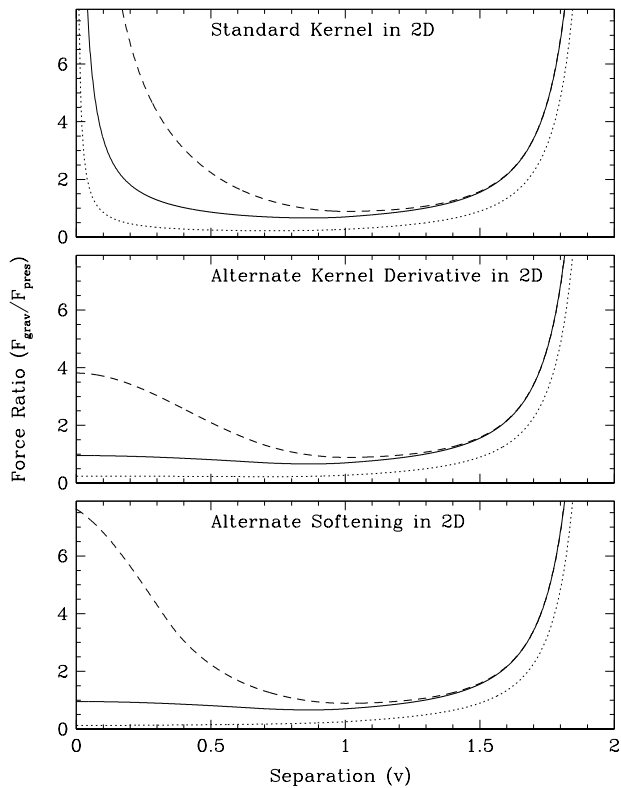}}
\caption{\label{fig:force-ratios} 
Ratios of the gravitational to the pressure forces for the three
scenarios we test, each as a function of the normalized separation,
$v=r/h$, between two particles. The top panel shows ratios using the
standard kernel derivative for the pressure force (top) and the
standard kernel softening for the gravity. The middle panel shows the
ratios using pressures obtained with the alternate kernel derivative
of equation \ref{eq:mod-kerderiv} with standard softening. The bottom
panel shows ratios using the modified kernel softening with the
standard kernel derivative. The solid lines in each panel refer to the
condition where the softening and smoothing lengths are equal, while
the dotted and dashed lines denote the condition where the softening
is twice and half of the smoothing, respectively. } 
\end{figure}

Figure \ref{fig:force-ratios} shows the ratio of the gravitational and
pressure forces between two particles located in a disk where the
Toomre $Q$ value set to unity, realized in 2D. As for the 3D case,
forces are nearly balanced when softening and smoothing are set to
identical scales, but large imbalances are present when they are not
identical. Unlike the 3D case however, the force ratios for the
standard kernel softening/smoothing option do not remain near unity as
the separation, $v$, goes to zero, but instead approach infinity. Due
to the reduced dimensionality, the gravitational force approaches a
finite, non-zero value rather than decreasing to zero as it does in
the 3D case. Force imbalances such as these, whether caused by unequal
softening and smoothing lengths, or by the form of the softening or
smoothing themselves (or both), may influence the susceptibility of
the simulation to artificially induced fragmentation. When the
imbalance favors gravity, particles are not only passively drawn to
each other, but are also actively attracted. Given an imbalance of
the opposite sense, particles may be artificially repelled from each
other, preventing fragmentation of physical origin from occurring.

Given the intrinsic force imbalances present in 2D simulations using
the standard kernel for smoothing, even when used with identical
length scales, simulations may therefore be susceptible to
artificially induced fragmentation. On the other hand, the modified
kernel derivative of \citet{TC92} allows the force ratio to remain
near unity all the way to zero separation because the pressure
approaches a finite, non-zero value in step with softened gravity. 
Similarly, the modified gravitational softening defined by
equation \ref{eq:mhat-2d} also results in force ratios near unity for
equal softening and smoothing lengths, but in this case, both forces
approach zero as their limiting value rather than a finite quantity.
In both cases, large force imbalances can develop when the softening
and smoothing are unequal. With the modified softening variant, force
imbalances of up to a factor $\sim7.5$ at $v=0$ develop when the
lengths are not equal, nearly twice the imbalance present with the
alternate kernel derivative. Because the actual magnitudes are
smaller, the net influence of the imbalance might be smaller. 

We will investigate this question, the influence consequences of using
the alternate kernel derivative formulation and the alternate
softening formulation and the importance of using fixed
softening or variable softening for which force imbalances may or may
not develop respectively, in our work below.

\subsection{The relative merits and shortcomings of 2D and 3D
treatments of the disk}\label{sec:2d-vs-3d}

As computational facilities have become more and more powerful,
simulations of greater and greater complexity have become possible to
perform at acceptable cost. An important step in increasing complexity
is to increase the dimensionality, first from one to two dimensions,
and then to three. In the context of circumstellar disks, the full
transition to 3D remains incomplete. Some workers prefer to limit
their simulations to two dimensions, while others have begun work in
3D. A number of significant consequences derive from each choice. 

Of primary importance is the computational cost and its relation to
the spatial resolution affordable to the simulation. For a grid based
method, computational cost in 3D will be approximately proportional to
the fourth power of the number of cells in any one dimension, while
for 2D, the proportionality drops to the third power. A similar
proportionality will hold for particle simulations as well. Therefore,
simulations in 3D will be able to employ fewer total cells or
particles per dimension: linear resolution is intrinsically coarser in
3D than in 2D for simulations of comparable cost. In contrast, while
2D simulations allow dramatically higher linear resolution in the two
dimensions they actually model, they do so at the cost of requiring
assumptions be made about the character of the system and its behavior
in the third dimension.

For a 3D simulation, if it is to be truly three dimensional, the
physical extent the system in each direction must be resolved by some
number of particles or grid cells greater than one. The exact minimum
number will be a function of both the problem and of the method
employed to evolve the system. In the context of circumstellar disks,
the fact that the disk is spatially thin means that the available
resolution must be allocated inhomogeneously in space if the cost of
calculation is not to become too exorbitant. In a grid simulation for
example, many more grid cells must be allocated to the same physical
length in the $z$ coordinate than either the $x$ or $y$ coordinates,
to resolve the vertical extent of the disk. Problems may arise from
asymmetric grid resolution, including incorrect gravitational forces
\citep{Pick03}, or incorrect hydrodynamic evolution.

The problem is especially acute for SPH simulations because the kernel
used to reconstruct the hydrodynamic quantities is spherically
symmetric. Although experiments with non-spherical kernels have been
published, they are not common, due in part to the difficulty of
implementing them in a form that retains angular momentum conservation
\citep[e.g.][]{ful95}. With a spherical kernel and in the low resolution
limit, particle smoothing lengths may actually exceed that of the
disk's vertical extent, so that only a single fluid element spans the
entire system in that coordinate. Similar conditions may occur in grid
based simulations, if extremely asymmetric zone dimensions are not to
be encountered. Any 3D simulation carried out under such conditions
will effectively model only two dimensions, but will not include any
supporting assumptions present in a simulation explicitly limited to
2D.

One such assumption will be in the description of the fluid itself. In
most `normal' fluids, pressure is by definition a scalar quantity
whose gradient causes a force to be exerted in all directions. The
procedure used in SPH to construct the hydrodynamic quantities however
assumes a roughly spherical distribution of particles. Otherwise, the
interpolations at the heart of SPH are no longer interpolations but
instead extrapolations in directions where few particles exist. As a
result, and as is well known to its practitioners, the method becomes
very inaccurate at boundaries. In circumstellar disk simulations,
performed with SPH at resolution where only one or few particles span
the entire vertical extent of the disk, essentially {\it all}
particles will be located near boundaries.

Whether or not a particular simulation with one dimensionality will be
more physically meaningful than a comparable one of the other depends
on the quality of the assumptions made about the third dimension in a
2D simulation and whether those assumptions offset the loss of linear
resolution affordable in 3D. In section \ref{sec:disks-3d}, we will
show that 3D simulations of disks require much higher resolution than
one might naively expect in order to reproduce basic hydrodynamic
features of the flow correctly. Such high cost means that a given
study will be able to perform many fewer simulations, severely
constraining the possibility of performing the large parameter studies
often required to fully explore the implications of a given physical
model. We have therefore concentrated on the study of disks in 2D
throughout the rest of this paper.

\subsection{The interpretation of hydrodynamic quantities and gravity
in 2D}\label{sec:2d-interp}

Because disks are in fact truly three dimensional, in spite of our
approximation that they are thin, it will be important for correctly
interpreting the results of the simulations to understand the
consequences of a 2D approximation and where it may break down. For 2D
simulations, two fundamentally different assumptions about the third
coordinate are possible. The modeler may either assume that the
simulation is modeling an infinite cylinder or that the simulation is
modeling a thin system in which the system's dynamics and morphology
are either negligible in the third dimension or some approximation is
made regarding their behavior. Each assumption leads to quite
different treatments of the hydrodynamics and gravitation in the
simulation.

In the case of the infinite cylinder assumption, each point in the
plane actually corresponds to a line extending to infinite distance in
the positive and negative third coordinate, which we will assume to be
the $z$ coordinate. For simple hydrodynamical problems, the
interpretation poses no particular conceptual difficulty since
hydrodynamic quantities will carry over from 3D to 2D unaltered. The
consequence for gravitational or electrostatic forces however, is that
they become inversely proportional to the separation in the $xy$
plane, rather than to the inverse square of the separation. For a thin
system, the opposite situation holds. Inverse square law forces retain
their familiar 3D form, but hydrodynamic quantities must be altered,
specifically into integrals of their true 3D forms. For example,
surface density may replace volume density.

The circumstellar disks in this study are spatially thin, and the most
natural interpretation of a 2D simulation of such a disk is that of a
thin, vertically integrated model. In keeping with this characteristic
and with many previous analytic and numerical treatments of disks, our
2D disk simulations are performed in this context. In order to allow
readers to evaluate the results of our study more thoroughly, we now
discuss several factors important for their interpretation and similar
work by others.

In addition to the requirement for 2D simulations that state variables
be integrated over the $z$ coordinate, a more subtle modification must
also be made to other hydrodynamic quantities. \citet{GGN86} and
\citet{OSA92} each discuss the modifications in the effective value of
its polytropic index, $n$, of the gas, corresponding to a degree of
freedom corresponding to the disk `puffing up' in the third
coordinate. Both conclude that a value of $\gamma$ slightly reduced
from that expected for the 3D case should be used, due to the
additional freedom. When an isothermal equation of state is employed,
the $\gamma$ value will retain its limiting value of unity, so no
affect will be present in the work here. The change will also not be
required in a truly `razor thin' 2D model, where motion is restricted
in the third dimension entirely. In this case, the effective $\gamma$
would increase instead, due to a decrease in the number of internal
degrees of freedom for the gas.

Vertically integrated quantities also require special consideration in
realization and interpretation of gravitational forces in the system,
even though they retain their familiar inverse square form. As noted
in section \ref{sec:our-softening}, gravitational forces obtained from
a straightforward carryover of the 3D method of kernel softening
exhibit a non-zero force at zero separation. From a physical
perspective, such a condition will simply be erroneous, since in
reality the disk mass described by the particles is spread over some
vertical extent: it is no longer `thin' compared to interparticle
separations. Where in reality, the force of one vertical column on
another falls to zero at zero separation, the force based on a
vertically integrated mass located at the disk midplane does not. 

From a numerical perspective, this condition challenges the assumption
that the particles are collisionless, in turn the assumption that
softening was introduced to ensure. While considerably weakened,
consequences will be less severe then in the 3D case because the
interaction force does not become infinite at any separation, as it
would if two truly collisional particles were to interact. We may
therefore attempt to salvage the gravitational forces via some simple
modification, as we suggest with the altered softening prescription
defined in equation \ref{eq:mhat-2d}, or the altered description of
pressure forces derived from equation \ref{eq:mod-kerderiv}.

Care must still be taken using either modification in any 2D
simulation. Interparticle forces will deviate from their vertically
integrated forms over spatial scales comparable to the scale height of
the disk. At extremely high resolution, where particles correspond to
very thin columns of finite extent in the vertical direction, the
softening or kernel derivative modifications will affect the
calculated forces only over a small fraction of that distance. At
separations between the particle size, $h$, and the scale height, $H$,
forces will therefore be overestimated. Numerical experimentation has
shown that deviations will not be large at resolutions where
$h\ga0.1H$, but become much more significant when $h\sim0.01H$.
Simulations presented here do not fall into the latter category.

\section{Test problems for determining the required resolution of
simulations obeying the Toomre criterion}\label{sec:testing}

What resolution is required (i.e. what values of $T$ and \nreso\ from
equations \ref{eq:Toom-cond} and \ref{eq:max-surfdens}) to ensure that
a simulation that produces collapsed objects in a disk is producing
numerically valid results? As was done to develop criteria for Jeans
collapse, we will define a specific problem on which to compare the
results of a numerical code at different resolutions. In parallel, 
we will investigate the influence of three different strategies for
the treatment of small scale interactions between particles: a `base'
version and two variants that modify the kernel derivative in one
case or the kernel softening in the other.

We use a small variation (see below) of a simulation discussed in
\citet{DynI}, who used an SPH code to model the evolution of disks in
two dimensions. Simulations using SPH are especially sensitive to
violation of a resolution criterion because resolution is dynamically
allocated. Particle smoothing lengths are ordinarily considered to be
functions of the local flow variables so that in high density regions,
they shrink in an attempt to follow the small scale motions of the
fluid there. In most respects, this feature can be extremely desirable
because there is no {\it a priori} reason to expect fragmentation in
one or another part of a given simulation. On the other hand and as we
show below, insufficient care in its use can lead to numerically
induced fragmentation.

We use the VINE code (Wetzstein \etal, Nelson \etal, in preparation)
in its `SPH only' mode and using its leapfrog integrator to perform
our simulations. An earlier version of this code, with a second order
Runge-Kutta integrator, was used in the original calculations in
\citealt{DynI}. Exploratory tests with VINE using this same integrator
in the present simulations showed that similar results were obtained
from both. Most calculations were performed with a single, global time
step for all particles in order to assure that our results were
unaffected by as few systematic effects as possible. Some tests were
made with individual time steps in order to explore effects of
numerical stability due to this source. These calculations required
$\sim3-5$ times less computer time to complete and, while some
differences between global and individual time step versions were
present in the results, none materially affect the conclusions made
from them. VINE uses a binary tree to organize particle data, so that
they may be accessed efficiently for use in both the hydrodynamic and
the gravitational force calculations. In order to avoid calculation
times for the gravitational forces of order ${\cal O}(N^2)$,
sufficiently distant particles are approximated as nodes in the tree,
resolved to quadrupole order in the actual calculation. The
acceptability criterion for the nodes was set so that forces on
$\ga99$\% of particles would be accurate to $\la0.1$\%. 

VINE employs an artificial viscosity with both bulk and
von~Neumann-Richtmyer terms to stabilize the evolution and to convert
kinetic energy into thermal energy in shocks. The coefficient for each
term were set to $\alpha=1$ and $\beta=2$, which are values standard
in the literature. Using these values, simulations of disks using SPH
are afflicted with a quite large and unphysical level of shear
viscosity. In order to minimize such effects, the simulations here
were run with the shear viscosity reduction switch of \cite{Bals95},
which reduces the magnitude by a factor $\sim3-5$.

Important parameters from the simulations presented here and in the
following sections are listed in Table \ref{tab:sims}\footnote{Note to
MNRAS latex programmer/copy editor: It would be nice if the reference
to the table automatically came out correctly: There is only one table
in this paper, but latex calls it table 3 here and elsewhere in the
text, and table 1 in its definition.}. The columns of the table show
the resolution of each simulation, the type of gravitational softening 
used (for SPH simulations) and, finally, the time at which the time 
at which the first clump is formed and the duration of the simulation.
Simulation names ending in `TC' and `grv' denote simulations run with
the modified kernel derivative of equation \ref{eq:mod-kerderiv} or
the modified gravitational softening of equation \ref{eq:mhat-2d},
respectively, while those suffixes represent identical initial
conditions but run with the unaltered kernel derivative or softening.
All simulations are performed in 2d and include the effects of self
gravity, except the {\it sgoff2d4} simulation in 2D which does not,
and the remaining {\it sgoff} simulations, discussed in section
\ref{sec:disks-3d}, which also do not include self gravity and are
done in 3D.

\begin{table}\label{tab:sims}
\caption{Simulation Parameters}
\begin{tabular}{lrclll}
\hline
Label  & Resolution  &  Softening  & 
                                    T$_{\it first}$  & T$_{\it end}$  \\
\hline
mod1    &  7944                     &  Var.                         & 0.83\td           & \phantom{0}1.2\phantom{0}\td \\
mod2    &  32122                    &  Var.                         & 1.07\td           & \phantom{0}1.6\phantom{0}\td \\
mod3    &  129384                   &  Var.                         & 5.91\td           & \phantom{0}6.0\phantom{0}\td \\
mod4    &  260213                   &  Var.                         &  ---              &           12.0\phantom{0}\td \\
mod1TC  &  7944                     &  Var.                         & 1.15\td           & \phantom{0}1.5\phantom{0}\td \\
mod2TC  &  32122                    &  Var.                         & 1.28\td           & \phantom{0}1.75\td           \\
mod3TC  &  129384                   &  Var.                         &  ---              &           12.0\phantom{0}\td \\
mod4TC  &  260213                   &  Var.                         &  ---              &           12.0\phantom{0}\td \\
fix1TC  &  32122                    &  \phantom{0}.055AU            & 0.06\td           & \phantom{0}0.17\td           \\
fix2TC  &  32122                    &  \phantom{0}.40\phantom{0}AU  & 2.15\td           & \phantom{0}3.0\phantom{0}\td \\
fix3TC  &  32122                    &  \phantom{0}.65\phantom{0}AU  &   ---             & \phantom{0}5.0\phantom{0}\td \\
fix4TC  &  32122                    &  1.1\phantom{00}AU            &   ---             & \phantom{0}5.0\phantom{0}\td \\
m1fxTC  &  260213                   &  \phantom{0}.025AU            & 0.52\td           & \phantom{0}0.87\td           \\
m4fxTC  &  260213                   &  \phantom{0}.14\phantom{0}AU  &  ---              &           12.0\phantom{0}\td \\
mod1grv &  7944                     &  Var.                         & 0.95\td           & \phantom{0}1.27\td           \\
mod2grv &  32122                    &  Var.                         & 2.28\td           & \phantom{0}2.5\phantom{0}\td \\
mod3grv &  129384                   &  Var.                         &  ---              &           12.0\phantom{0}\td \\
mod4grv &  260213                   &  Var.                         &  ---              &           12.0\phantom{0}\td \\
fix1grv &  32122                    &  \phantom{0}.055AU            & 0.07\td           & \phantom{0}0.20\td           \\
fix2grv &  32122                    &  \phantom{0}.40\phantom{0}AU  & 1.42\td           & \phantom{0}1.9\phantom{0}\td \\
fix3grv &  32122                    &  \phantom{0}.65\phantom{0}AU  &   ---             & \phantom{0}5.0\phantom{0}\td \\
fix4grv &  32122                    &  1.1\phantom{00}AU            &   ---             & \phantom{0}5.0\phantom{0}\td \\
m1fxgrv &  260213                   &  \phantom{0}.025AU            & 0.29\td           & \phantom{0}0.50\td           \\
m4fxgrv &  260213                   &  \phantom{0}.14\phantom{0}AU  & 2.20\td           & \phantom{0}2.50\td           \\
sgoff2d4&  260213                   &  Var.                         &   ---             &           12.0\phantom{0}\td \\
sgoff3  &  129384                   &  Var.                         &   ---             & \phantom{0}2.0\phantom{0}\td \\
sgoff4  &  260213                   &  Var.                         &   ---             & \phantom{0}2.0\phantom{0}\td \\
sgoff5  &  502089                   &  Var.                         &   ---             & \phantom{0}2.0\phantom{0}\td \\
sgoff6  &  994740                   &  Var.                         &   ---             & \phantom{0}2.0\phantom{0}\td \\
Boss    &  100$\times$23$\times256$ &  Grid                         & 345yr             & 359~yr  \\
\hline
\end{tabular}
\end{table}

\subsection{The definition of our test problem}\label{sec:2dtest-defn}

\citet{DynI} modeled the evolution of self gravitating disks in 2D
with masses between 0.05 and 1.0 times the mass of the central star,
using SPH. Other simulations from that work, performed using PPM, are
not considered here because they could not be carried out far enough
into the high amplitude regime to make fragmentation likely. The
specific model we consider here (labeled `scv2' in that work), had an
assumed disk mass of \mdisk=0.2$M_*$ and a minimum Toomre $Q$ value
defining its stability of \qmin=1.5. We believe that model will be a
particularly challenging test of the criterion because collapse was
observed after only about one orbit of the outer disk edge,
corresponding to about 11 orbits in the region where clumps first
started to form. By the conclusion of the run at 1.6\td, more than
30 clumps had formed. As in the originals, we have simulated the
evolution in two dimensions, so that the surface density is directly
available from the calculation. 

The temperature and surface density of the gas were defined in that
model with softened power laws as:
\begin{equation}\label{eq:dyn1-temp}
T(r) = T_0\left[ 1 + \left( {{r}\over{r_c}}\right)\right]^{-q/2}
\end{equation}
and
\begin{equation}\label{eq:dyn1-sdens}
\Sigma(r) = \Sigma_0\left[ 1 + \left({{r}\over{r_c}}\right)\right]^{-p/2}
\end{equation}
where $q=1/2$ and $p=3/2$, respectively, and the core radius for both
power laws was set to $r_c=1$. The constants $T_0$ and $\Sigma_0$ were
determined from the assumed Toomre stability parameter and the radial
dimensions of the disk, defined at its inner edge by $R_I=0.5$ and its
outer edge by $R_D=50$. Particles were laid out on concentric rings
and an equilibrium state accounting for stellar and disk self gravity
as well as pressure forces defined the velocities of each particle.
The gas was evolved under the forces of stellar gravity, self gravity
and gas pressure, which was computed using an isothermal equation of
state (i.e. $\gamma=1$ with fixed temperature as a function of
radius). Because of the simple physical assumptions, the dimensions of
the system were scalable. Given a star of one solar mass and a disk
radius of 50~AU, one orbit at the outer edge of the disk requires
approximately 353~yr\footnote{Note that these units are slightly
different than those used in \citet{DynI}, where a star of 0.5\msun\
was assumed in order to correspond more closely to typical T Tauri
star masses.}.

During the preparation of this manuscript, we determined that
simulations evolved using initial conditions identical to those in
\citet{DynI} were not possible because at high resolution the large
pressure gradient at the disk's outer edge caused unphysical behavior
in the system. In order to sidestep this problem, we modified the form
of the surface density power law near the outer disk edge so that the
discontinuity is spread over a larger radial range. In this work, the
surface density is distributed according to:
\begin{equation}\label{eq:sdens}
\Sigma_{mod}(r) = S\Sigma(r)
\end{equation}
where the factor, $S$, is a linearly decreasing function near the disk
boundary and is defined by:
\begin{equation}\label{eq:sdens-alter}
S = \cases{  1                          & for $r<R_D-\delta$ ;\cr
             1 - {{(r-(R_D - \delta))}\over{2\delta}}
                                        & for $R_D-\delta<r
                                                     <R_D+\delta$;\cr
             0                          & for $r>R_D+\delta$ ;\cr }
\end{equation}
With this definition, the disk edge is smoothed in the region within a
distance $\delta$ inward and outward of the nominal disk radius. In
the simulations here, we define the smoothing parameter $\delta=5$~AU.
The other initial conditions and physical assumptions remained the
same.

\subsection{Evolution of our test simulations}\label{sec:2dtest-evo}

\begin{figure*}
\includegraphics[width=16cm]{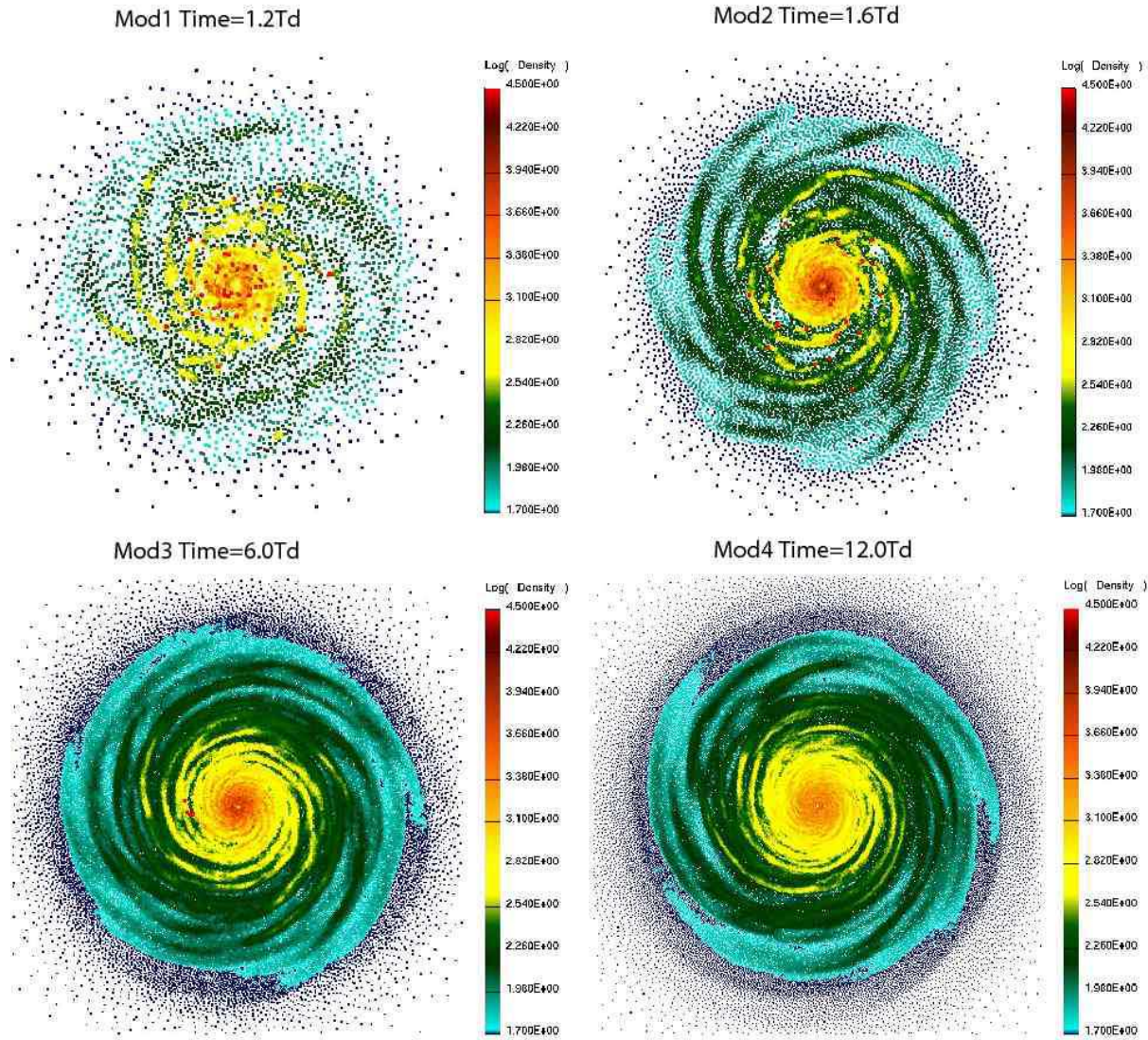}
\caption
{\label{fig:Dyn1-morph}
The particle distribution for the {\it mod} series of simulations at
the end of each simulation. At the times shown, filamentary spiral
structures have developed throughout the disks and are visible as
azimuth varying blue, green and yellow color variations in the images.
The three lowest resolution realizations have also produced a number
of clumps (visible as red dots in the image) typically containing as
many as several hundred individual SPH particles. Clumps in the {\it
mod3} realization are visible at about the 9 o'clock position in the
figure, while in the two lowest resolutions, they are distributed
throughout. The highest resolution realization produced no clumps. The
color scale shown defines the base 10 logarithm of the surface
density.}
\end{figure*}

\begin{figure*}
\includegraphics[width=16cm]{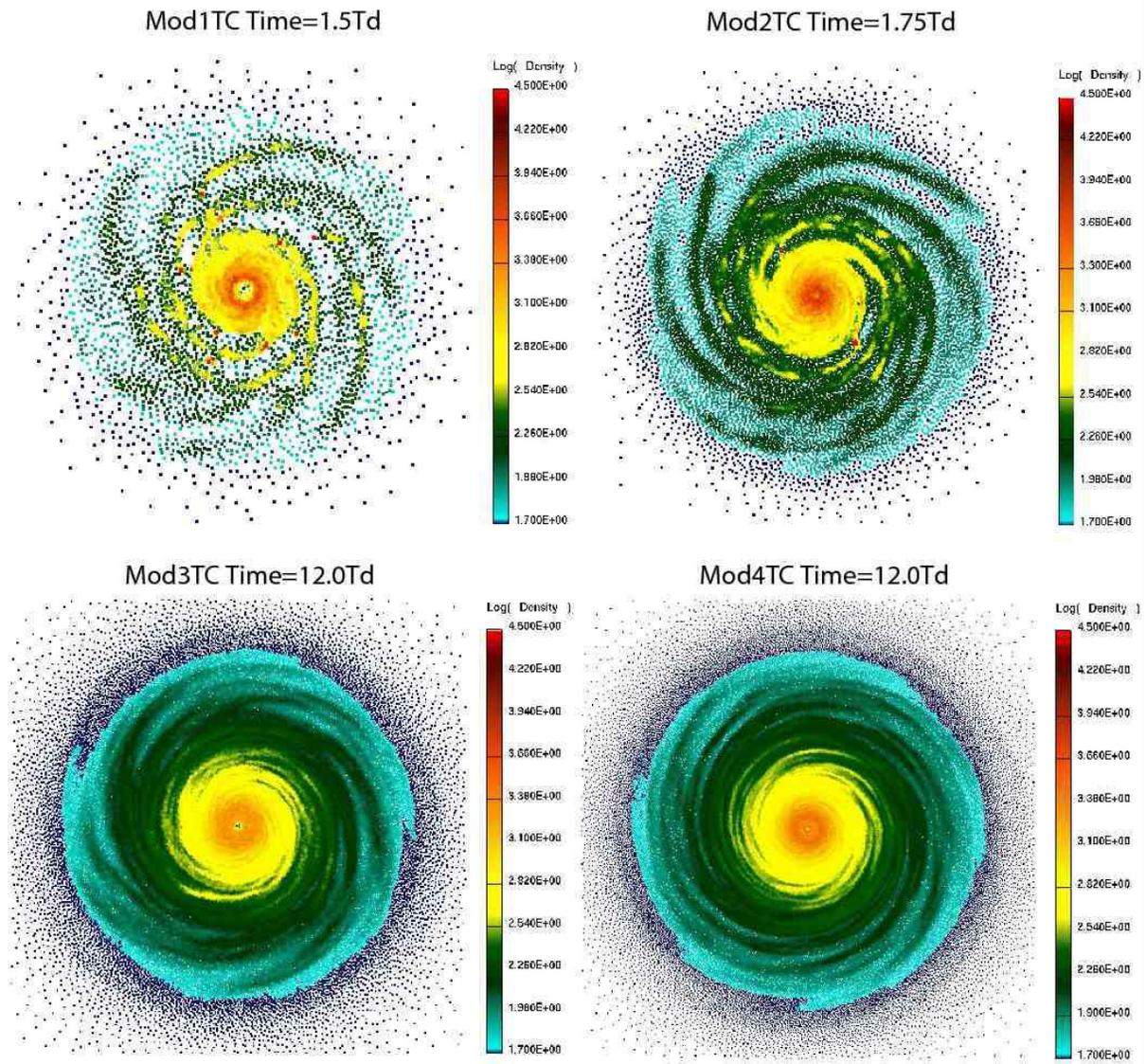}
\caption
{\label{fig:Dyn1-morph-modk}
The particle distribution for the {\it mod} series of simulations with
the TC92 kernel derivative, each at the end of each
simulation. As in figure \ref{fig:Dyn1-morph}, filamentary spiral
structures have developed throughout the disks and are visible as
azimuth varying blue, green and yellow color variations in the images,
with clumps (in the two lowest resolution realizations) visible as
red dots in the image. The color scale is the same as in
\ref{fig:Dyn1-morph} }
\end{figure*}

\begin{figure*}
\includegraphics[width=16cm]{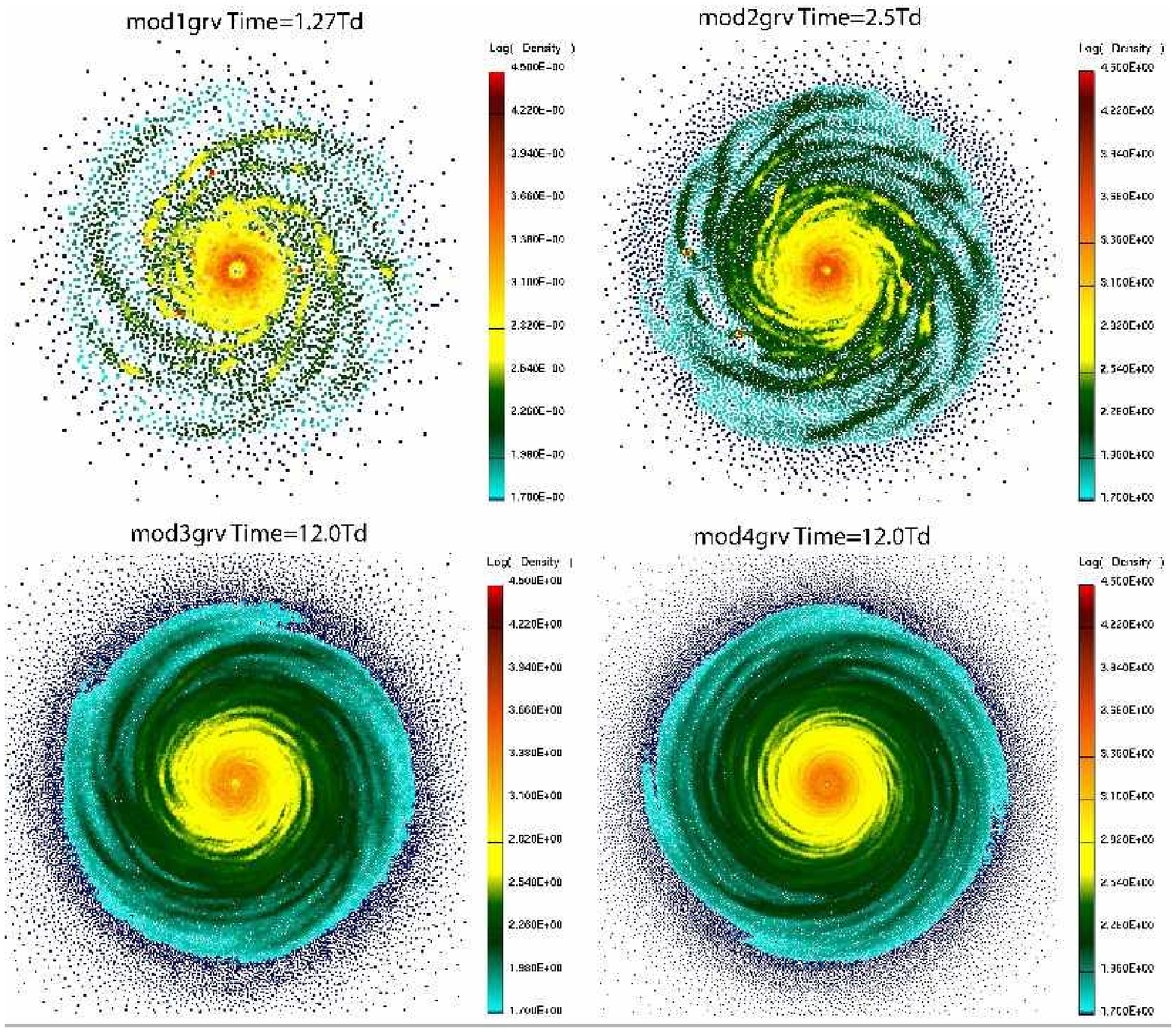}
\caption
{\label{fig:Dyn1-morph-modsoft}
The particle distribution for the {\it mod} series of simulations with
the alternate gravitational softening, each at the end of each
simulation. As in figure \ref{fig:Dyn1-morph}, filamentary spiral
structures have developed throughout the disks and are visible as
azimuth varying blue, green and yellow color variations in the images,
with clumps (in the two lowest resolution realizations) visible as
red dots in the image. The color scale is the same as in
\ref{fig:Dyn1-morph} }
\end{figure*}

Using the conditions defined above, we ran a set of four simulations
with varying resolution using each of three treatments for the small
scale interactions between particles, but otherwise identical. The
first treatment, with simulations labeled {\it mod1-mod4} in table
\ref{tab:sims}, uses gravitational softening of the form defined by
equation \ref{eq:mass-kernsoft} and the standard kernel gradient
defined derived directly from equation \ref{eq:SPHkern}. The second
treatment, with simulations labeled {\it mod1TC-mod4TC}, uses the
modified kernel derivative defined by equation \ref{eq:mod-kerderiv}
to determine mutual pressure forces between particles, again with the
standard kernel softening. The third treatment, with simulations
labeled {\it mod1grv-mod4grv}, uses the modified gravitational
softening defined in equation \ref{eq:mhat-2d} with the standard
kernel gradient. Figures \ref{fig:Dyn1-morph},
\ref{fig:Dyn1-morph-modk} and \ref{fig:Dyn1-morph-modsoft} show these
four realizations of the model, each at the termination of the
simulation.

Each model develops multiple armed spiral structures over most of its
radial extent, as in the previous \citet{DynI} work. The spiral
structures are filamentary and change the details of their appearance
as the simulation proceeds. Evolution after the formation of the
spiral structure however, was strongly dependent on the resolution
employed. In evolution of the three lower resolution realizations of
the {\it mod} series, spiral structures eventually fragmented into
multiple clumps, as in the \citet{DynI} work. The time at which clumps
begin to form, measured from the beginning of the simulation, was
later at higher resolution than at lower. Moreover, fewer clumps
formed in the higher resolution realizations than in the lower, and in
the highest resolution realization, no clumps were produced at all:
the delay before clumps begin to form has become longer than the
duration of the simulation (12\td, or about 4200~yr). 

Similar statements are true of both the {\it modTC} and {\it modgrv}
runs, though there are a number of important differences as well. Of
particular interest is the fact that the structure seen in figures
\ref{fig:Dyn1-morph-modk} and \ref{fig:Dyn1-morph-modsoft} is
substantially smoother than in the corresponding panels of figure
\ref{fig:Dyn1-morph}. The overall relative smoothness is also
reflected in the number of clumps that form in each realization. Where
the {\it mod3} simulation formed several clumps, simulations with the
two modified treatments {\it mod3TC/mod3grv} did not. Although each of
the corresponding lower resolution realizations did form clumps, the
number that formed was smaller in each case and delayed in time
relative to the standard kernel versions. 

\begin{figure*}
\rotatebox{-90}{
\includegraphics[width=13cm]{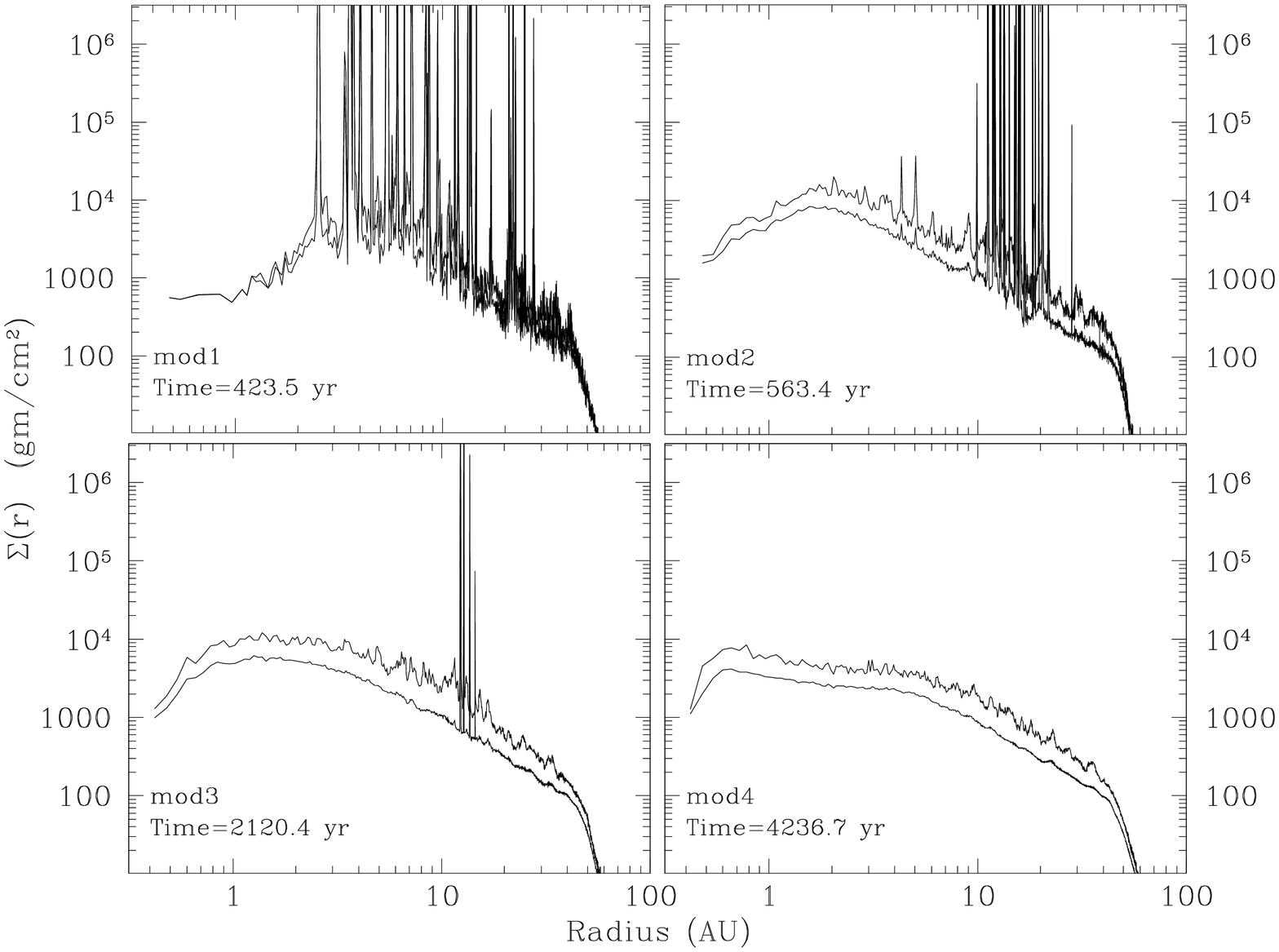}}
\caption
{\label{fig:Dyn1-dens} 
The azimuth averaged and instantaneous maximum surface densities
plotted as functions of distance from the star for the test model
derived from a simulation from \citet{DynI}, realized at four
resolutions (the `{\it mod} series of simulations--see table
\ref{tab:sims}), each at the end of the simulation. Each panel in this
figure corresponds to the same panel as in figure
\ref{fig:Dyn1-morph}.} 
\end{figure*}

\begin{figure*}
\rotatebox{-90}{
\includegraphics[width=13cm]{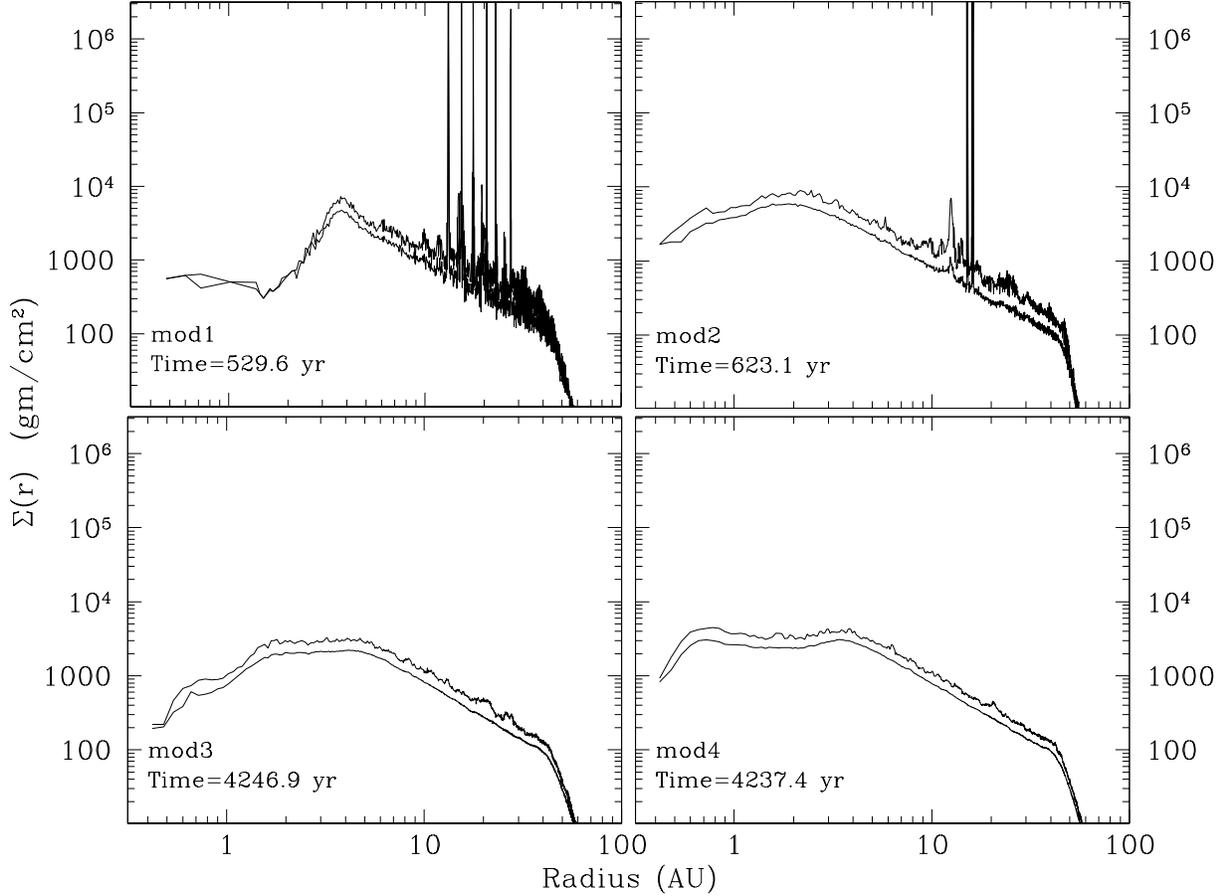}}
\caption
{\label{fig:Dyn1-dens-modk} 
The same as figure \ref{fig:Dyn1-dens}, but for the TC versions 
of the simulations. Each panel in this figure corresponds to the 
same panel as in figure \ref{fig:Dyn1-morph-modk}. } 
\end{figure*}

\begin{figure*}
\rotatebox{-90}{
\includegraphics[width=13cm]{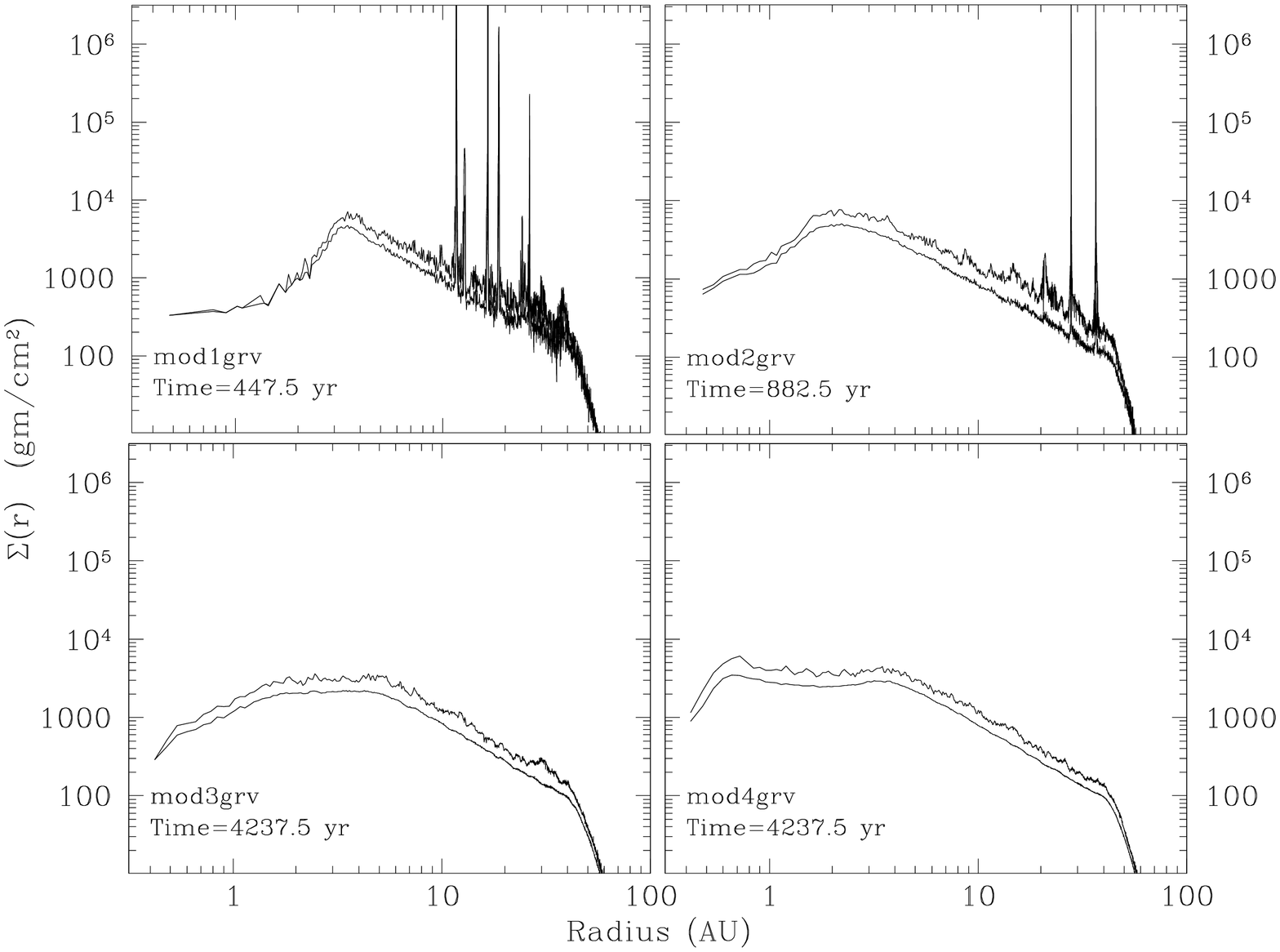}}
\caption
{\label{fig:Dyn1-dens-modsoft} 
The same as figure \ref{fig:Dyn1-dens}, but for the alternate
gravitational softening versions of the simulations (`grv'). Each
panel in this figure corresponds to the same panel as in figure
\ref{fig:Dyn1-morph-modsoft}. } 
\end{figure*}

Figures \ref{fig:Dyn1-dens}, \ref{fig:Dyn1-dens-modk} and
\ref{fig:Dyn1-dens-modsoft} show azimuth
averaged, radially binned surface density profiles of the same models
as shown in figures \ref{fig:Dyn1-morph}, \ref{fig:Dyn1-morph-modk}
and \ref{fig:Dyn1-morph-modsoft}, at the same times. Also shown are
instantaneous maximum surface densities, defined as the maximum in
each ring seen at the time shown. The average in each bin is weighted
by the number of particles in the bin, with the width of each bin set
to 0.02~AU. At small radii, accretion onto the star has depleted the
initial power law density distribution, in each case by a differing
amount due to the differing magnitude of the dissipation derived from
the artificial viscosity included in the simulation (see \citet{DynII}
for a discussion of this effect). Further out, a number of spikes are
visible in the profiles, each corresponding to the radial location of
a clump in the disk. The lowest resolution realizations display a very
large number of clumps, while successively higher resolution
realizations produce fewer, or none. Both the number of clumps that
formed and the radial extent over which they are found are smaller
when we use the two modified kernel treatments, compared to the
standard form. Though not possible to see in snap shots of a single
time, an important difference between the evolution with the standard
and either of the modified kernel treatments is that the cumulative
maxima in the latter case remain far lower than the former and also do
not appear to increase over time. We believe that this near steady
state behavior represents the correct evolution of the system, even if
it were to be evolved further in time. 

\begin{figure*}
\rotatebox{-90}{
\includegraphics[width=13cm]{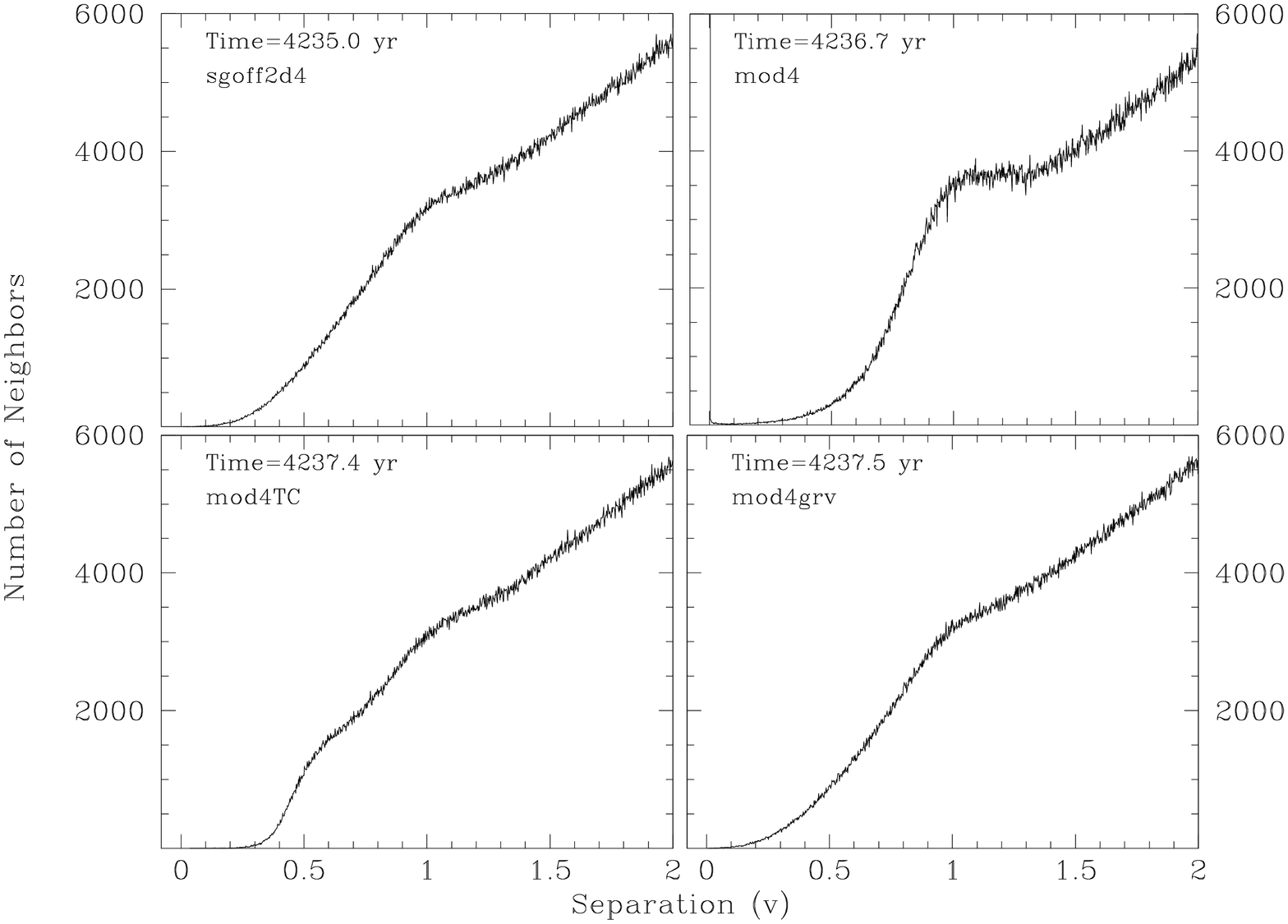}}
\caption{\label{fig:neigh-hist} 
A histogram of the number of SPH neighbor particles found at a given
separation, for four variants of the {\it mod4} simulation, as
labeled. The separation is normalized to the dimensionless variable,
$v=r/h$, and the histograms account for all neighbors of all particles
in each simulation.}
\end{figure*}

What is the origin of the differences in the morphology between each
series and the others? We postulate that it lies in the treatment of
the small scale interactions between particles, and specifically in
force imbalances between pressure and gravity that may be present. If
so, we may expect distortions in the distribution of interparticle
separations, relative to what may be expected and to what may be
present with other realizations. Figure \ref{fig:neigh-hist} shows
histograms of the distances between all neighbor particles for each of
the three realizations of the {\it mod4} simulations, as well as a
simulation with the same resolution and initial condition, but without
self gravity. Each histogram bin is defined to be of width $\delta v=
0.002$, and all neighbors for all particles in each simulation are
accounted for in the histograms. On the scale of the smoothing length,
we can approximate the surface density as constant. For a uniform
density distribution in a 2D simulation, realized with a random
distribution of particles, we expect that the number of neighbors to
be found at a given separation to increase linearly with the
separation itself. At large separations, we recover the linear
behavior. At smaller separations, all four distributions depart from
the linear proportionality because the assumption underlying the
linear proportionality neglects the influence of interparticle forces.
Due to the different treatments of the small scale interactions, some
differences also appear in the neighbor distributions.

The most striking feature of the plot however is the behavior near
$v=0$. In the standard case, a large number of particles actually
coincide in space, while in each of the other three variants, none do.
This coincidence represents an extreme example of Herant's pairing
instability discussed above, that develops due to the interparticle
pressure vs. gravitational force imbalance at small separations. The
pairing instability seen here is clearly not identical to that
discussed by Herant however, since no pairing is present in the non
self gravitating realization for which conditions are most similar to
those discussed by Herant. The pairing is also not similar to the
numerical instabilities seen by \citet{II02} for the same reason.

\begin{figure}
\rotatebox{-90}{
\includegraphics[width=6.5cm]{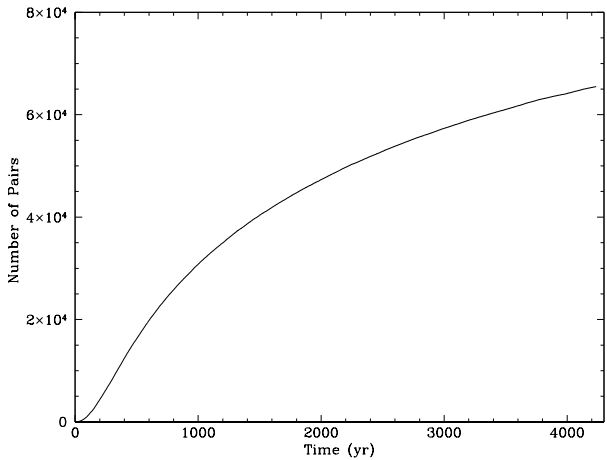}}
\caption {\label{fig:neig-pairs}
The number of particle pairs in simulation {\it mod4} as a function of
time. Corresponding figure for the other {\it mod4} variants are not
shown, since no particles become paired during their evolution. } 
\end{figure}

Figure \ref{fig:neig-pairs} shows that the number of paired particles
continues to grow as the simulation proceeds. After 12\td, more than
65000 pairs have formed from $>130000$ particles: of the $\sim 260000$
original particles, about half have become paired. The existence of
paired particles means first that the effective resolution of the
simulation decreases with time as more and more particles become
paired, and is consistent with our qualitative observation above that
the cumulative maxima in these simulations also increased. Their
presence also means that even though the force contribution from a
single neighbor particle is small relative to the individual
contributions from the rest of the system, when the particles approach
coincidence the pairwise contribution can have an important and large
scale influence on the behavior of the simulation. Based on these
results, and the impact they have on the large scale disk morphology
and its tendency to fragment, we conclude that using the standard
kernel based gravitational softening in combination with the standard
kernel derivative is likely to produce results contaminated with
numerical artifacts in 2D simulations, and should not be used. 

\subsection{Determining the resolution criterion}\label{sec:test-crit}

For each of the three series' of simulations, the propensity of
simulations to fragment decreased as resolution increased, until at
sufficiently high resolution no fragmentation occurred at all. This
behavior reflects that seen by \citet{Truelove97}, where fragmentation
was enhanced when resolution was insufficient, rather than that seen
by BB97, where fragmentation was delayed. This is fortunate because it
allows us to use the change in behavior of realizations of the same
initial condition but differing resolution to determine empirically
the approximate resolution (in number of particles) necessary to
obtain `correct' evolution. Specifically, we can apply the criterion
to two simulations which straddle the boundary between those that
produced fragments and those that did not. We can then note the value
of \nreso\  at which the criterion succeeds for the entire simulation
at the higher resolution, but fails for the lower resolution version.
BB97 scaled the value of the Jeans resolution criterion through the
value of \nreso\  as a multiple of the average number of neighbors,
\nneigh. It will be convenient to scale the two dimensional criterion
similarly here although in both cases, the correct measure will be a
quantity independent of the specific neighbor count. We also note that
in two dimensional SPH simulations, it is usual to use a smaller
number of neighbors for each particle due to the lower dimensionality.
In this work, we have used a number ranging between 10 and 30,
depending on the local flow \citep[see][for details]{Benz90}. We
therefore scale by a factor \nneigh$=20$ for these simulations.

\begin{figure}
\includegraphics[width=8cm]{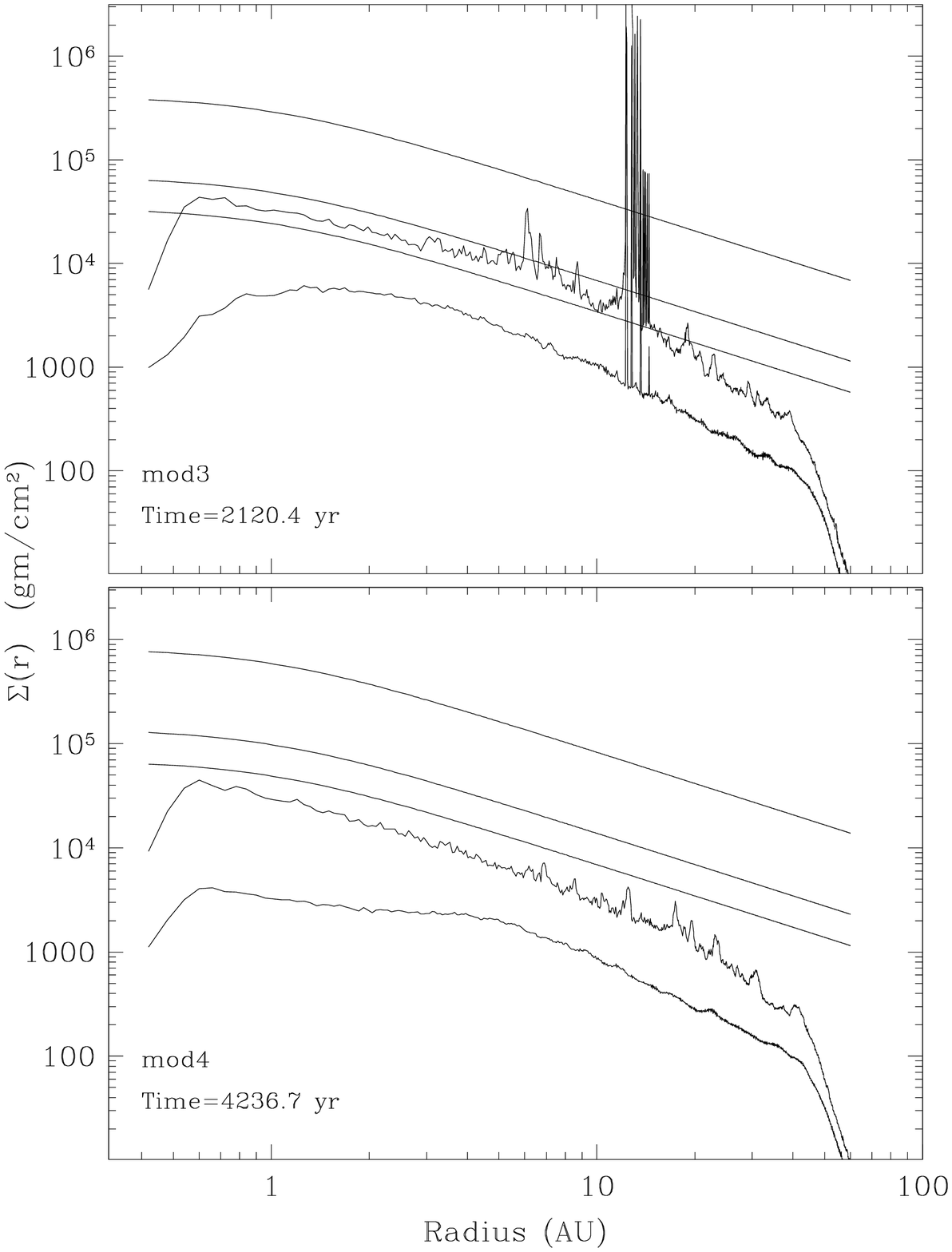}
\caption
{\label{fig:Dyn1-crit} 
The cumulative maximum surface density reached at each point in the
disk, binned as a function of radius in the {\it mod3} and {\it mod4}
simulations. For reference, the azimuth averaged surface density shown
in figure \ref{fig:Dyn1-dens} as well (lower curve). The three smooth
curves show the condition from equation \ref{eq:max-surfdens} with a
value of \nreso\  of 20, 120 and 240 particles (top to bottom in each
panel).} 
\end{figure}

Figure \ref{fig:Dyn1-crit} shows the cumulative maximum surface
density binned as a function of radius, along with the maximum
resolvable surface densities determined from equation
\ref{eq:max-surfdens} using three values of \nreso\ equal to 1, 6, and
12 times the average number of neighbors for SPH particles evolved in
2D (i.e. 20). The cumulative maximum for each given bin is defined as
the maximum surface density achieved by any particle in that bin over
the entire course of the simulation up until the time shown.

For the {\it mod3} realization, the cumulative maximum exceeds the
critical value using the \nreso$=$12\nneigh\  criterion for all radii
inside $\sim20$~AU. At the same time, the \nreso$=$6\nneigh\ criterion
is not violated except at a few localized radii. Interestingly, a
small density spike relatively early in the simulation near 7~AU did
not lead to clump formation, but later interactions near 10-15~AU did.
For the {\it mod4} simulation, the \nreso$=12$\nneigh\ condition is
satisfied over the entire radial range for the life of the simulation,
and no clumps formed. We can make only a tentative assignment of the
value of \nreso\  from this result however because of the importance
particle pairing may have on the effective resolution. 

\begin{figure*}
\rotatebox{-90}{
\includegraphics[width=13cm]{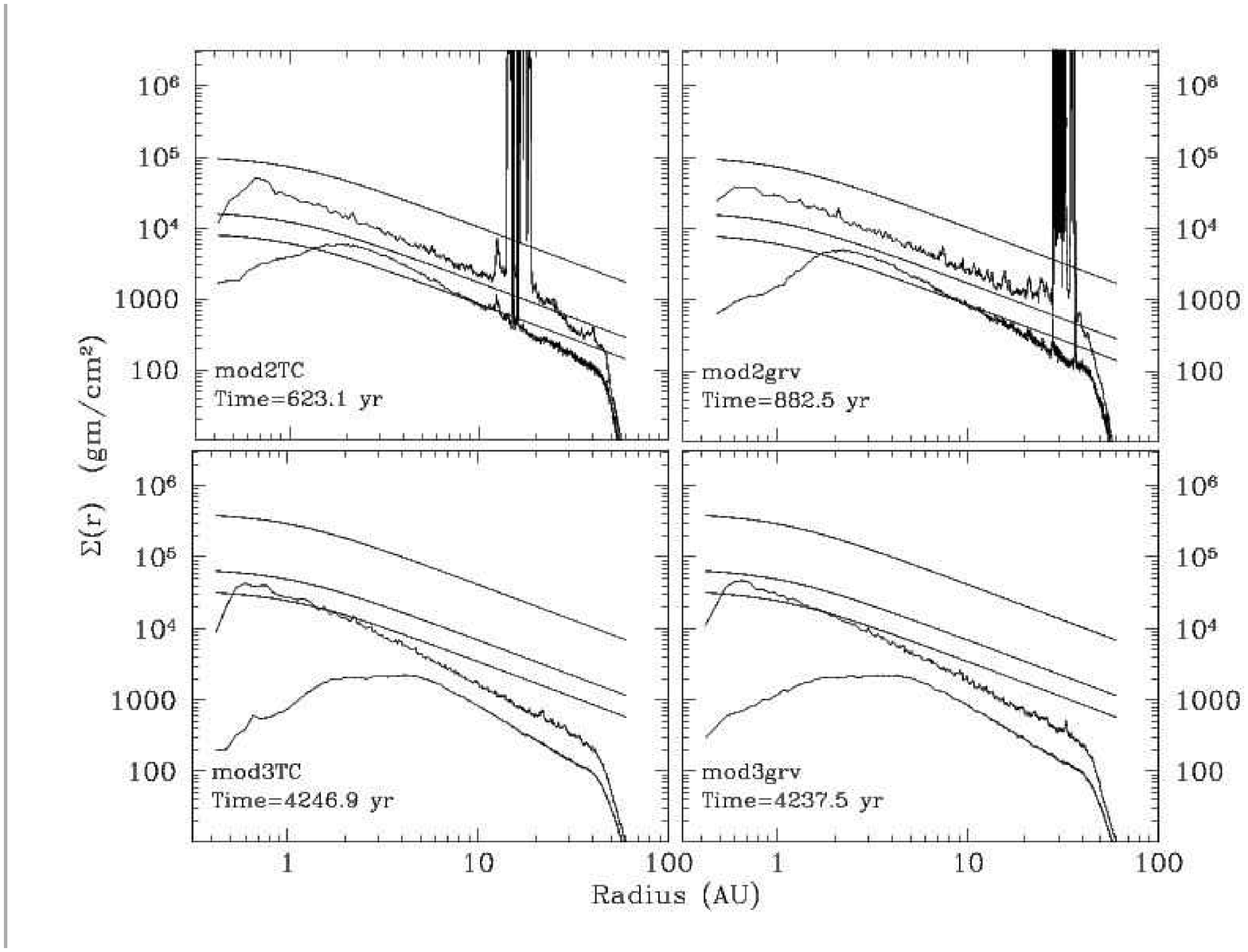}}
\caption
{\label{fig:Dyn1-crit-TCgrv} 
The cumulative maximum surface density reached at each point in the
disk, binned as a function of radius in the {\it mod2TC/mod3TC} and
{\it mod2grv/mod3grv} simulations. As in figure \ref{fig:Dyn1-crit},
the azimuth averaged surface density shown in figure
\ref{fig:Dyn1-dens} as well (lower curve) and the three smooth curves
show the condition from equation \ref{eq:max-surfdens} with a value of
\nreso\  of 20, 120 and 240 particles (top to bottom in each panel).} 
\end{figure*}

Figure \ref{fig:Dyn1-crit-TCgrv} shows the cumulative maximum surface
densities for two of the simulations using the TC92 kernel derivative
and the modified gravitational softening. In this case we plot the
curves for {\it mod2TC/mod3TC} and {\it mod2grv/mod3grv} rather than
for {\it mod3} and {\it mod4} in order to maintain the straddle of the
fragmenting/non-fragmenting outcomes in these simulations. In both
variants, the cumulative maximum densities in the {\it mod2}
realizations exceed the critical value using the both the
\nreso$=$6\nneigh\ and \nreso$=$12\nneigh\ criteria, while the {\it
mod3} realizations obey the 6\nneigh\ criterion but violate the
12\nneigh\  criterion. Since no clumps formed during the evolution,
the latter must be considered too conservative, and we conclude that
the resolution required to avoid fragmentation due to unphysical
growth of self gravitating structures in the disk is \nreso$=$6\nneigh. 

At first sight, the required resolution for the Toomre condition
appears much larger in comparison to that for the Jeans condition
analysis discussed by BB97. The apparent paradox is resolved if we
note, as in section \ref{sec:consistent}, that their assumed value of
the Jeans mass was quite low and resulted in a relatively lower
neighbor requirement. Our definition of the Toomre mass corresponds to
an analogue of the larger definition of the Jeans mass defined as in
equation \ref{eq:sphere-wavel-Mjeans}.

The origin of the large required value of \nreso\  becomes clear when
we observe that the values of the cumulative maxima (or indeed, also
the slightly lower instantaneous maxima not shown) are typically a
factor of several higher than the averages. Most of the difference can
be accounted for by the existence of spiral structure in the disk,
however fluctuations due to the effect of the exact, time varying
positions of particles relative to each other on the calculation of
the density make up a smaller, but still significant component of the
difference. Such fluctuations are intrinsic to the SPH method itself
and, while a small decrease is observable between the low and high
resolution simulations in figure \ref{fig:Dyn1-crit}, their existence
will be a part of all SPH simulations.

It is interesting to note that the densities can exceed the resolvable
maximum in the lower resolution realizations for some time before
clump formation begins. Further, the amount of time before the onset
of clump formation is a resolution dependent quantity, with higher
resolution leading to more time before clump formation. It is
therefore clear that resolution studies are particularly important for
particle simulations showing evidence of clump formation. Applying
this statement to all three variants of our own {\it mod} series' of
simulations, it is clear that if clumps do indeed form in disks
similar to those studied, the process requires a time scale than longer
the 12\td\ ($\sim$4200 yr) for which we have evolved the system.

In the two lowest resolution realizations of all three variants (i.e.
{\it mod1/TC/grv} and {\it mod2/TC/grv}), the numerical stability
criterion is violated even in the initial condition. Since these
simulations are only slight modifications from those presented in
\citet{DynI} and at the same resolution, the same statement applies to
those simulations as well. Regarding the clump formation seen in the
simulations from that work, we originally concluded that although the
physical model (in particular the isothermal equation of state) was
insufficient to correctly model clump formation, if it was to occur at
all, it would be most likely in between about 10 and 40~AU. The
conclusion that the physical model in \citet{DynI} may be insufficient
may indeed remain valid \citep[see e.g.][]{DynII}). However, the
location and existence of clump formation is definitely invalid: it
was due purely to failure of the Toomre criterion and not to any
physical process, however modeled.

\section{Fixed vs. dynamically variable gravitational
softening in particle simulations}\label{sec:softening} 

After investigating the effects of resolution and choice of kernel on
disk simulations, in this section, we turn to an investigation of
gravitational softening. We first discuss factors that motivate the
selection of the value of a fixed softening length, then perform a
side by side comparison of otherwise identical simulations employing
either fixed or variable softening. Due to our conclusion in the last
section that the standard softening/smoothing combination should not
be used, we limit our investigations here to the TC92 and modified
gravitational softening variants.

\subsection{Choosing a fixed softening length}\label{sec:choose-fixed}

Unlike the case of variable softening, where the softening length is
set locally and dynamically, with fixed softening the modeler must
make a specific choice of the length that is appropriate for the
entire simulation at all times. Choosing an appropriate value in disk
models is substantially less trivial in the case of disk models than
in the cloud collapse scenario discussed by BB97 because, simply as a
consequence of the initial conditions (i.e. the density gradient as a
function of distance from the central object), the smoothing lengths
of particles are relatively steep functions of position. This is
important because for any constant choice of softening, the relative
magnitudes of softening to smoothing will also vary with position,
perhaps artificially suppressing fragmentation in one region while
enhancing it in another. Moreover, because the Toomre wavelength will
itself be a function of radius through both density and temperature,
both lengths will vary relative to it as well.

Naively, one might expect that variable softening could lead to
results that are quite susceptible to small scale fragmentation. For
example, if a region begins to collapse and the particles representing
it approach each other, their mutual gravitational attraction
continues to increase as their softening and smoothing lengths
decrease, perhaps instigating the very problem softening is meant to
avoid. It is not clear that such a condition exists in practice
however. Variable softening was implemented for both of our {\it mod}
series of simulations discussed in section \ref{sec:2dtest-evo}, as
well as our previous work in \citet{DynI,DynII}. These models did not
produce clumps, except as a consequence of insufficient resolution. On
the other hand, fixed softening has been strongly advocated by
\citet{Mayer02,Mayer04} and both their lower and higher resolution
simulations do produce clumps.

There are a number of questions that we must answer in order to
understand the implications of the softening choice and magnitude on
the outcome of a simulation. First, since we have no {\it a priori}
knowledge of where in the disk we may expect fragmentation to occur,
if it is to occur at all, we must decide how to choose a value for the
gravitational softening that allows the best reproduction of the real
system. Given such a choice, what is the difference between the
influence that fixed softening has on the simulation compared to
variable softening? For example, to what extent can one or the other
choice suppress fragmentation in models where it should not occur, but
which may be under resolved? To what extent are physically valid
evolutionary signatures also suppressed? Can one or the other choice
actually instigate fragmentation in models that are otherwise stable?
Can some values of fixed softening length both artificially suppress
and enhance fragmentation in different parts of the same disk? 

\subsection{The influence of fixed softening on fragmentation in
simulations with insufficient resolution}\label{sec:suppression}

In this section, we investigate the influence of particular fixed
values of the softening with a series of simulations (denoted the {\it
fix TC} and {\it fix grv} series in Table \ref{tab:sims}) that are
identical to the {\it mod2TC} and {\it mod2grv} simulations which we
have shown to be unstable to numerically induced fragmentation with
variable softening. These simulations implement a fixed softening
using the same SPH spline kernel as is used for the hydrodynamic
evolution, but are run, in the first case, with the modified kernel
derivative for the pressure force calculation or, in the second case,
with the modified 2D softening. Each of the four pairs of simulations
uses a different softening length corresponding to the magnitude of
the initial smoothing length at different locations in the disk. These
are, respectively, the inner disk edge (0.5AU), the orbit radius
corresponding to the region most susceptible to clump formation
(15AU), and near the middle of the radial span of the disk (30AU), and
its outer edge (50AU). 

\begin{figure*}
\includegraphics[width=16cm]{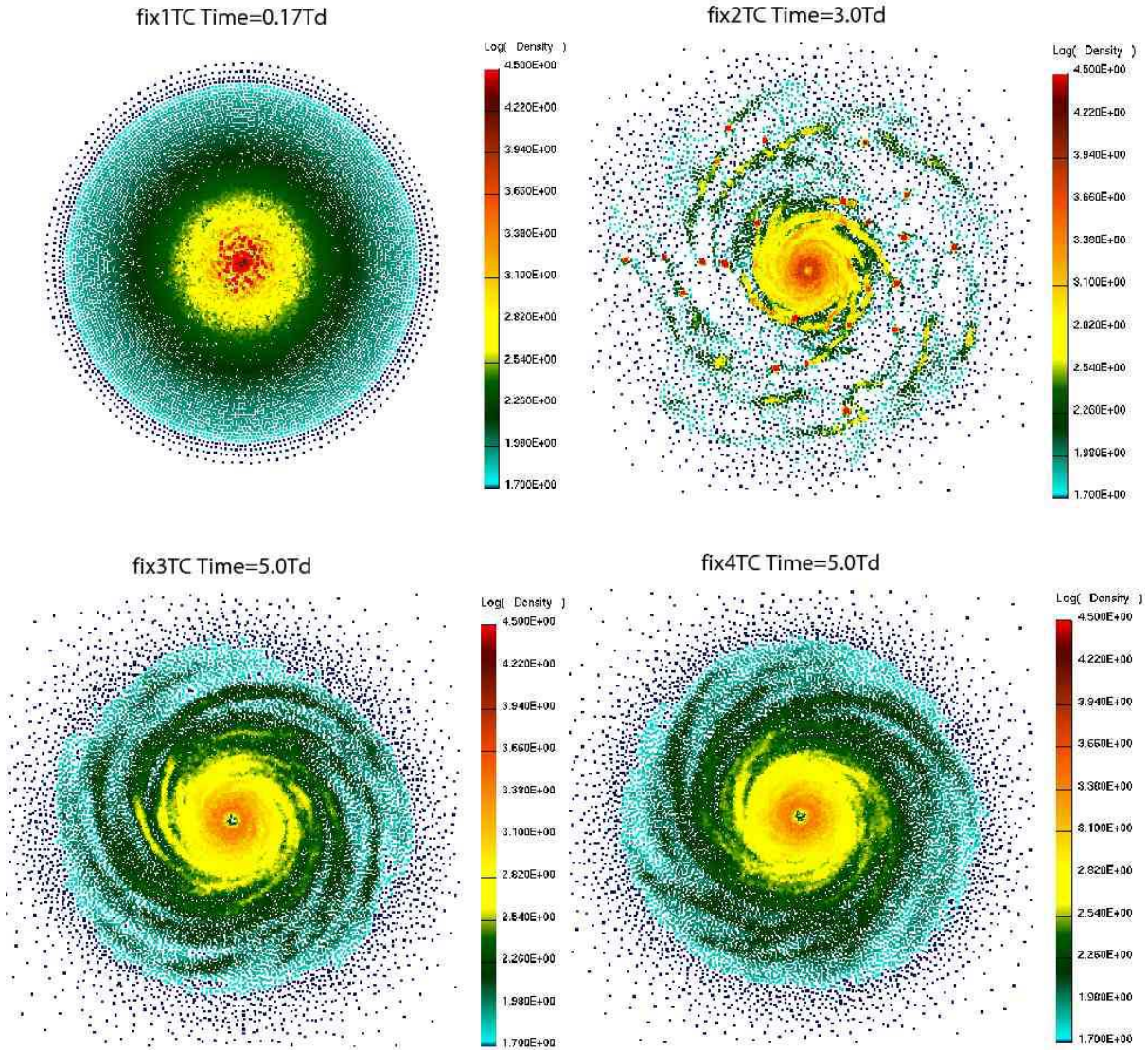}
\caption
{\label{fig:fix-morph-modk}
The particle distributions for the {\it fix TC} series of simulations,
at the end of each simulation.} 
\end{figure*}

\begin{figure*}
\includegraphics[width=16cm]{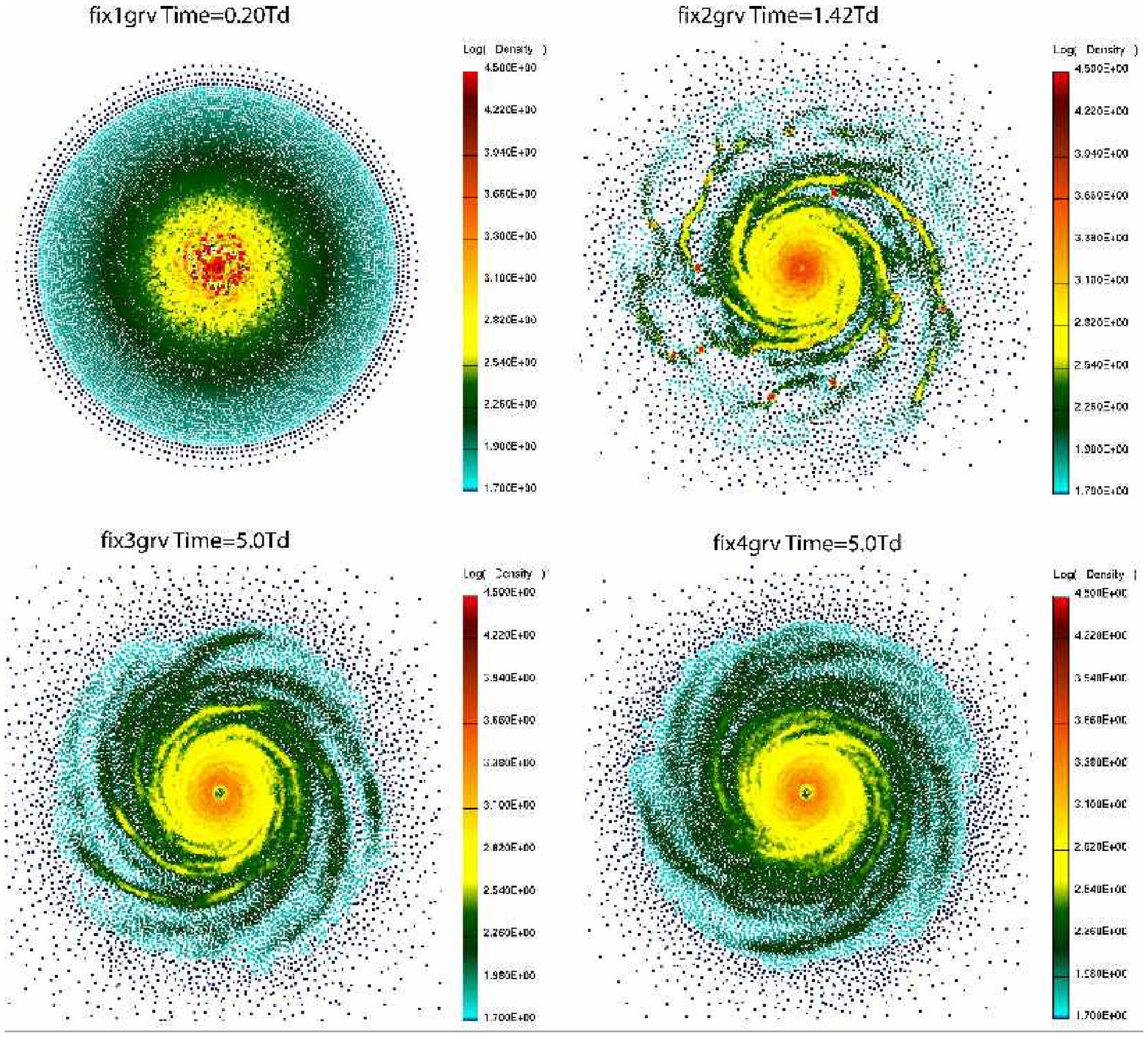}
\caption
{\label{fig:fix-morph-grv}
The particle distributions for the {\it fix grv} series of simulations,
at the end of each simulation.} 
\end{figure*}

\begin{figure*}
\rotatebox{-90}{\includegraphics[width=120mm]{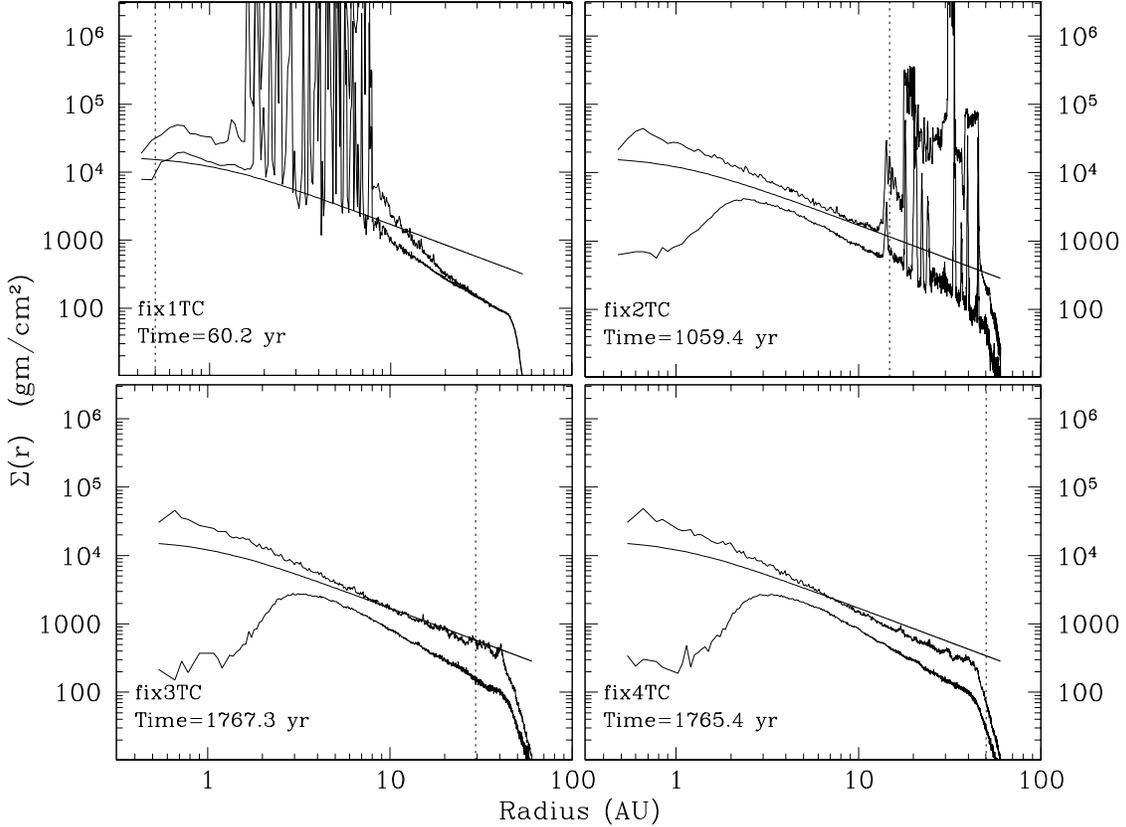}}
\caption
{\label{fig:fix-crit-modk}
Azimuth averaged and cumulative maximum surface densities (as defined
for figures \ref{fig:Dyn1-dens} and \ref{fig:Dyn1-crit}) for the {\it
fix TC} simulations, each defined with different, fixed gravitational
softening lengths, and which are unstable to numerically induced
fragmentation. The smooth solid curve represents critical surface
density defined by the equation \ref{eq:max-surfdens} with
\nreso=6\nneigh. The vertical dotted line in each figure corresponds
to the radius at which the gravitational softening and the initial
hydrodynamic smoothing lengths for the particles are equal. } 
\end{figure*}

\begin{figure*}
\rotatebox{-90}{\includegraphics[width=120mm]{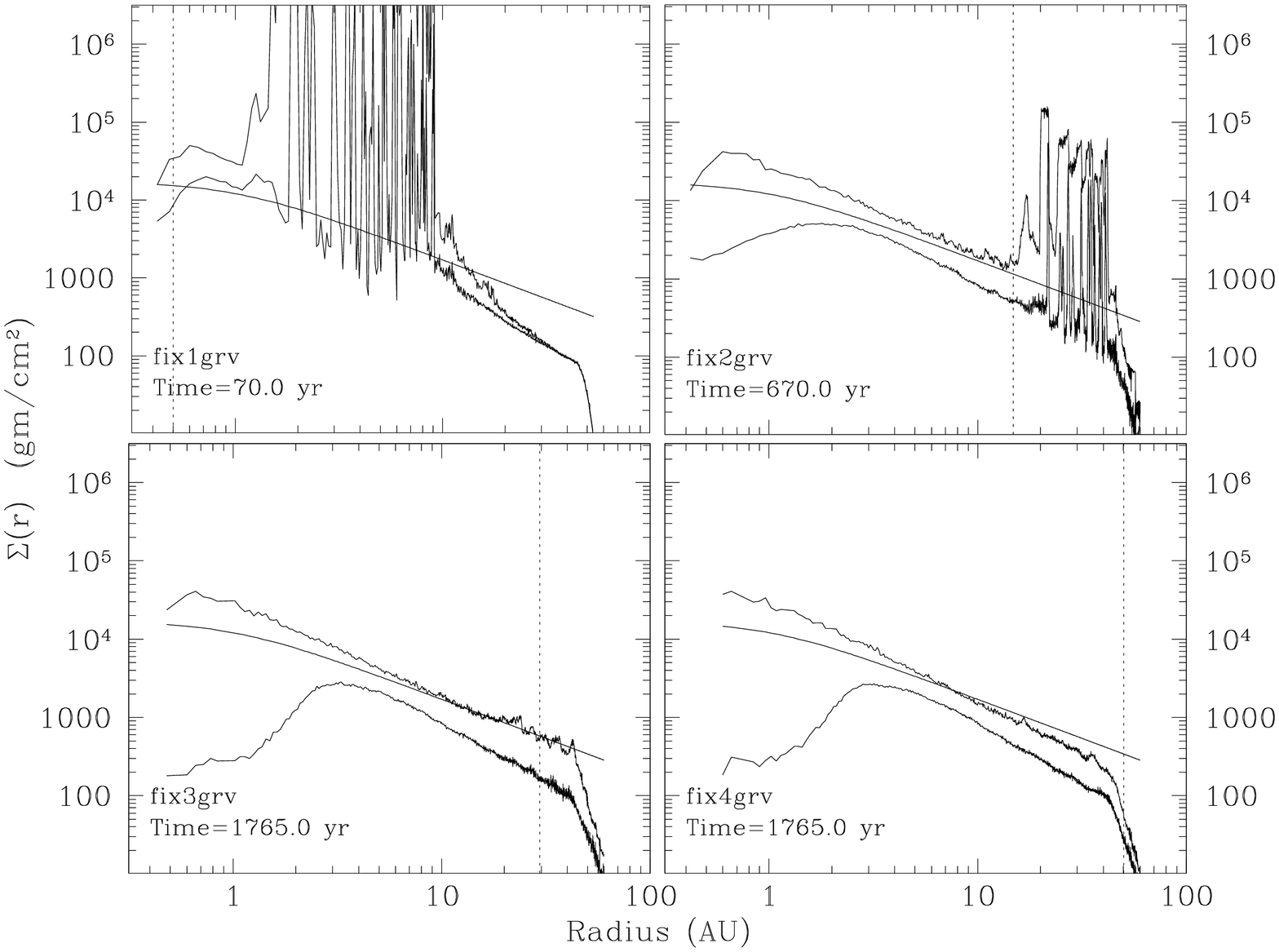}}
\caption
{\label{fig:fix-crit-grv}
The same as figure \ref{fig:fix-crit-modk}, but for the {\it fix grv}
series. } 
\end{figure*}

We show the mophologies of these simulations in figures
\ref{fig:fix-morph-modk} and  \ref{fig:fix-morph-grv}, and the
azimuth averaged surface densities in figures \ref{fig:fix-crit-modk}
and \ref{fig:fix-crit-grv}. Overall, the behavior and morphology
observed for each variant is quite similar to that of the other. With
small softening ({\it fix1TC/fix1grv}) fragmentation occurs at
essentially all orbit radii, on the same time scale as required for
the disk itself to become active. Due to this rapid onset, we
terminated it very soon after it began, so that radially more distant
parts of the disk simply had insufficient time to become active. We
believe that fragmentation would occur there too if the simulation
were to be evolved further. Only at the inner disk edge, where the
softening and smoothing lengths become comparable, is fragmentation
suppressed. At all other locations, it is actually enhanced relative
to the case of variable length softening shown in the {\it
mod2TC/mod2grv} simulations above, where clumps only occurred at much
larger radii. The {\it fix2TC/fix2grv} pair of simulations were
evolved for longer in time than the corresponding {\it mod2TC}
version, but ultimately they too produced clumps. Fragmentation
occurred only in the region exterior to $\sim15-20$~AU, where
softening was smaller than smoothing.

In the {\it fix3TC/grv} pair, evolved with a still larger softening,
strong spiral structures developed in the outer half of the disk,
while comparatively weak spiral structure developed closer to the
star. Significantly, the disk was evolved for 5\td\  and did not
produce any clumps over that time. For simulations {\it
fix4TC/fix4grv}, with the largest softening, spiral structure is
largely suppressed throughout the entire disk, relative to their
smaller softening cousins and to {\it mod2TC}. The gross structure of
the spiral arms that form also appear somewhat weaker than those in
the {\it mod2TC/mod2grv} simulations, however in the absence of a
quantitative analysis of the pattern amplitudes (beyond the scope of
this work) this statement remains somewhat subjective. 

From these simulations, we can make the rather unsurprising conclusion
that with a large enough softening value, clumping can be suppressed
in simulations where it would otherwise occur due only to insufficient
resolution. While not surprising, it is still important to quantify
the both the value of `large enough' and what changes in behavior
occur as a function of gravitational softening, in order to be able to
separate out real and artificial effects. In this case, `large enough'
gravitational softening appears to be larger than the hydrodynamic
smoothing lengths in each part of the disk. In order to completely
suppress clumping, softening must therefore be set comparable to the
smoothing values in the outer part of the disk.

\subsection{Enhancement of fragmentation in simulations with
sufficient resolution}\label{sec:enhancement}

While we have seen that large enough softening can alter the behavior
of an already numerically suspect (under resolved) simulation, it is
also important to determine its effect on what might otherwise be
considered a well resolved simulation. To investigate this question,
we have run four simulations, denoted {\it m1fxTC}, {\it m4fxTC}, {\it
m1fxgrv} and {\it m4fxgrv} in table \ref{tab:sims}, with identical
initial conditions and resolution to our {\it mod4TC/mod4grv}
simulations, but with fixed gravitational softening. As in the last
section, the value of the softening is set approximately to the value
of the hydrodynamic smoothing length at a predetermined radius in the
disk. For the {\it m1fxTC/grv} pair of simulations, we set the
softening equal to the smoothing at the inner edge of the disk
(0.5~AU), so that they correspond to higher resolution versions of the
{\it fix1TC/grv}\  pair. For the {\it m4fxTC/grv} pair, we set the
softening equal to the smoothing at $\sim15$~AU. These simulations
also correspond to high resolution versions of the {\it fix2TC/grv}
simulations. 

\begin{figure*}
\rotatebox{-90}{\includegraphics[width=120mm]{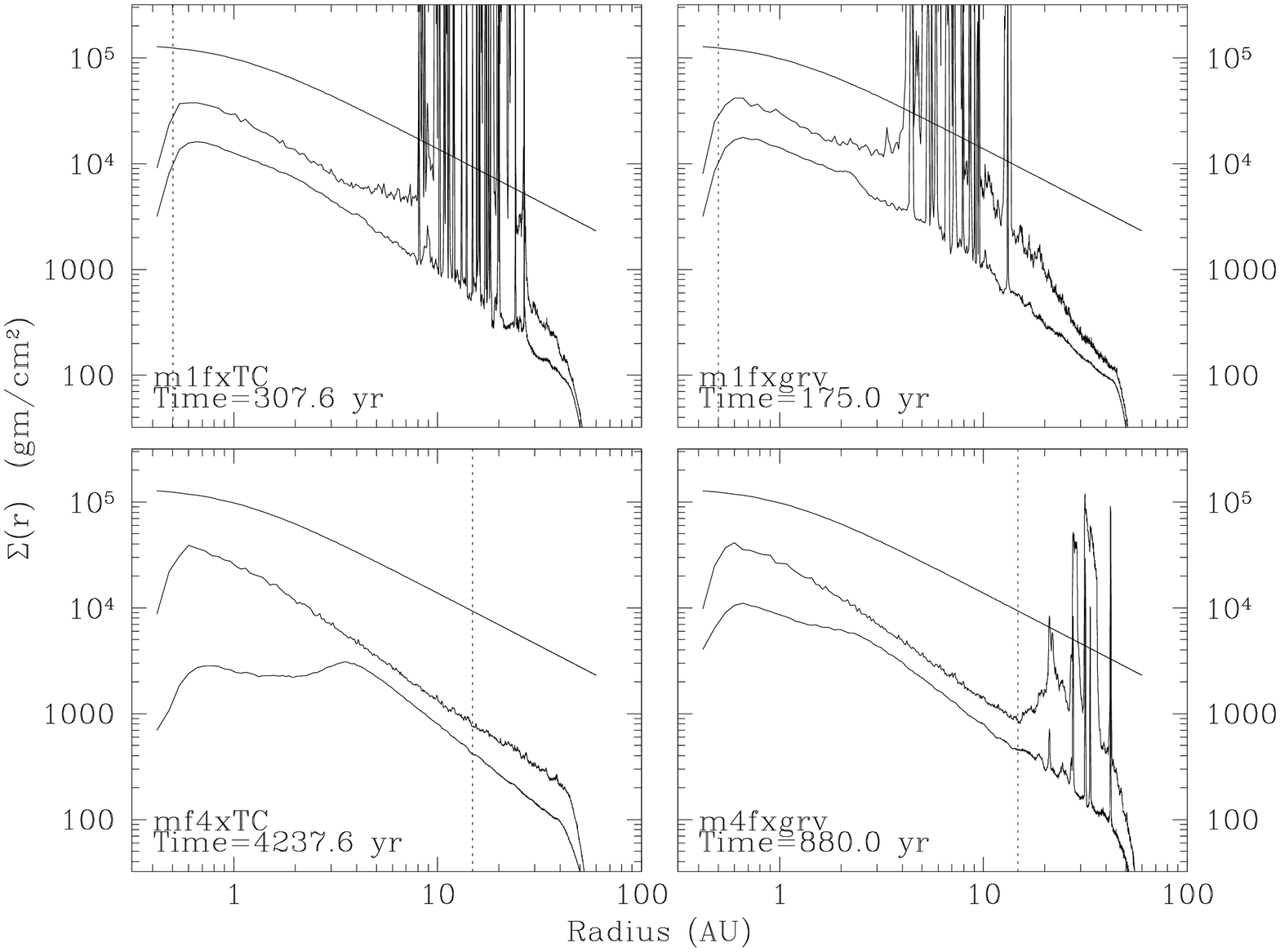}}
\caption
{\label{fig:stable-fixcrit}
Azimuth averaged and cumulative maximum surface densities for the two
simulations using the TC kernel derivative (left) and the two with 2D
softening variant (right). In each case the gravitational softening
length is fixed to the smoothing length at the inner edge (top) of the
disk or at 15~AU (bottom), and all realizations are expected to be
stable according to the Toomre criterion defined by equation
\ref{eq:max-surfdens} with \nreso=6\nneigh. The vertical dotted lines
correspond to the radius at which the gravitational softening and the
initial smoothing lengths for the particles are equal.} 
\end{figure*}

Figure \ref{fig:stable-fixcrit} shows the azimuth averaged and
cumulative maximum surface densities for the two TC92 simulations. In
contrast to the {\it mod4TC/grv} realizations, the two {\it m1fx}
realizations produce fragments in the disk in much less than 1\td. As
for their lower resolution cousins, {\it fix1TC/fix1grv}, we
terminated the simulations quite soon after they began due to the
fragmentation. We expect that had evolution proceeded further in time,
fragmentation would have occurred in the outermost parts of the disk
as well. Consistent with our finding in section \ref{sec:2dtest-evo},
that higher resolution realizations produced fragments later than
lower resolution realizations of the same initial condition,
fragmentation occurred later than in the {\it fix1} variants.
Fragmentation was delayed longer in the TC92 variant and was
suppressed over a slightly larger portion of the inner disk.

In both cases, fragmentation only occurred in regions where the fixed
gravitational softening was smaller than the hydrodynamic smoothing,
consistent with the expectation from the discussion in section
\ref{sec:softsmooth-limits} and in BB97 for 3D simulations. Even
though the relative size of the softening and smoothing lengths were
the same in the low and high resolution realizations, clumps appeared
only at somewhat larger orbit radii than at lower resolution. In all
cases, the maximum surface densities in and near the clumps exceed the
critical value determined from equation \ref{eq:max-surfdens} and must
therefore be considered to be numerically induced rather than due to
physical processes. 

Results in the {\it m4fxTC/m4fxgrv} simulations, with a much larger
softening length, were quite different from each other: fragmentation
did not occur with the modified kernel, but did with the modified
softening, again in the outer part of the disk where interparticle
force imbalances favored gravity over pressure. Although we have not
shown the earlier evolutionary history of the {\it m4fxgrv}
simulation, we note in passing that its character was markedly
different than any other discussed in this work. High amplitude,
filamentary spiral structures developed in the outer disk as early as
$\sim1$\td, much earlier than fragmentation occurred as defined by a
violation of the Toomre criterion. Although the structures developed
with some regularity, they were unable to remain distinct, instead
becoming progressively more sheared out, as they disappeared into the
background flow. Based on this behavior, we conclude that the
softening length chosen for this simulation was very near the boundary
between enhancing and suppressing fragmentation, as we saw for the 
{\it fix2grv/fix3grv} simulations at lower resolution.

\begin{figure}
\rotatebox{-90}
{\includegraphics[width=65mm]{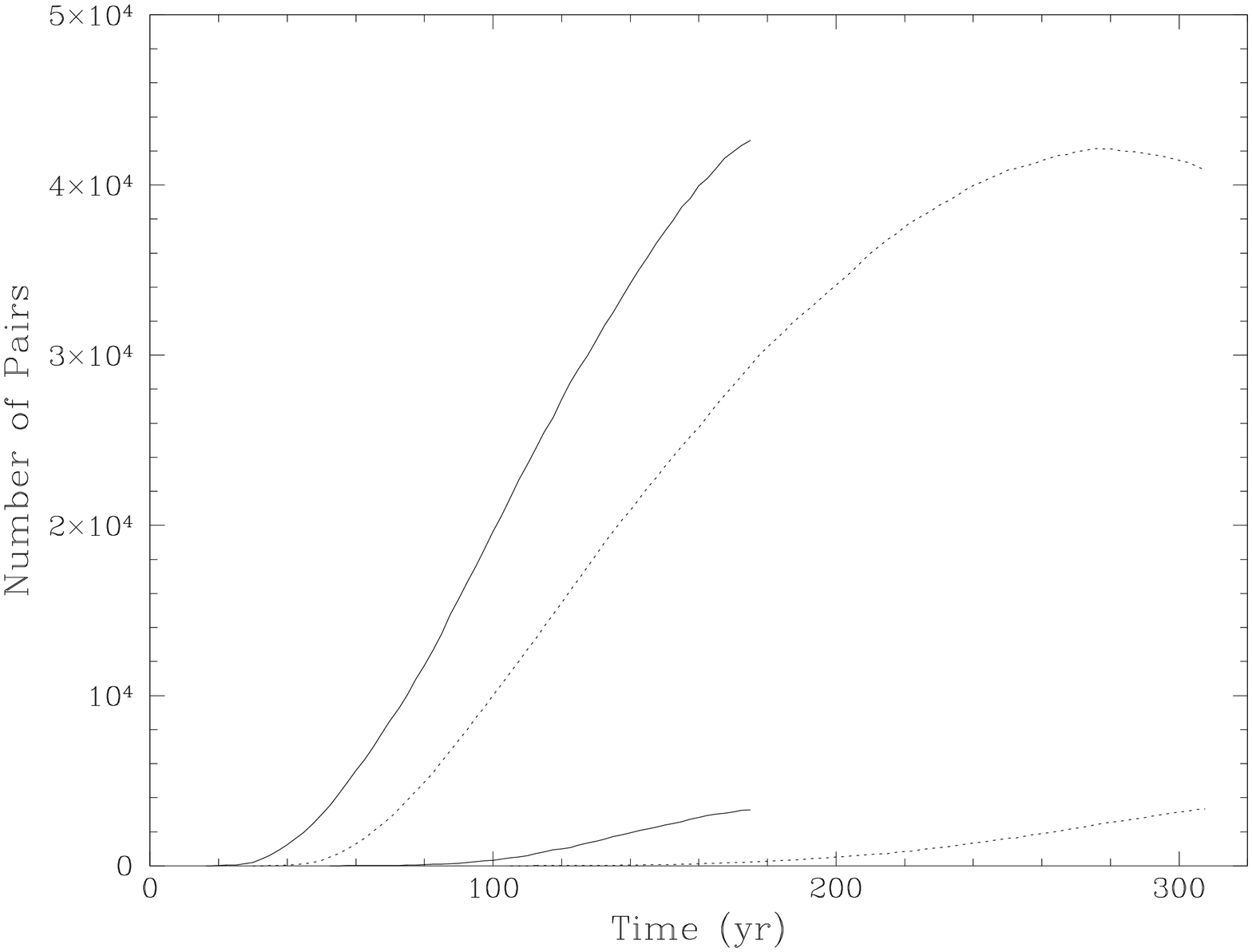}}
\caption
{\label{fig:m1pairs}
The number of multiple particles in the {\it m1fxgrv} (solid) and {\it
m1fxTC} (dotted) simulations as a function of time. The two curves
ending near $4\times10^4$ represent the number of particle pairs,
while the lower pair of curves represent particle triples. } 
\end{figure}

If, as we have suggested, small scale interparticle force imbalances
enhance fragmentation when the imbalance favors gravity, then we
should expect simulations with fragmentation to contain a significant
number of both paired particles and particles at small mutual
separations, but few or none in simulations that do not fragment.
Figure \ref{fig:m1pairs} demonstrates that indeed, this supposition is
true for the two smallest softening length variants. The number of
paired particles rises immediately and dramatically from zero to over
$4\times10^4$ over the short span of each simulation. Further, while
the number of pairs appears to level out or even decrease at late
times, the feature is actually a consequences of the increasing
prevalence of higher order multiple particle groupings, such as
triples or quadruples, as demonstrated by the lower pair of curves in
the figure, representing the population of triples.

\begin{figure}
\includegraphics[width=85mm]{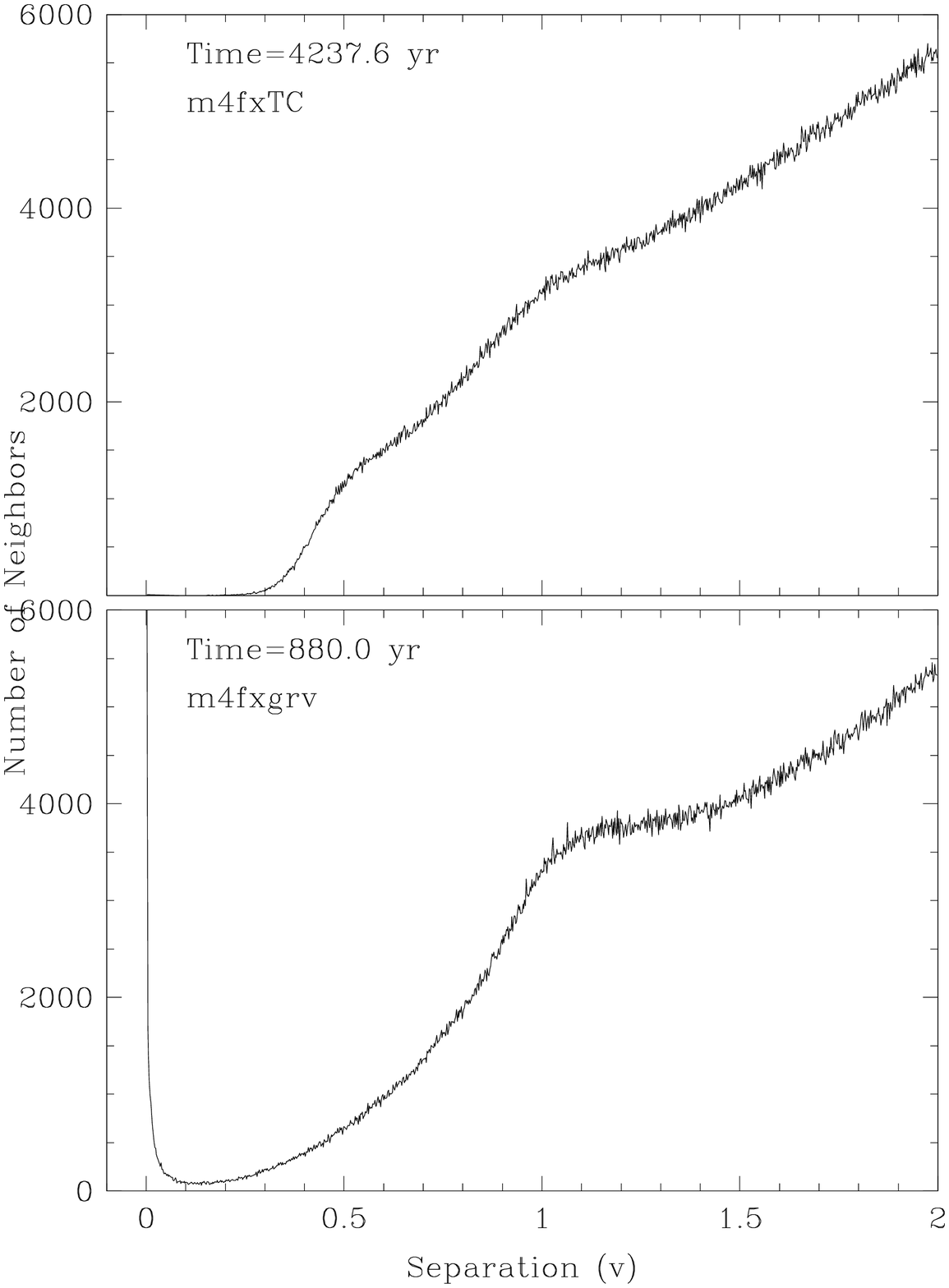}
\caption
{\label{fig:m4neigh}
Histograms of the neighbor separations for all neighbors of
all particles in the {\it m4fxTC} (top) and {\it m4fxgrv} (bottom) 
simulations. } 
\end{figure}

\begin{figure}
\rotatebox{-90}
{\includegraphics[width=65mm]{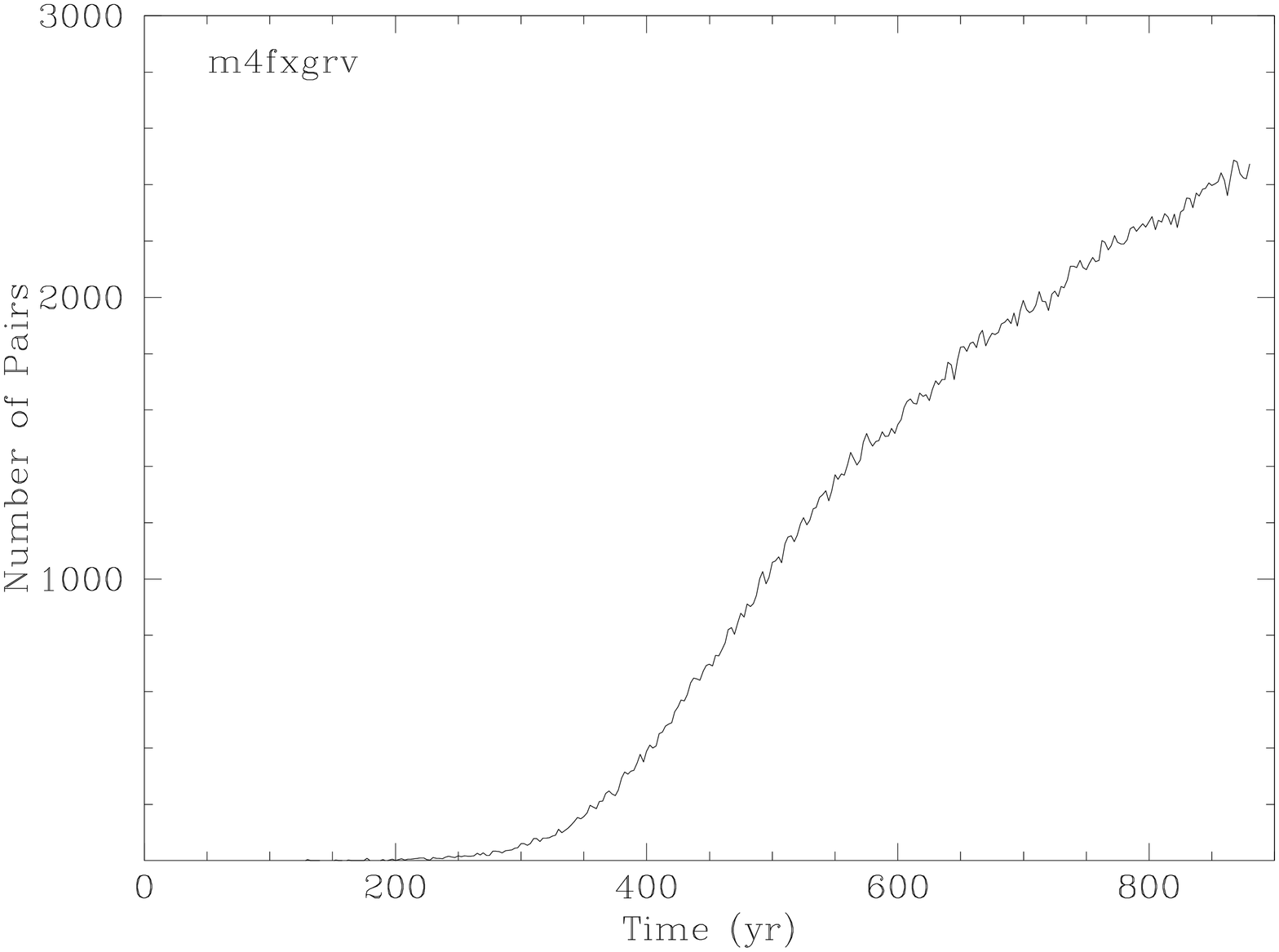}}
\caption
{\label{fig:m4pairs}
The number of particle pairs present in the {\it m4fxgrv} simulation
as a function of time.} 
\end{figure}

Figure \ref{fig:m4neigh} shows histograms of the full neighbor
distributions of the two larger softening variants and figure
\ref{fig:m4pairs} shows the number of pairs as a function of time for
the {\it m4fxgrv} simulation. As we suppose, a significant population
of pairs and near pairs (i.e. with $v\la0.1$) develops in the modified
softening variant, which fragmented, while few do in the TC92 variant.
Over the course of 12\td, only a tiny population of near pairs
developed compared to its cousin, and we observed only $\sim5$
particles to become paired, a small enough number to be unimportant
overall. While force imbalances must have been present at small
separations in both models, we conclude that the enhanced
interparticle pressure forces in the TC92 model, coupled with the
higher resolution, were of sufficient magnitude to eliminate the
numerically induced fragmentation that occurred when the softening
length was smaller.

In an apparent contradiction of our supposition, the total number of
paired particles generated in the three simulations here, which
fragmented, is smaller than in the {\it mod4} simulation, which did
not fragment. Although we have made no quantitative comparison, a
brief inspection indicates that the contradiction may be resolved by
accounting for the spatial distribution of pairs and of near pairs.
Because the small softening simulations were not evolved for a long
enough time to become fully active, the distribution of pairs is quite
concentrated in the regions did become active and did fragment,
relative to the distribution in the {\it mod4} realization where pairs
are found spread much more evenly throughout the system. The {\it
m4fxgrv} simulation, which did become fully active, generated many
fewer pairs but did generate a large population of near pairs,
indicating again the consequences of force imbalances favoring
gravity. Given that the effective resolution in the {\it mod4} run is
decreasing with time, we expect that if it were to be evolved further,
and the concentration of paired particles increased further,
fragmentation would occur. 

Disks realized with small fix softening fragmented in both the TC92
and the modified softening variants, and both in their low resolution
realizations and at resolution high enough to satisfy the Toomre
condition defined in section \ref{sec:test-crit}. An immediate
conclusion from the similarities might be that, since the
fragmentation occurred each realization, the simulations are in fact
converged. While this conclusion may in fact be correct, it would also
be seriously misleading. Due to flaws in their design, the simulations
converge to an incorrect result. Interpretation of that result in
terms of the physical behavior of the system and the importance of
other correctly implemented physical processes becomes difficult or
impossible. While the result and its invalidity may be clear in cases
like our test simulations, the same statement may not be true in other
simulations where the results are more difficult to verify. As a
trivial example, the high and low resolution variants with larger
fixed softening for which both low resolution variants fragmented, but
only one at higher resolution. Our conclusion in this case must be
limited only to the fact that the TC92 kernel modification is more 
likely to suppress fragmentation than the softening modification. 

Two other, more technical aspects of simulations with fixed
gravitational softening deserve mention. First, the cpu time required
to run simulations with large, fixed softening to some specified time
is far longer than with either the variable softening case or the
cases with small, fixed softening. The reason is that with a large
softening length, the interparticle spacing requires that a much
larger fraction of the total number of gravitational interactions
between particles be calculated as `atoms' rather than grouped
together as nodes, as is ordinarily done in tree-based gravity solvers
in general use for particle simulations. Secondly, in cases where
fragmentation does occur, a simulation with large fixed softening can
proceed for much longer in time, because the time step size does not
decrease nearly as much as occurs in a variable softening simulation
at the same resolution. Continuing a simulation for a long time after
clumping occurred and the resolution criterion has been violated would
be of limited utility however, since the clump may cause large
perturbations to the rest of the system that would not occur
otherwise.

\subsection{Recommendations for choice of
softening}\label{sec:recommend}

We conclude from our models that a fixed softening may either enhance
or suppress density inhomogeneities depending on the specific choice
of softening value relative to the hydrodynamic smoothing. No single
value of softening may be considered `optimal' in a disk with a large
radial extent. In the extreme case, an incorrect (too small) softening
value may induce fragmentation to occur in a simulation that would
not, given another fixed choice or a variable softening set to the
hydrodynamic smoothing length. Given an incorrect but too large value,
suppression of physically correct structure may occur. In contrast,
while a variable gravitational softening length may produce some
violation of energy conservation, in the examples shown here the
outcomes of the simulations are not affected in a manner as
drastically as with fixed softening. We therefore recommend that
simulations involving self gravitating hydrodynamic systems
incorporate a spatially and temporally variable gravitational
softening, whose length scale is the same as that over which the
hydrodynamic quantities are smoothed.

Our recommendation is similar to, but stronger than that made by BB97
because we observe that fragmentation may develop in simulations
whether or not the Toomre criterion (equation \ref{eq:Toom-cond} or
\ref{eq:max-surfdens}) is satisfied. Rather than signaling a temporal
endpoint, beyond which BB97 conclude that evolution towards some final
collapsed state is under resolved, we find that force imbalances at
small particle separations can alter the evolutionary trajectory of a
system which would otherwise never approach gravitational instability
into one that does. Given that BB97 did not specifically study systems
starting in a dynamically stable equilibrium condition for which
fragmentation is known {\it not} to occur, as found in our disks, the
behavior we observed in our simulations may not have been observable
in their work. We suggest that it is independent of the specific
system being simulated, and in fact applies to the Jeans collapse
situation as well. Our conclusion is also consistent with that of
\citet{Dehnen01}, who made a detailed study of the effects of various
choices of softening on the force errors produced by a static
distribution of particles, and who recommends spatially variable
softening in that case as well. Since in nearly all particle based
hydrodynamic simulations, however, the smoothing length is not fixed
in time, our recommendation goes beyond Dehnen's study of a static
distribution to including temporal variation as well.

We base our recommendation on the results of our simulations and on
the following argument. Completely independent of whether or not
either the gravitational or pressure force actually can be resolved on
better or worse length scales than the other, is the consequence of
the assumption underlying both softening and smoothing. Namely, an
error in the net force is present {\it by assumption} when the
contribution from either source is not fully resolved. We therefore
have only the choice of how to handle this error most gracefully in
the code. We submit that avoidance of large force imbalances in
regions that are by definition under resolved is a trait much to be
desired in the properties of the code because pathological behavior
that results from such imbalances--artificial fragmentation or
suppression of physically realistic fragmentation--are avoided.

Such a recommendation is not without cost however because it is
equivalent to recommending that simulations using SPH should not
require that energy be conserved. Although we have not attempted to
quantify the magnitude of the violation, in general we believe it to
be a relatively small component of the total energy of a given
particle since only neighbor particles will contribute. Also, the
violation will be smaller in higher resolution simulations because
those neighbors will represent a proportionally smaller component of
the total mass in the system, from which the potential energy of a
given particle is derived. Finally, the fact that the gravitational
forces and potential are approximated by a summation of terms derived
from a tree search already violates strict energy conservation. Given
the large body of literature using such approximations, often checked
for veracity against other methods or observations, it would appear
that few, if any, unacceptable consequences originate from a low level
of violation. Assertions such as these are no replacement for a
measurement however, and we refer the reader to \citet{Benz90} for
discussion of a method to make an approximate quantification. We are
also aware of current work, developing on the techniques discussed in
\citet{PM04} and by the same authors (Price and Monaghan 2006, submitted),
to create an SPH formulation that allows variable softening, while 
still conserving energy. Adapting SPH codes in existence to utilize 
this technique would appear to be highly desirable, since it would 
remove the most serious drawback of variable softening. 

We have considered and discarded a third, intermediate softening
option, in which the gravitational softening of each particle is set,
for example, to the {\it initial} value of the hydrodynamic smoothing
length, and is thereafter fixed. Although no specific tests have been
performed to study this question, we believe that using such a
prescription will yield results similar to the fixed softening case
with a small softening length. For example, a particle initially
located in a dense region (small softening value), which later moves
into a less dense region (larger softening value) could act as an
attractor for other particles in its current neighborhood. An exactly
opposite effect might be seen for a particle initially in a low
density region which moves to a high density region. 

Secondarily, we question the aesthetic utility of such a prescription
because it implies knowledge of the internal mass distribution of each
particle that is both independent of any hydrodynamic expansion or
contraction of the gas (the assumption underlying a temporally
variable softening) and needlessly complex. The complexity is due to
the fact that using constant, individual softening lengths suggests
that information is known about not only the internal mass
distribution of particles taken as an ensemble, but also the
information that each particle has a different internal mass
distribution from every other. We can postulate no circumstances where
such exquisitely detailed information is likely to be available. 

\section{Simulations in 3 dimensions and/or using grid
methods}\label{sec:disks-3d}

The wave analysis underlying our resolution criterion makes the
assumption that the flow is two dimensional: an integration over the
$z$ coordinate is assumed. The test problem defined in section
\ref{sec:testing} was chosen to be similar to the models originally
employed by \citet{DynI}, and was therefore performed in the same two
dimensional approximation as the originals. This approximation has
historically been common in other works as well. In this section, we
examine methods for applying the criterion to fully three dimensional
models, and to the question of whether 3D simulations require
additional constraints on resolution in order to ensure their
accuracy.

\subsection{Calculating surface density in 3D
simulations}\label{sec:3d-surfdens}

For fully three dimensional simulations, where a surface density is
not directly available, the Toomre criterion cannot be applied without
some additional computation. For grid based simulations, the
computations are trivial and involve only an integral of the volume
density in the $z$ direction. In cylindrical or Cartesian coordinates,
for example the integral would pass to a summation of $\rho_k \delta
z$ over a set of grid zones with spacing $\delta z$, with $\rho_k$
being the volume density in the $k$th grid zone in the $z$ direction. 

For 3D SPH simulations, the computation of a surface density is more
complex because there is no grid. Instead, we recommend the creation
of a temporary grid onto which the volume density can be mapped. Then
the surface density can be computed as a summation in the $z$
direction as in the case for grid based simulations. In test models,
we have found good results by using a mapping to the grid based on the
same smoothing technique used to determine the volume densities at the
particle locations, for the mapping. At the location of each grid
zone, we determine a list of `neighbor' particles which may contribute
to the density of that zone, then we use the formula
\begin{equation}
\rho(i,j,k) = \sum_{\rm neighbors} m_p W(r,h)
\end{equation}
where $m_p$ is the mass of each neighbor particle, $W$ is the
smoothing kernel, $r = | {\bf r_p} - {\bf r_{\rm zone}}|$ is the
distance between the particle and the grid zone at location $(i,j,k)$
and $h$ is the mutual smoothing length of the particle and zone,
defined as $\max(h_{\rm zone}, h_p)$. The grid smoothing length is
defined as 
\begin{equation}
h_{\rm zone} = \sqrt{\delta r_i^2 + (r_i \delta \phi_j)^2 + \delta z_k^2}
\end{equation}
where the components are the grid spacing in each of $r$, $\phi$ and
$z$. We include a smoothing length for the grid itself in order to
ensure that every particle contributes to at least one grid zone even
in the case of vanishingly small smoothing lengths.

For the test problem discussed in sections \ref{sec:3dsph-nosg-test},
\ref{sec:3dsph-nosg-result} and \ref{sec:3dsph-nosg-signif} below, we
map the volume density onto a cylindrical coordinates grid where the
radial coordinate is logarithmically spaced and the azimuth coordinate
is uniformly spaced so that its spacing is $\delta r \approx r \delta
\phi$ everywhere. In order to resolve the smallest scale features of
the flow, we have chosen our grid resolution so that its spacing is
comparable to the smallest smoothing lengths in the calculation. This
grid spacing will in practice over resolve the features of most of the
flow, and will not strictly be required in most circumstances. Our
choice is motivated by the desire to ensure that the choice of grid
(i.e. $h_{\rm zone}$) will contribute negligibly to the derived
densities and the quality of the fitted quantities in our analysis
that depend on them. In any case, the computational expense of
choosing of a finely spaced grid for the mapping is small in terms of
the time to run a simulation because the mapping is not required for
the simulation itself, but only for post-processing.

A conceptually simpler mapping would be to use a Particle in Cell
technique to assign the mass of each particle to the grid cell that
contains it, thus determining a volume density for that cell.
Unfortunately, we cannot recommend this method because it is
susceptible to large errors due to the fact that SPH particles may
have smoothing lengths much different than the grid spacing. For
example, when a particle's smoothing length is larger than the grid
spacing, it should contribute to the density over a larger region than
assumed by the PIC technique. 

\subsection{An unwarranted approximation:
$\rho=\Sigma/2H$}\label{sec:bad-approx}

When two dimensional simulations are performed, it is common to obtain
an approximation to the volume density by dividing from the disk's
surface density by some factor multiplying the disk's scale height.
For an non-self gravitating isothermal gas, the factor will be
$\sqrt{2\pi}\approx2.5$ (see section \ref{sec:3dsph-nosg-test} below),
but will vary slightly depending on the exact input physics. For
purposes of this section and for simplicity, we shall assume a factor
of two exactly. As discussed at the end of section
\ref{sec:applicability}, the Jeans or Toomre stability criteria may be
applied using this conversion to obtain one or the other of surface or
volume density as required. In this section, we will investigate the
impact that this conversion may have on the indirectly obtained values
used in the Jeans criterion. We will refer here to the criteria
directly applied through the use of equations \ref{eq:Jeans-wavel} or
\ref{eq:Toom-wavel} as the Jeans or Toomre criteria, while the
criteria applied through the use of the disk scale height to convert
surface to volume density (or vice versa) as the approximate Jeans or
Toomre criteria.

In order to proceed, it is necessary to examine the results of a
simulation modeled fully in 3D, so that both volume density and
surface density are known. In one of a series of papers studying
fragmentation in disks, \citet{Boss02} discusses 3D disk models and
\citet{Boss04} continues this study with a close examination of the
properties of a few models (primarily the model named `edh') from the
earlier work. We will use these models as a test bed for our
investigation. Since our study has not specifically defined a value
for $T$ in equation \ref{eq:Toom-cond}, we will assume that its value
is $T=1/4$, which is identical to the value of $J$ required for Jeans
collapse simulations. While we believe this to be a reliable
assumption, only by examining the results of a specifically chosen
test problem can we be fully confident that the value is sufficient.
The consequences of the uncertainty in specifying $T$ are that
comparisons between the magnitudes of $T$ and $J$ and the critical
lengths associated with them, may be more difficult to interpret. In
this section however, we seek primarily to test the veracity of the
approximate forms of the criteria compared to the exact forms, and the
absolute magnitudes are less important.

Boss's model consists of a circumstellar disk modeled between 4 and
20~AU, evolved on a grid of $100\times23\times256$ (or $\times 512$ in
some simulations) zones asymmetrically distributed primarily near the
disk midplane. He includes an ideal gas equation of state and
radiative cooling in the diffusion approximation, but omits an
artificial viscosity, thereby omitting the viscous and shock heating
modeled by it. With this model, he evolves a disk for 345~yr, at which
time the simulation begins to violate the Jeans criterion due to onset
of clump formation, and is terminated. We will apply our stability
criteria to this model using each of the four relevant versions
discussed in section \ref{sec:sizes}.

\begin{figure}
\rotatebox{0}{\includegraphics[width=80mm]{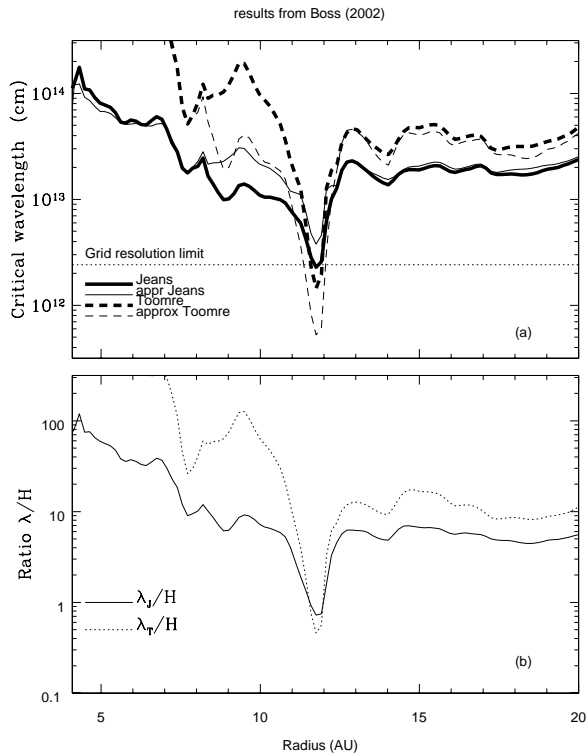}}
\caption
{\label{fig:Boss-crit}
(a) Critical wavelengths for the \citet{Boss02} simulation
along the radial column including the forming clump. The exact Jeans
and Toomre criteria are shown with heavy solid and dashed lines
respectively, while the approximate Jeans and Toomre criteria (as
defined in the text) are shown with light solid and dashed curves,
respectively. The horizontal solid line, defines the spacing of the
grid on which the system was evolved. (b) The ratio of the Jeans and
Toomre wavelengths to the isothermal scale height ($H=c_s/\Omega$) as
functions of orbit radius.}
\end{figure}

The top panel of figure \ref{fig:Boss-crit} shows the value of the
critical wavelengths at the time the simulation was terminated, using
each of the four applicable implementations of the stability criteria.
The failure of the Jeans criterion was used to define the end of the
simulation, so its minimum value matches the grid spacing very
closely. The exact Toomre criterion yields a critical wavelength that
is about 30\% smaller than that from the Jeans analysis. Using it and
assuming that $T=J$, the model has already passed the boundary beyond
which it becomes numerically suspect. Due to the speed at which
clumping occurs, once underway, we do not believe that this small
difference is very significant however, except to note that the
simulation would have been terminated slightly earlier and at a lower
density. At all other locations, the Toomre wavelength exceeds the
Jeans wavelength and the evolution of the Boss simulation leading up
to its termination passes the resolution test. The conclusions made
from it stand on valid numerical grounds according to the Toomre
resolution criterion.

On the other hand, the approximations of the Jeans (light solid) and
Toomre (light dashed) criteria lead to critical wavelengths in the
forming clump that differ by a factor of 1.56 larger than and 4.6
smaller than the grid spacing, for the approximate Jeans and
approximate Toomre criteria, respectively. Moreover, using the
approximate Toomre criterion, the simulation is already well past the
time for which its evolution is numerically valid. Each of the
approximate forms uses the isothermal scale height to render the
conversion. Does this quantity remain relevant in a collapsing region?

The bottom panel of figure \ref{fig:Boss-crit} shows the ratios of the
Jeans and Toomre wavelengths to the scale height. Except near the
forming clump, the Jeans wavelength is many times the disk scale
height, making its application as a resolution criterion unclear
according to the argument in section \ref{sec:sizes}. The Toomre
wavelength is also many times the disk scale height, but this only
makes its application to a resolution criterion more appropriate,
according to the same arguments. We therefore recommend that it be
used in favor of the Jeans criterion in simulations of disks. In and
near the collapsing region, both wavelengths become significantly
smaller than the scale height. Since those wavelengths loosely define
the physical size of the collapsing body, we can conclude that the
structure of that object is no longer well represented as a slightly
perturbed disk structure under which assumption the scale height was
originally defined. Thus, the approximate criteria should not be used
because their values become inaccurate near the clump, due to the fact
that the disk vertical profile becomes distorted.

\subsection{A simple 3D test problem to test the accuracy of the
hydrodynamics in disk simulations}\label{sec:3dsph-nosg-test}

In section \ref{sec:2d-vs-3d}, we discussed the possibility that a
disk simulation performed in 3D may not in fact be truly three
dimensional because the disk thickness may be comparable to the
smoothing lengths of the particles in and SPH simulation, or the grid
dimension in a grid based simulation. Here we explore the consequences
of such simulations using a simple test problem. Specifically, the
problem of a non-self gravitating disk evolved with an isothermal
equation of state. For this system, an analytic expression for the
vertical structure of the disk can be derived exactly, and direct a
comparison between the analytic result and theory can be made. This
test problem is equally applicable to grid based simulations and to
particle simulations, and we propose it as a general test for
interested numericists.

The vertical structure of the gas in a non-self gravitating disk will
obey the equation
\begin{equation}\label{eq:vert-diffeq}
{{d p} \over{d z}} = \rho {{GM_* z}\over{ (r^2 + z^2)^{3/2} }}
\end{equation}
where $p$ and $\rho$ are the pressure and volume density, $z$ is the
altitude above the disk midplane and $r$ is the cylindrical radius.
For an isothermal equation of state, the pressure is related to the
density and sound speed through the definition $p=\rho c_s^2$. Using
the equation of state to replace pressure with density, the solution
to the differential equation is
\begin{equation}\label{eq:vert-profile}
\rho(z) = \rho_0 e^{-z^2/(2H^2)}
\end{equation} 
where $H=c_s/\Omega$ defines the isothermal scale height and
$\Omega=\sqrt{GM_*/r^3}$. For any given location in the disk, the
surface density can be obtained by integrating equation
\ref{eq:vert-profile} over all $z$, giving the relation
\begin{equation}\label{eq:3dtest-surfdens}
\rho_0 = {{\Sigma}\over{\sqrt{2\pi}H}}.
\end{equation}

We use a modification of the 2D test problem described in section
\ref{sec:2dtest-defn} for this test. Particle layout, temperature and
surface density are defined as in the 2D problem, such that a minimum
Toomre $Q$ value of $\sim1.5$ would be obtained if self gravity were
included. To obtain an initial $z$ coordinate for each particle, we
use a Gaussian pseudo random number generator whose width is defined
by the local value of the disk scale height, consistent with equation
\ref{eq:vert-profile}. Definition of the initial velocities takes
account only of stellar gravity and pressure forces rather than disk
self gravity as well. Velocities in the $z$ coordinate are set to
zero.

Before proceeding further, it is significant to note that the specific
model we propose is a step removed from those of greatest relevance
for actual systems because vertical density distributions depend on
the details of the physical model employed. Also, as a practical
matter, the exact definition of `scale height' itself can lose meaning
in such models. To the extent that it does remain meaningful, any
resolution requirement may be sensitive to the details of the model
if, for example, a larger fraction of the mass were in an extended
envelope or were concentrated closer to the midplane than with an
isothermal structure, as is the case for the isentropic structure
discussed in section \ref{sec:3dsph-RT} below. In the context of the
spatial distribution of particles or grid cells that represent the
mass and at a coarse level of examination, models that result in a
higher midplane mass concentration will be somewhat similar in
structure to colder isothermal disks in which the scale height is
simply not as large. We therefore believe that any modifications to
the critical resolution appropriate for other models will be not be
large and proceed to define a criterion for isothermal structure that
we believe will be generally applicable. In the following two
sections, we will however, proceed to define a resolution criterion in
two complimentary ways: as a requirement per scale height at the disk
midplane, and as a resolution per vertical column, in order to
facilitate its use more generally.

\subsection{Results from simulating our 3D test problem with
SPH}\label{sec:3dsph-nosg-result}

We have run a series of simulations modeling the 3D test problem at
different resolutions, from $1.3\times10^5$ to 10$^6$ particles, each
labeled {\it sgoff3-6} and defined in table \ref{tab:sims}. We allow
each simulation to run for two orbits of the outer edge of the disk in
order to ensure that the configuration represents the true system as
realized by the hydrodynamic code rather than any peculiarity of the
initial configuration. We then map the volume density and velocities
onto a three dimensional grid and calculate the surface density at
each location. We pose the following question for the configuration,
which must be answered affirmatively before the results can be
validated: does the configuration generated by the simulation
correctly reproduce the correct midplane densities and scale heights
everywhere in the disk? 

To answer the question, we generate least squares fits to the
densities as mapped onto a temporary grid to compare to the analytical
values defined by equation \ref{eq:vert-profile}. Each fit returns
three parameters corresponding to the values of $\rho_0$ and $H$ in
equation \ref{eq:vert-profile}. We limit the fits to the radial region
between 1~AU and 45~AU in order to eliminate the possibility that
conditions at the inner or outer boundary of the disk distort the true
picture of the structure. Inspection of the disk structure shows that
these limits ensure that the fits are limited to the regions where the
rotation curve is unaffected by either the gravitational softening of
the star at the inner boundary or large density gradients near the
outer boundary. In order to remove any potential systematic errors
derived from the resolution of the grid or simulation itself, we
average the volume densities at each altitude over patches of
approximately one scale height on a side, so that the same number of
fits are generated for each simulation. Finally, in order to retain
self consistency in the face of any evolution of the disk away from
the initial surface density profile, we derive the analytic values for
$\rho_0$ and $H$ using the locally determined value of the surface
density and rotation velocity, through the relation $H=c_s/\Omega$ and
equation \ref{eq:3dtest-surfdens}.

Patches are defined in successive radial rings of the disk by the
condition that a ring's radial extent is the local, analytically
determined value of the scale height. Their azimuthal extent is
determined by the same condition, so that the total number of patches
in a ring is defined by its circumference divided by the scale height.
At the grid resolution chosen for the analysis, each patch covers a
region of $\sim 200$ or more cells in the vertical coordinate, and
$10\times10$ radial and azimuthal grid cells near the inner disk edge
where the scale heights are small, and $15\times20$ near its outer
edge where they are much larger. In total, approximately 14000 patches
are required to cover entire disk, so that 14000 separate fits of the
vertical structure in the disk are performed. 

\begin{figure}
\begin{center}
\rotatebox{0}{\includegraphics[width=80mm]{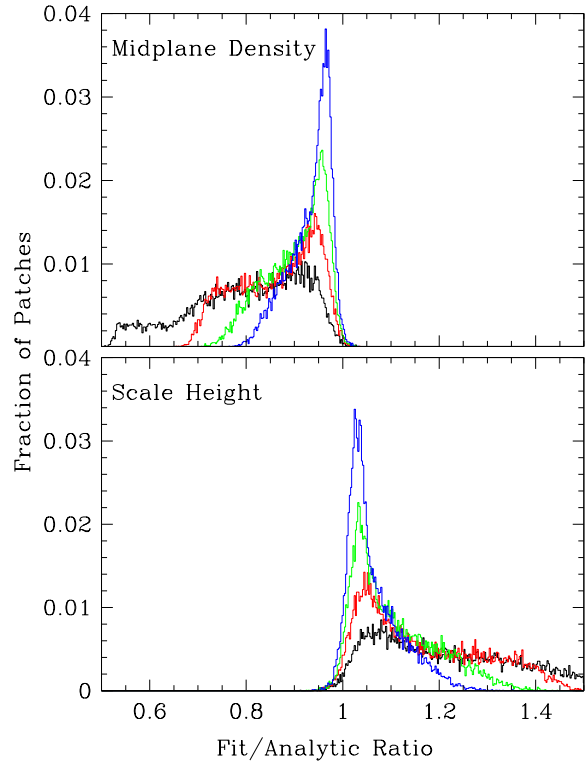}}
\end{center}
\caption
{\label{fig:bestfit-vstruc}
Histograms of the best fit midplane density, $\rho_0$ (top) and scale
height, $H$ (bottom). The black, red, green and blue curves correspond
respectively to the progressively higher resolution simulations {\it
sgoff3-sgoff6}. The bins of the histogram are of width 0.25\%.}
\end{figure}

Figure \ref{fig:bestfit-vstruc} shows histograms of the ratios of the
fitted quantities $\rho_0$ and $H$ to their analytic predictions,
normalized to the total number of patches. At the lowest resolution
shown, the vertical structure in the simulation is clearly more
extended vertically than predicted, with the distribution of scale
heights extending more than 50\% larger than the analytic values and
the midplane density distribution extending nearly 50\% below the
analytic values. Both distributions are spread relatively evenly over
the range extending from the most extreme under or over estimates on
one end, to nearly the correct analytic value on the other. The
distributions become narrower for each of the each of the higher
resolution variants of the model, and appear in fact to be converging
to the analytically expected values. However, even at a resolution of
one million particles, the highest resolution in our study, the peaks
of the distributions deviate from the analytical values by about 5\%
and a significant population of patches are best fit with midplane
densities and scale heights as much as 20\% below or above the
analytic values, respectively.

\begin{figure*}
\rotatebox{-90}{\includegraphics[width=130mm]{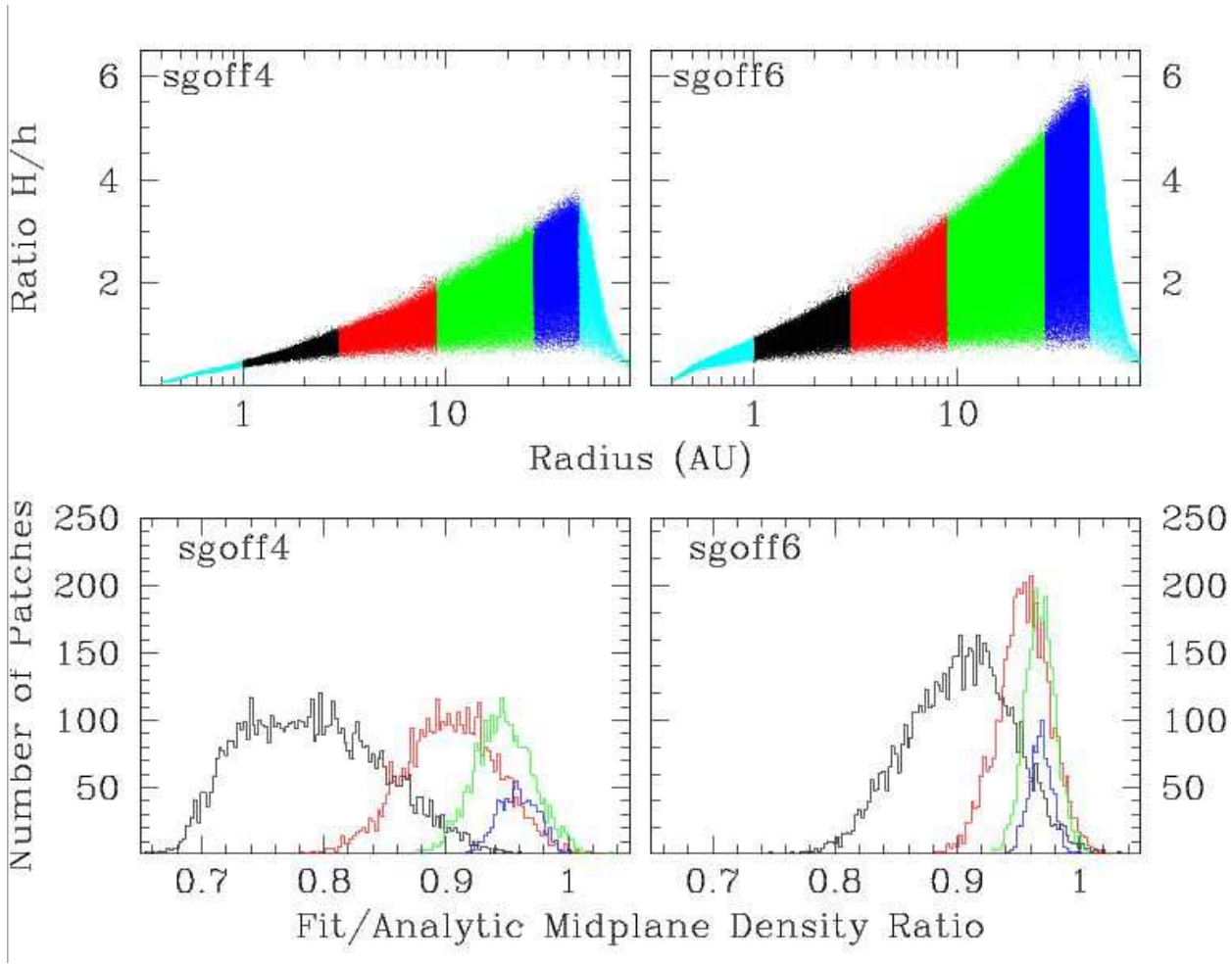}}
\caption
{\label{fig:hiterat-rhodist}
Top panels: Scale height to smoothing length ratio for all particles
in the {\it sgoff4} and {\it sgoff6} simulations, as labeled.
Bottom panels: Histograms of the ratio of the best fit to
analytic midplane densities expected for all patches in the disks
between 1 and 45 AU. The color of each histogram corresponds to to patches
in the ranges 1--3~AU (black), 3--9~AU (red), 9--27~AU (green) and
27--45~AU (dark blue), as shown in the top panels. Patches at orbit
radii smaller than 1~AU or larger than 45~AU (light blue) are excluded
from the histograms in order to minimize distortions due to decreased
resolution at the inner and outer disk edges. As in figure
\ref{fig:bestfit-vstruc}, histogram bin widths are set to 0.25\%.}
\end{figure*}

We can convert the correspondence between simulation and theory into a
resolution requirement if we can quantify the minimum number of
particles required in a vertical column for which the fit parameters
are accurately reproduced. The top panels of figure
\ref{fig:hiterat-rhodist} show the vertical resolution of the disks
quantified as a ratio between the expected scale height and the
smoothing length of each particle. The bottom panels show the quality
of the reproduction of the analytically determined disk midplane
densities as in figure \ref{fig:bestfit-vstruc}, broken out into
separate histograms for each of four radial regions of the disk. 

The vertical resolution varies between less than one smoothing length
per scale height in the inner disk to as many as six near the outer
edge of the high resolution simulation, {\it sgoff6}. The ratio takes
values over a range from less than unity up to a orbit radius
dependent maximum value because particles located near the midplane at
some radius, where densities are high, will have correspondingly
smaller smoothing lengths. Particles located at higher altitudes where
densities are lower will have larger smoothing lengths. Dividing each
by the scale height (constant at a given radius) produces the range.
The maximum value of the ratio varies as a function of orbit radius
primarily because the scale height itself is a function of radius,
through the combination of the predefined temperature profile
(equation \ref{eq:dyn1-temp}) and the rotation curve.

In the 1--3~AU region of the disk, the entire vertical structure of
the disk is represented by the smoothing length of only a single
particle. The exact positions of neighboring particles relative to
each other therefore cause a correspondingly larger influence on the
densities, as demonstrated by the very wide distribution of midplane
densities present in both the {\it sgoff4} and the {\it sgoff6}
realizations, relative to their analytic values. In each successively
more distant region, and at both resolutions, the distribution becomes
narrower and of better quality, demonstrating the value of the 
increased vertical resolution. For example, the vertical structure is
resolved by $\sim1-2$ smoothing lengths per scale height at $r\la
3$~AU, and the fitted midplane densities fall between 65 and 95
percent of their analytic value for the 260000 particle simulation and
75 and 100 percent in the 1 million particle run. At the other end of
the spectrum, density fits fall within $\sim5$\% of their predicted
values only in the two outermost regions in the high resolution
simulations, corresponding to orbit radii $>9$~AU (i.e. the green and
dark blue histograms in the figure), and only part of the distribution
in the single outermost region of its lower resolution counterpart,
corresponding to radii $>$27~AU.

If we specify that midplane densities within 5\% of their analytic
values are sufficiently accurate for the purpose of simulating the
evolution of a circumstellar disk, then only these outer regions of
the disk have sufficient accuracy. In the two best resolved regions of
simulation {\it sgoff4}, ratios of fit to analytic midplane density
values peak near $\sim94-95$\% and extend to as low as 90\%, resolving
one scale height with at most $\sim3-4$ particles. In the same regions
of {\it sgoff6}, The peaks of the distributions increase to near
$\sim96-97$\%, with tails extending to as low as $\sim92-93$\%,
resolving one scale height with as few as 3.5 particles at 9~AU and as
many as 6 further out. We conclude that midplane densities within
$\sim5$\% of their analytic values are achieved only in regions where
the vertical structure of the disk is resolved with at least $\sim4$
smoothing lengths per scale height at the disk midplane.

\subsection{Interpreting the vertical resolution requirement at the
disk midplane in the context of the full thickness of the
disk}\label{sec:fullthickness}

It is important to note that the requirement for $\ga 4$ smoothing
lengths per scale height near the disk midplane is sufficient only for
obtaining reasonably accurate {\it midplane} densities, and only for a
mass distribution described by equation \ref{eq:vert-profile}.
Quantities at higher altitudes may still be significantly under
resolved because smoothing lengths are larger there due to the
correspondingly lower densities. Errors due to insufficient resolution
at high altitudes will be of lesser importance for simulations
modeling gravitational instabilities because the magnitudes of the
pressure and gravitational forces important for fragmentation will be
largest where the mass is concentrated, close to the midplane.
Nevertheless, some particles must be present at high altitudes in
order to compress those lower down to the required densities. How many
are enough? 

In this section we quantify the minimum number of particles required
per vertical column needed to obtain good correspondence with
theoretical expectations for density near the midplane in our models.
Because it is a metric that is effectively an integral over a full
vertical column, this quantification is likely to be much less
sensitive to the exact distribution of mass over the column, and so
more generally applicable than a quantification per scale height. It
also eliminates the possibility of misperceptions based on the idea
that the total mass outside the midplane is small enough to be
neglected in estimates of the total resolution required per vertical
column. 

For the isothermal disks in this study, 98\% of the disk mass will
reside within three scale heights of the midplane, according to
equation \ref{eq:vert-profile}, meaning that the mass is effectively
distributed over a total of six scale heights accounting for symmetry
above and below the midplane. At the simplest level, we might
therefore approximate the required vertical resolution to be $\sim24$
smoothing lengths per vertical column to avoid serious errors in the
midplane densities. A somewhat more accurate quantification might also
account for the fact that mass is preferentially located near the
midplane, with fewer particles present at high altitudes.

In principle, we can derive an approximate number of particles
required per vertical column if we simply quantify the particle
density over all altitudes in a single vertical column. At the minimal
resolution required to obtain accurate midplane densities however, we
must presume that the high altitudes remain inaccurate due to the fact
that densities are much lower there and particle separations larger,
and they may therefore remain under resolved. Any direct
quantification of the resolution required based on the actual
distribution of particles in a simulation would therefore suffer from
the inaccuracies in their high altitude distribution.

A fact that allows the analysis to proceed is that, for the purpose of
ensuring accurate midplane densities, only the total weight of
particles at high altitudes is needed rather than their distribution.
This is important because weight is a quantity integrated over the
vertical column of material and will therefore take the same value
whether or not the mass distribution responsible for it is well
resolved. Any high altitude mass distribution yielding the same weight
will yield the same midplane density so that, given a correct midplane
density, we may infer that the total weight is approximately correct
as well. We may therefore assume that the high altitude mass
distribution derived from our simulations is correct (even though it
may not be), for the purpose of converting densities at a given
altitude to a corresponding number of particles needed to supply the
correct compressional forces to midplane material.

We can quantify the total particle count per vertical column in the
simulations using a modification of the patch averaging strategy used
above. Rather than creating full 3D density distributions for each
patch, it is sufficient only to quantify the averaged number of
particles located in each, and to create a coarse vertical
distribution of the particles by assigning each particle to a
histogram bin according to its altitude. The result is the patch
averaged volume density of particles in each histogram bin, from which
a linear (vertical) particle density per scale height can easily be
obtained by taking a cube root. A final link is to note that
interparticle separations are $\sim h$ and that, by definition, the
histogram bins are of width $H$, so that the linear particle density
is equivalent to the ratio of smoothing length to scale height.

\begin{figure}
\begin{center}
\includegraphics[width=85mm]{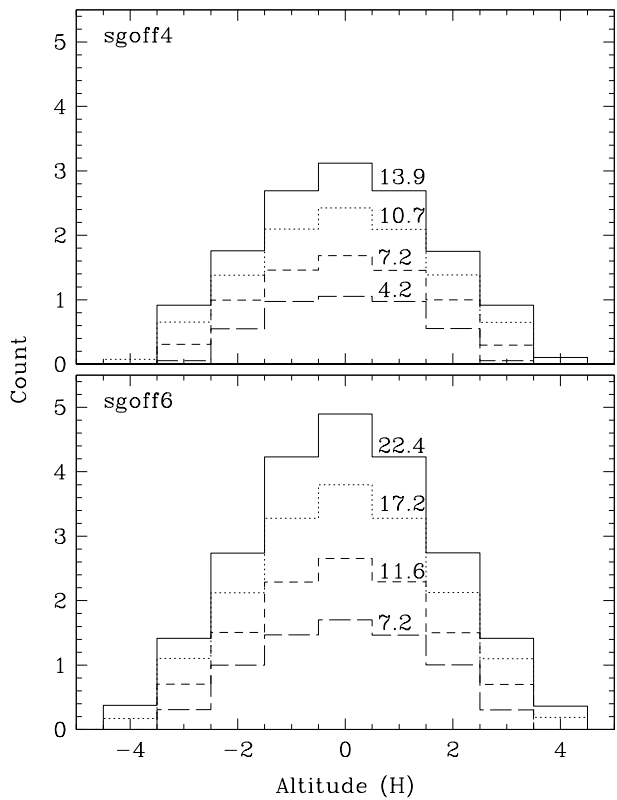}
\end{center}
\caption
{\label{fig:num-vert}
The patch averaged vertical, linear particle density binned in
histograms of width $H$, for the {\it sgoff4} and {\it sgooff6}
simulations. From top to bottom, the histograms in each panel
correspond to the dark blue (solid), green (dotted), red (short
dashed) and black (long dashed) regions as defined in figure
\ref{fig:hiterat-rhodist}. The numerical values associated with each
curve are sums over all histogram bins thus defining the averaged
total particle counts per vertical column.}
\end{figure}

Figure \ref{fig:num-vert} shows histograms of the patch averaged
ratios of $H/h$ defined using this procedure, separated into the same
radial regions defined in section \ref{sec:3dsph-nosg-result}.
Consistent with expectations based on the physical model and on the
fit results above, each histogram displays a maximum at the midplane
and non-zero populations to as high as $\sim3-4$ scale heights in each
direction. Histograms with larger net populations exhibit a
progressively more peaked structure, reflecting the better agreement
with the analytical model provided by the higher resolution. For the
present purposes, it is interesting to note that in every region of
the disk, the total resolution per vertical column is several times
the resolution assuming that only the midplane contributes
significantly to the total. 

In section \ref{sec:3dsph-nosg-result}, we determined that only the
two outermost regions of the {\it sgoff6} simulation were well
resolved in our simulations. Here, we see that on average, the
resolution of vertical columns in these regions is $\sim17$ and
$\sim22$ particles per column for the 9--27~AU and 27--45~AU regions,
respectively. For the lower resolution {\it sgoff4} simulation, the
highest overall vertical resolution reaches $\sim14$ particles per
column, and all other regions in both simulations fall progressively
farther below this value. Since fits in all of these regions deviate
from the analytic expectation by more than 5\%, we therefore conclude
that a minimum of $\sim 17-20$ particles per vertical column are
required to adequately reproduce the density structure at the disk
midplane.

\subsection{Significance of the results of the 3D test
problem}\label{sec:3dsph-nosg-signif}

In self gravitating systems, the exact balance of pressure and
gravitational forces will tip that system towards or away from two
dramatically different outcomes: fragmentation or continued smooth
evolution. In marginally stable systems of interest to modelers, the
exact balance between the two forces will be determined from terms of
nearly equal magnitude, but of opposite sign. If one of those
quantities is incorrectly calculated, the outcome will be dramatically
different. 

The results of the test above demonstrate that simulations with too
few particles tend to underestimate the midplane densities, in the
case of 2.6$\times 10^5$ particles, by as much as 30-35\%. This is
important because the pressure will be underestimated by a similar
factor through the equation of state. Fragmentation will therefore be
enhanced in an under resolved simulation compared to that of either a
well resolved simulation or, more importantly, a real, physical
system. Because we have not performed self gravitating simulations, we
cannot specify the exact effects the errors will have in simulations
or the level of enhanced fragmentation that may occur however. Such
details are difficult to specify with precision because any
enhancements may be mitigated in part by the fact that the cause of
erroneously low densities in SPH is simply that particles separations
are greater than they should be, so that gravitational attraction
between them is correspondingly less. Any possibility of mitigation of
this sort provides little comfort relative to simply solving the
hydrodynamic equations accurately in the first place however. 

For the disk morphology discussed here, resolution sufficient to
accurately reproduce the disk's vertical structure and midplane
density beyond $\sim10$~AU requires at least one million particles.
Although we have not attempted to quantify the differences through
simulations, colder disks closer to the fragmentation boundary will
have correspondingly smaller scale heights, and will therefore require
still more total particles to ensure adequate resolution.

For example, the disk scale height and $Q$ values both depend directly
on the sound speed, so a change in one will be reflected
proportionally in the other. A disk with minimum $Q=1.3$ or $Q=1.1$
will decrease the scale height below that in our simulations by a
factor $1.3/1.5\approx0.87$ or $1.1/1.5\approx0.73$ respectively. In
order to retain adequate resolution of the vertical structure,
smoothing lengths must be decreased by a similar factor by increasing
resolution. In 3D, the magnitude of the increase will be a factors of
$1/0.87^3\approx1.5$ or $1/0.73^3\approx2.5$ corresponding to
$\sim1.5$ or $\sim2.5$ million particles. No simulations of
circumstellar disks in the context of planet formation have yet been
performed at such high resolution. 

Even at a resolution of 1 million particles, the vertical structure in
our simulations is not accurately reproduced inside 10~AU, where
midplane densities drop well below their analytic values. This is
important for simulations of gravitational fragmentation because the
Jovian planets in our own solar system formed at such radii. The
erroneously low densities may lead to enhanced fragmentation in those
regions, even if the structure is accurately modeled further out.
Also, the resolved and unresolved parts of the disk are in no way
isolated from each other. As waves or other structures propagate
through the disk, their subsequent evolution will be perturbed by the
change in resolution, so that the evolution throughout the entire
simulation become suspect.

Both the small values of the scale height at small orbit radii and the
insufficient resolution there are consequences of the $r^{-1/2}$
temperature profile assumed for the disk, and commonly used throughout
the literature. Through the sound speed and rotation curve, the scale
height will exhibit a proportionality $H/r\propto r^{1/4}$ and, for
minimum $Q=1.5$, will take values of $\sim0.025$ near 1~AU increasing
to $\sim0.065$ near 45~AU. The temperature profiles in our simulations
reflect the stage of active disk evolution during which gravitational
instabilities are most likely to be present. A temperature profile
inversely proportional to radius will yield a flat profile and so
avoid the issues of variable vertical resolution, but such steep
profiles are not observed in real systems \citep{BSCG}, and so we
discount them here.

\subsection{Additional requirements on vertical resolution for
simulations including radiative transfer}\label{sec:3dsph-RT}

Above, we showed that the vertical structure of an accretion disk must
be resolved with some minimum number of particles if mass densities in
the disk midplane are not to be substantially underestimated. Although
this requirement may be sufficient to ensure accurate simulations in
the case when comparatively simple physical models are employed, it
will be less so when radiative transfer is included. Here we describe
an exercise meant to illustrate requirements for resolution of the
high altitude structure of the disk, if serious errors in the cooling
rates derived from radiative emission from the disk photosphere are
also to be avoided.

At the orbit radii most relevant for planet formation ($\la20$~AU),
disks will be optically thick in the sense that the optical depth,
$\tau$, calculated from infinite distance to the disk midplane is
large. This is important because it means that the cooling rate of
packet of disk material will be well modeled by a blackbody cooling
law whose temperature is defined at the disk photosphere. If the
altitude of the photosphere is known only approximately, due to low
resolution or simple miscalculation, the cooling rate will be
similarly approximate because the temperature structure near the
photosphere is not known precisely. Comparatively small errors in
temperature will lead to much larger errors in the local cooling rate
because of the $T^4$ proportionality of the blackbody cooling
function.

Accurate knowledge of the cooling rate and its associated photosphere
temperature is important because the dynamical balance of heating and
cooling processes in disks is quite precarious: most of the heating
and cooling sources are capable of removing or replacing essentially
all of the disk's thermal energy over the course of only a few orbits
\citep{DBMNQR_PP5}. Specifically for the question of disk
fragmentation, \citet{gammie01} shows that cooling rates faster than
$\sim3/\Omega$, where $\Omega$ is the local orbit frequency, lead to
disk fragmentation, but longer cooling rates may not in a local
calculation. Later global calculations of \citet{RABB} confirmed this
result for low mass disks, where they found that a cooling time of
$5/\Omega$ was long enough to prevent fragmentation, but a cooling
time of $3/\Omega$ was not. Simulations of higher mass disks with
cooling rates of 10 and 5 $\Omega^{-1}$ showed similar changes in
behavior. From these results it is clear that an error in the cooling
rate as large as a factor of two will be sufficient to suppress or
enhance fragmentation in a simulation which would otherwise follow a
very different evolutionary path. A error of factor of two in the
cooling rate will be equivalent to an error of $\sim20$\% in the
photosphere temperature, due to the $T_{eff}^4$ dependence of the
cooling on the temperature. Since an error of such magnitude may  
seriously alter the balance between cooling and heating, a more
restrictive limit must be set to ensure that any fragmentation that
does occur is not subject to numerical error. If we set an arbitrary
standard that a cooling rate accurate to 25\% will be sufficient to
ensure results are not contaminated by numerical errors, then we will
in turn require a photosphere temperature accurate to $\sim5-6$\%.

\begin{figure}
\rotatebox{0}{\includegraphics[width=80mm]{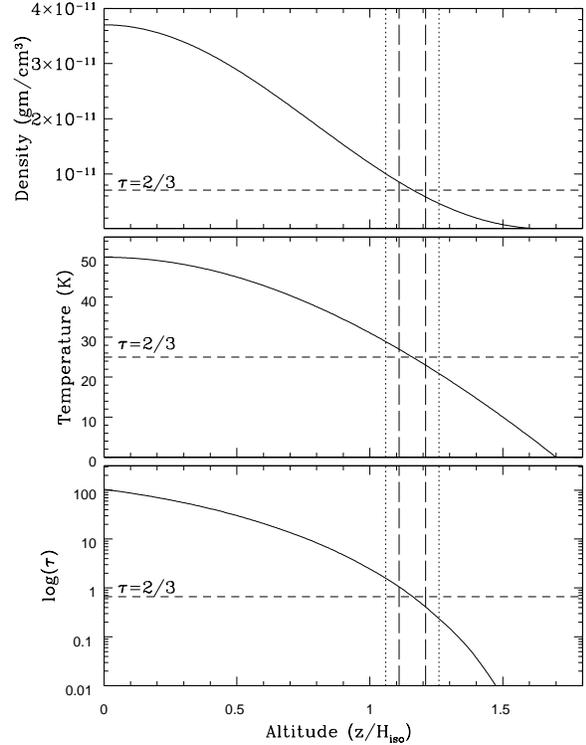}}
\caption
{\label{fig:vert-struct}
Vertical density (top), temperature (middle) and optical depth
(bottom) profiles of a disk at 10~AU, with local surface density of
500 gm/cm$^2$ and a midplane temperature of 50~K, using the model of
\citet{DynII} to determine structure. The altitude coordinate is
normalized to the value of the isothermal scale height $H_{\rm iso}$
determined at the midplane. Dotted and long dashed vertical lines on
each curve are placed to straddle the photosphere surface at
$\tau=2/3$ (horizontal dashed line), at distances of $H_{\rm iso}/10$
and $H_{\rm iso}/20$, respectively. Over these distance, the
temperature deviates $\sim15$\% and $\sim7$\% above or below the true
photosphere temperature.} 
\end{figure}

Given the requirement as stated, it remains to convert the constraint
on the photosphere temperature determination to a constraint on
required spatial resolution of the vertical disk structure. One
approximate model of the vertical structure of a disk is available
from the work of \citet{DynII}, in which an isentropic vertical
structure is assumed, for a locally plane parallel, self gravitating
disk. Figure \ref{fig:vert-struct} shows temperature and optical depth
profiles as a function of altitude for a region of a disk expected to
be interesting for disk fragmentation, derived from the
\citealt{DynII} model. Several points are of interest in evaluating
the figure. First, although the density structure follows an
approximately Gaussian structure, at least in coarse outline, it is
dramatically compressed relative to that expected for an isothermal
disk. No material is present above an altitude of $\sim1.7H_{\rm
iso}$, due to the different physical assumptions employed in deriving
the mass distribution. Second, this configuration is optically thick,
with an optical depth to the midplane of $\sim100$, meaning that a
disk photosphere surface may in fact be defined.

For the purposes of evaluating the radiative energy losses, a third
point will be of most importance. Namely, that near the disk
photosphere surface both the density and temperature are changing
rapidly with altitude. A calculation that resolves these profiles too
coarsely will therefore also only coarsely resolve the photosphere
surface itself, possibly exposing the hot disk interior to space. For
example, if the particle density near the photosphere for an SPH
simulation is such that smoothing lengths are $h\approx H_{\rm
iso}/10$, then to a first approximation, the photosphere surface
itself will be resolved on the same spatial scale because
interparticle separations are themselves $\sim h$. A particle actually
located anywhere within $h$ of the actual photosphere surface may be
determined to define the photosphere surface, simply because no other
particle happened to be at a higher altitude there. An identical
argument holds for grid based simulations, with the size of a zone at
the photosphere surface replacing smoothing length.

Figure \ref{fig:vert-struct} shows that uncertainties in the altitude
of the photosphere surface of $\sim H_{\rm iso}/10$ results in an
increase or decrease of the derived photosphere temperature of
$\sim15$\%. Doubling the linear resolution to $\sim H_{\rm iso}/20$
decreases the uncertainties to $\sim 7$\%, comparable to our
requirement of errors no larger than $\sim-5-6$\%. We therefore
conclude that simulations including radiative cooling must resolve the
disk photosphere surface at a spatial scale no coarser than $\sim
H_{\rm iso}/20$, when structure models similar to ours are employed.

After completing the exercise and deriving this requirement, we must
also immediately point out that the quoted required resolution is not
universally applicable without modification. Any constraint on spatial
resolution in the context of radiation will be considerably
complicated by the fact that the vertical temperature profile will be
significantly influenced both by preferential dissipation of shocks at
high altitudes \citep{Pick03}, by radiative heating by external
sources like the central star \citep{CG97} and, if any remains at the
time of the simulation, the surrounding molecular cloud. As
\citealt{CG97} also note, wavelength dependent opacities will also
play a role, to the extent that long wavelength radiation is able to
cool the disk interior effectively, even while shorter wavelength
radiation remains trapped. 

In spite of these cautions, we remain convinced that the above
exercise remains a valuable illustration. It is based only on the
condition that the transition between optically thick and optically
thin material is well resolved and, although it is applied to a
particular model of the vertical structure, it is not tied to it
specifically. Given the overall gross vertical structure of disks, in
which densities are high at the midplane and drop steeply as altitudes
increase, optical depth profiles similar to that in figure
\ref{fig:vert-struct} will occur in other models as well. Because the
optical depth is a very steeply falling function of increasing
altitude, the transition region will be narrow, as in the model above.
Whether that narrowness can be translated directly into a constraint
on the spatial resolution will depend on the details of the
temperature profile there, and its influence on the accuracy of
radiative energy transfer. The specific criterion as applied above
presumes a temperature profile decreasing with altitude as the density
decreases, $T\propto\rho^{\gamma-1}$. If instead a much flatter
profile exists across the transition, a less restrictive criterion may
still permit sufficiently accurate evolution.

\section{Concluding remarks}\label{sec:remarks}

After exploring the conditions required for simulations to produce
numerically valid results, we turn now to summarizing those
conditions, to comparing the work done here with previous work in the
literature, and to commenting on additional requirements beyond
numerics that are required for simulations to accurately reproduce
the evolution of real systems.

\subsection{Summary of our main results}\label{sec:summary}

Our first result is to extend the numerical criterion for the validity
of simulations involving collapse in clouds and based on the analytic
Jeans collapse formalism (BB97, \citet{Truelove97}), to the case of
disk systems where the geometry of the mass distribution is both
flattened and rotating so that the Jeans treatment is inapplicable.
Whereas a minimum spatial resolution per Jeans wavelength or minimum
number of particles per Jeans mass is required in the cloud collapse
case for grid or particle based simulations, respectively, in
simulations of disks we require a minimum spatial resolution per
`Toomre wavelength' or minimum number of particles per `Toomre mass',
for grid based or particle base simulations, respectively. The two
forms of the new criterion are given in equations \ref{eq:Toom-cond}
and \ref{eq:max-surfdens}. 

We also note that simulations failing to resolve one or the other
condition yield quite different outcomes. In the case of the Jeans
condition, BB97 conclude that insufficient resolution will delay or
suppress numerically induced fragmentation with softening and
smoothing equal, and only enhance it when smoothing is much larger
than softening. Our results demonstrate that a simulation that fails
to resolve the Toomre condition is characterized by enhanced
fragmentation compared to one that is well resolved, even when
softening and smoothing are equal. This fact allows an empirical
determination of the minimum resolution required in a simulation in
order to ensure fragmentation that develops is not of numerical
origin. For 2D SPH simulations, we have determined the minimum number
of particles per Toomre mass as six times the average number of
neighbors ($\sim20$) used for the realization of the hydrodynamic
quantities themselves.

We have not determined a value of \nreso\  for 3D models,
but its seems unlikely to be dramatically different from 2D, since the
same physical system is being modeled with the same numerical
technique. To the extent that differences may be present, we expect
that the criterion will be more conservative (more particles required)
based on two arguments. First, unstable waves in the Toomre analysis
are intrinsically two dimensional, so resolving their third dimension
is meaningless: small particle separations or small grid spacing in
that coordinate do not contribute to increasing the resolution of the
waves themselves. Using the same empirical
fragmentation/non-fragmentation limiting condition used in section
\ref{sec:test-crit} to establish a required particle count, will
therefore overestimate the required resolution because some neighbors
will be counted only for purposes of establishing a value for \nreso,
but will not actually serve to increase the resolution insofar as the
Toomre condition is concerned. Secondly, although the Toomre criterion
implicitly assumes validity of the hydrodynamic equations, in practice
the numerical method may fail to do so (section \ref{sec:disks-3d}),
with the consequence that pressure forces are underestimated. Although
a vertical resolution requirement is nominally an entirely different
condition from resolving instability wavelengths, we see no practical
manner in which they might be separated from each other in the
empirical tests we describe. Therefore, calibration of the
requirements for the Toomre condition in 3D will implicitly
incorporate both requirements from the wave analysis and from the
vertical resolution condition, yielding a resolution requirement
defined by contributions from both sources.

A second important result of our work is to reconcile a variety of
mutually inconsistent definitions of the Jean collapse stability
criteria made by different authors, due to the use of different
definitions of the Jeans mass itself. We show that using the same
definition of the Jeans mass, a particle based simulation will require
approximately 600 particles to resolve one Jeans mass, while a grid
based simulation will require about 64 zones (i.e. $1/J^3$ where $J$ is
defined in equation \ref{eq:Jeans-cond}, and $J\sim 1/4$). The minimum
resolution of 600 particles per Jeans mass corresponds to
approximately 12 times the average number of neighbor particles
($\sim50$) used by BB97 in their SPH code. This value is a factor of
two higher than the factor of six times the average number of
neighbors as derived from the 2D disk simulations explored in section
\ref{sec:2dtest-evo}. 

Third, we discuss the importance of the implementation of
interparticle forces in particle based simulations of self gravitating
disks. We show that a naive translation of the 3D method for
gravitational softening in SPH into 2D leads to an unphysical, finite
gravitational force at zero separations between particles. In
consequence, significant particle `pairing' occurs as simulations
proceed, resulting in an effectively time dependent resolution. We
consider two alternatives to remedy the behavior. First, we
artificially modify the derivative of the SPH kernel according to the
TC92 prescription, generating a finite, repulsive pressure force at
zero separation which compensates for the attractive gravitational
force. Second, we artificially modify the gravitational softening
itself, to produce instead a zero force at zero separation, as is
physically appropriate.

Both alternatives are effective in reducing the likelihood of disk
simulations to produce fragmentation and of particles in those
simulations to become paired. Although the TC92 variant appears to be
slightly more effective (see e.g. section \ref{sec:enhancement}), we
view it as marginally the less preferable of the alternatives because
it too produces a finite force at zero separation, albeit of opposite
sign so that a sum of gravitational and pressure forces yields a small
net force. Finite forces at zero separation, whether or not they are
combined with others to produce a near zero sum, must be considered a
less physical and therefore less desirable alternative to a solution
which retains the collisionless nature of the model more directly. 

Fourth, we show that force imbalances on the spatial scale of the
smoothing and softening develop when the two length scales are not
equal. While acting only at very small spatial scales, these
imbalances can induce large scale changes in the outcome of
simulations where they exist, including artificially induced
fragmentation in otherwise stable systems. To avoid such induced
outcomes, we conclude that the gravitational softening length and the
hydrodynamic smoothing length of each particle must be very nearly
identical. As noted above, our conclusion for disk simulations is
similar to that made by BB97 in their study of collapsing clouds, but
stronger in the sense that it holds even when the systems are, and are
expected to remain, properly resolved according to the Toomre
criterion.

In neither the disk or cloud case however, does our conclusion come
without price. Specifically, in order to better resolve the flow in
both high and lower density regions of the same simulation, most modern
SPH codes set the smoothing length dynamically as a function of the
local flow. Setting the smoothing and softening lengths equal means
that the softening length also must vary in time. This is important
because allowing temporally varying softening is equivalent to
explicitly allowing a violation of conservation of energy in the
simulation. While we believe the magnitude of violation to be small,
its effect must be considered when evaluating self gravitating
particle based simulations. 

Finally, we discuss the importance of adequately resolving the
vertical structure of 3D disk simulations. We find that insufficient
resolution of the vertical structure leads to substantial deficits in
the realized midplane densities compared to the predictions of
analytical models. We find that resolution comparable to at least
$\sim4$ smoothing lengths per disk scale height in the disk midplane
are required to accurately fit the coefficients in analytic formulae
for the vertical structure of non self gravitating isothermal disks in
particle simulations. Restated in terms of the full thickness of the
disk, the criterion is equivalent to the statement that at least
$\sim17-20$ particles are required per vertical column. While we have
not studied the effects of resolution in 3D grid based simulations, we
expect similar criterion to hold for them as well. 

The consequences of failures to resolve the vertical structure in self
gravitating systems will be that pressure forces will be incorrectly
determined. Any balance between them and self gravitation will
therefore be falsely biased in favor of gravitational fragmentation.
Presumably, the bias will distort the evolution of waves in the disk
(whether self gravitating or not) even when fragmentation does not
occur, because the density and pressure deviations imply a loss of
fidelity of the solution to the equations of hydrodynamics provided by
the numerical method. 

We discuss the fact that while satisfying the criterion above is
necessary for simulations to accurately evolve circumstellar disks, it
may not be sufficient for them to do so alone, depending on the goals
of the researcher. Simulations for which the criterion is satisfied
everywhere may still under resolve high altitude structure of the
disk, a fact which will be of great importance for models in which
radiative transport is included. Such models must also resolve the
temperature structure at the disk photosphere in order not to be
contaminated by unacceptably large numerical errors. We provide an
example of the requirements when an isentropic vertical structure is
present, for which we derive the computationally extremely demanding
condition that the photosphere must be resolved at a spatial scale
smaller than $\sim H/20$ to avoid serious errors. We point out however
that the exact requirement will be highly sensitive to the details of
the model and its effects on the temperature, density and opacity
structures at high altitudes.

\subsection{Comparisons to particle based simulations in the
literature}\label{sec:part-comps}

Our test problem clearly shows that the fragmentation processes seen
in \citet{DynI} were due to insufficient resolution, and we believe
that many other works in the literature both before and since may be
similarly affected. Of the work discussing particle based simulations,
the simulations done by \citet{Mayer02,Mayer04} are among the highest
resolution simulations so far published, evolving either $2\times
10^5$ or $10^6$ particles (in various runs) in three dimensions. It
may therefore be particularly useful to contrast our test problems
with the conditions employed by them, their results and conclusions.

They perform a series of SPH simulations with several disk masses
between \mrat=0.075 and \mrat=0.125, each with minimum Toomre $Q$
values between 0.8 and 1.9. They show that the more massive disks
fragment on a time scale of a few hundred years (about 10 orbits in the
region where the clumps form) when the minimum $Q$ values in the disk
are below \qmin$\approx1.4$. They also find that clump formation is
more vigorous with higher resolution simulations of the same initial
condition that with lower. An isothermal gas equation of state was
used initially for most of the models, and after a critical over
density was exceeded, was switched over a small transition time to an
adiabatic evolution that included heating due to compression,
viscosity and shocks. Other models used an adiabatic equation of state
for all times, and resulted in fragmentation that was much weaker or
non-existent.

We first attempt to obtain some indications of the degree to which the
flow is well resolved by the condition in equation
\ref{eq:max-surfdens} at different times during the evolution, using
their discussion of the physical features of the flow and resolution.
For example, prior to clump formation the most unstable wavelength is
quoted as $\sim2$~AU, including a correction factor for 3D effects.
The flow is clearly well resolved by the simulation at this early
stage, but the value of the Toomre wavelength\footnote{Note that their
definition of the `Toomre wavelength' is that of equation
\ref{eq:disk-critwavel}, but that their later definition of the most
unstable wavelength matches that of equation \ref{eq:Toom-wavel}.} at
this time is only representative of the mildly non-linear regime of
the background flow. It is not indicative of the later, more strongly
non-linear regime near the forming clump, for which both the locally
applied Jeans based or Toomre based resolution criteria are most
useful.

Well after a clump has formed, when the clump has become a well
defined, condensed and independent entity rather than a part of the
disk, we expect that the Toomre based criterion will become invalid.
At this time, \citealt{Mayer04} state that the resolution employed in
their simulations was sufficient to resolve the Jeans mass (for which
the underlying analysis is valid if the clump is much smaller than a
disk scale height) with a few hundred to a few thousand particles,
even though the densities there reached as much as $\sim10^6$ higher
than the background. An important detail of the implementation of
\citealt{Mayer04} is that the number of neighbors used for realizing
the hydrodynamic quantities is fixed at 32, a value common throughout
the cosmological SPH literature, but significantly smaller than the
value of $\sim50$ used (but not specifically recommended) by BB97.
Correcting for this systematic difference would not appear to alter
the statement that the Jeans mass is properly resolved however,
because while the number of particles per Jeans mass would be lower,
it would still remain above the $\sim100$ quoted by BB97.

We can make relatively few direct inferences about intermediate times
when the clump has begun to form but is still unbound or only weakly
bound. At such times the local value of $Q$ may fall below one half,
and we expect a much stronger resolution constraint from the Toomre
criterion than from the Jeans criterion (see equation
\ref{eq:TJ-relative}), especially if the approximation that
$\Omega\approx\kappa$ is not made. Therefore, while the clump may be
well resolved by the Jeans criterion, it may not be similarly well
resolved by the Toomre criterion. Based solely on the number of
particles and the similarity of the initial conditions to our own test
problem, we conclude that the high resolution (1 million particle)
simulations of \citealt{Mayer04} appear to be well resolved by our
Toomre based resolution criterion. We can make a similar, but much
more tentative conclusion about the lower resolution ($2\times10^5$
particle) versions in the same work. A more detailed examination of
the simulations at both resolutions in the context of the Toomre
resolution criterion would be desirable, particularly in the context
of the sensitive dependence of the resolution requirement to the
initial minimum value of $Q$ and on the local value of $Q$ in the
fragmenting region.

\citealt{Mayer04} advocate a fixed gravitational softening whose
length scale is half of the initial hydrodynamic smoothing length of
the particles. With this ratio of softening to smoothing, BB97 show
that the gravitational force can exceed the pressure force by a factor
of as much as seven on size scales of $h/2$, and by a factor of 2--3
on scales of $h$. The actual length scale used is determined (Mayer
2004 private communication) as an average of the smoothing lengths in
the `flat' temperature region of the disk, corresponding approximately
to the outer half of its radial extent, meaning that the outer half of
their disks will have small scale force imbalances that originate in
the numerical treatment of softening and smoothing. 

The results of our simulations show the consequences of such
imbalances in disk simulations. In regions where the hydrodynamic
smoothing length exceeded the gravitational softening length,
fragmentation was strongly enhanced over that of the variable
softening case, even when sufficient numerical resolution was used
according to the Toomre criterion. Since the force imbalance is not
due to any physical origin, but only to a breakdown in the simulation
method itself, we conclude that clumping observed in the simulations
may not be an accurate depiction of an underlying physical system, and
should be reexamined to ensure that they are not subject to this
numerical flaw.

Of still greater concern is the resolution of their simulations in the
context of the discussion in section \ref{sec:3dsph-nosg-signif}. Both
the morphology of the system modeled by \citealt{Mayer04} and the
physics itself are different than those employed by our test models,
however it seems unlikely that these modifications will alter the
required resolution by the large factors needed to conclude that their
vertical structure is well resolved, especially for their most
unstable (lowest $Q$) models. The extent to which any violation of the
criterion actually affects fragmentation is unclear however. It is
likely that a definitive statement of the effects will require a
complete study of the problem at extremely high resolution, making the
question quite costly to address at present.

Several other recent papers have discussed fragmentation in disks.
Similar concerns regarding the vertical resolution will apply to many
of them. For example, in \citet{Boss04}, as well as his earlier,
similar work, the vertical coordinate (the spherical coordinate
$\theta$ in his case) is resolved with a total of 23 zones. Although
they are asymmetrically arranged to concentrate resolution where they
are needed most, only some $\sim$10-12 lie within ten degrees
($H/R\approx0.17$) of the midplane, where most disk matter will be
located. Although our vertical resolution requirement has been tested
only with SPH simulations, there is no reason to believe that an
analogous requirement will be required for grid based simulations as
well, since they must solve the same set of equations. Naively
applying the $\sim4$ smoothing lengths per scale height as a $\sim4$
grid cells per scale height criterion directly to the Boss
simulations, we observe that the vertical structure may be very near
the required limits. The exact number of cells will undoubtedly be
different than a straightforward carryover however, so the extent to
which Boss's or any other any grid based simulations will be affected
by the errors arising from deficient vertical resolution is unclear.

\citet{RABB} showed that for sufficiently fast cooling rates, clumping
could occur in disks resolved with $2.5\times10^5$ particles. Due to
their cooling prescription, disks in their work dropped to globally
averaged $Q$ values near unity, well below those in our simulations,
for which we determined erroneously low midplane densities developed.
In later work discussing similar systems (with overall morphologies
identical to those in our study, but with additional physics),
\citet{LR04} claim that their vertical structure is resolved with
$\sim5$ smoothing lengths per scale height. It is difficult to
reconcile this result with our own figure \ref{fig:hiterat-rhodist},
especially given the much lower $Q$ values (see their figure 3)
realized in their study and in the earlier work of \citealt{RABB}. It
appears likely however, that low vertical resolution has likely
influenced their results through the artificially low pressure forces
induced by density deficits.

Apart from the issue of vertical resolution, we note that their
overall resolution (in 3D) is similar to that in our highest
resolution 2D test problem. The authors have not discussed resolution
issues in the context of either Jeans or Toomre stability, so we can
draw few inferences about the viability of their results under either
the criteria of equation \ref{eq:max-voldens} or
\ref{eq:max-surfdens}. In the context of the latter condition however,
the $Q$ values throughout their simulations rapidly approach values
near unity. This is important because equation \ref{eq:max-surfdens-Q}
shows that our resolution requirement is quite sensitive to the value
of $Q$, so that much higher resolution will be required for later
evolution even if the initial condition is known to be well resolved
by the Toomre condition. 

Other simulations of \citet{Luf04} perform a large parameter study of
disks already containing one planet, in order to explore the
possibility that its present may trigger the formation of further
objects. For their parameter study, they use simulations with $10^5$
particles, but with varying minimum $Q$ or initial planet mass. The
initial conditions of these disks were in most respects quite similar
to those of Mayer \etal, however the resolution employed was far lower
in order to facilitate the large parameter study. Given the much lower
resolution, it appears likely that these authors have under resolved
the vertical structure of the disk, resulting in incorrectly low
densities and pressures and the possibility of artificially enhanced
fragmentation. While an attempt to duplicate their study is beyond the
scope of this work, we have run selected simulations sampling their
parameter space and have observed that they resolve (or fail to
resolve) the Toomre condition at similar points in the parameter space
for which they conclude that clumping was induced or not induced by
the passage of the seed planet. Due to the probable violation of both
criteria, we therefore believe that their conclusions have been
contaminated by insufficient resolution and must be reexamined.

\subsection{Beyond numerics: what is required for the viability of the
disk instability model for Jovian planet formation to be either
verified or falsified}\label{sec:beyond}

In this work, it has not been our purpose to discuss all of the
requirements for simulations modeling gravitational fragmentation in
disks to produce correct and physically relevant results. Such a
description of a real physical system requires not only numerically
valid simulations, but also relevant initial conditions and a correct
and complete physical model. Numerically valid simulations will be
interesting only to the extent that they model real systems with real
physics.

Of particular interest for the physical understanding of disk
evolution is the fact that our highest resolution SPH test model did
not produce {\it any} clumps although it was evolved for much longer
in time than its lower resolution cousins. This is an important
physical result because the test problem employed a locally isothermal
equation of state, which is usually (and erroneously-see the
discussion in \citet{DynII,Pick03}) equated with the strong cooling
limit where fragmentation is nearly inevitable. The fact that clump
formation did not occur in this limit, in a properly resolved
simulation, may place the disk instability model for planet and
brown dwarf formation on considerably less solid ground.

At the same time, the Boss simulation, implementing conditions that
are much less biased toward clumping, {\it did} produce clumps while
remaining numerically valid according to the Toomre condition, though
some doubt may remain regarding the resolution of the vertical
structure. His simulations include a description of radiative cooling
processes in 3D, and he concludes that the clumping is due both to the
efficient radiative cooling and to the fact that convection in the
disk's vertical direction is efficient. Convection is able to
transport thermal energy out of the optically thick disk midplane to
its photosphere surface at higher altitudes, so that even optically
thick regions like forming clumps cool efficiently. On the other hand,
previous 2D work of \citet{DynII}, which assumed efficient vertical
convection as a consequence of the radiative cooling prescription
employed, did not produce clumps. Moreover, the total radiated energy
emitted by the disks modeled in those simulations seriously
underestimated the values observed in real systems because they were
too cold. In a more complete physical model of the disk energy budget
of heating and cooling processes, the disks would have to be warmer in
order to reproduce the observations, making clumping even less likely. 

Other fully 3D radiative transfer \citep{MDPC} models confirm neither
the Boss conclusion about the efficiency of vertical convection nor
the clumping that results from it. The authors continue to investigate
the origin of the discrepancies, but no certainty yet exists.
Possibilities include both differences in the numerical treatments,
for example of the boundaries, the resolution of both the Poisson
solver and the hydrodynamic codes themselves, and of the physical
models, for example the treatment of external radiation field, what
opacities are used in the radiative transport and the equation of
state for the gas. The important point to take from these discrepant
results is that not only do questions of numerical validity remain to
be addressed, but also the physical models. Although progress has
certainly been made, to date an insufficient understanding of what
physical processes are important for fragmentation has been developed.
Determination of the true viability of the disk instability model for
Jovian planet formation therefore still lies in the future.

\section*{Acknowledgments}
Except for the simulation discussed in figure \ref{fig:Boss-crit},
generously shared prior to its publication by Alan Boss, the
computations reported here were performed using the UK Astrophysical
Fluids Facility (UKAFF), specifically the UKAFF Fellow guaranteed time
allocation, which also provided financial support. This work was
partially carried out under the auspices of the National Nuclear
Security Administration of the U.S. Department of Energy at Los Alamos
National Laboratory under Contract No. DE-AC52-06NA25396 and
W-7405-ENG-36, for which this is publication LA-UR-05-5347. We
gratefully acknowledge useful conversations and correspondence with
Alan Boss, Richard Durisen, Annie Mejia, Megan Pickett, Lucio Mayer
and Willy Benz during the evolution and preparation of this work.
Special note is given to Richard Durisen, with whom discussion of the
Toomre resolution criterion work shortly before IAU Symposium 200,
gave original impetus to its eventual publication. We also appreciate
the comments of the co-referees Matthew Bate and Daniel Price during the
contentious review process experienced by this paper over two years
and five revisions. Finally, we would like to acknowledge the support 
of the University of Edinburgh Development Trust.


\begin{thebibliography}{}

\bibitem[Armitage \& Hansen(1999)]{ArmHan99} Armitage, P. J., 
Hansen, B. M. S., 1999, Nature, 402, 633 

\bibitem[Athanassoula \etal(2000)]{Athan00} Athanassoula, E., Fady,
E., Lambert, J. C., Bosma, A., 2000, MNRAS, 314, 475 

\bibitem[Balsara(1995)]{Bals95}  Balsara, D., 1995, J. Comp. Phys,
121, 357

\bibitem[Bate \& Burkert(1997)]{BB97} Bate, M. R., Burkert, A.,
1997, MNRAS, 288, 1060 (BB97)

\bibitem[Beckwith \etal(1990)]{BSCG} Beckwith, S. V. W., Sargent, A. I., 
Chini, R. S. \& G\"usten, R., 1990, AJ, 99, 924

\bibitem[Benz(1990)]{Benz90} Benz, W. 1990 in  The Numerical Modeling
of Nonlinear Stellar Pulsations p. 269, J. R. Buchler ed.

\bibitem[Binney \& Tremaine(1987)]{GalDyn} Binney, J.,
Tremaine, S., 1987, Galactic Dynamics, Princeton University
Press: Princeton

\bibitem[Boss(1998)]{Boss98} Boss, A. P., 1998, ApJ, 503, 923

\bibitem[Boss(2000)]{Boss00} Boss, A. P., 2000, ApJ, 536, L101

\bibitem[Boss(2002)]{Boss02} Boss, A. P., 2002, ApJ, 576, 462

\bibitem[Boss(2004)]{Boss04} Boss, A. P., 2004, ApJ, 610, 456

\bibitem[Chiang \& Goldreich(1997)]{CG97} Chiang, E. I., Goldreich,
P., 1997, ApJ, 490, 368

\bibitem[Dehnen(2001)]{Dehnen01} Dehnen, W., 2001, MNRAS, 324, 273

\bibitem[Durisen \etal(2006)]{DBMNQR_PP5} Durisen R. H., Boss, A. P.,
Mayer, L., Nelson, A. F., Quinn, T. \& Rice, W. K. M. 2006, In
Protostars and Planets 5 in press, ed. Reipurth, B., Jewitt, D. \&
Keil, K. University of Arizona Press:Tucson 

\bibitem[Fletcher(1997)]{fletcher97} Fletcher, C. A. J., 1997,
Computational Techniques for Fluid Dynamics: Fundamental and General
Techniques, Second Edition, Springer-Verlag: Berlin

\bibitem[Fryxell, M\"uller \& Arnett(1991)]{FMA} Fryxell, B. A.,
M\"uller, E., Arnett, D., 1991, ApJ, 367, 619

\bibitem[Fulbright, Benz \& Davies(1995)]{ful95} Fulbright, M. S.,
Benz. W., Davies, M. B., 1995, ApJ, 440, 254

\bibitem[Gammie(2001)]{gammie01} Gammie, C. F., 2001, ApJ 553, 174

\bibitem[Goldreich, Goodman \& Narayan(1986)]{GGN86} Goldreich, P.,
Goodman, J., Narayan, R., 1986, MNRAS 221, 339

\bibitem[Herant(1994)]{Herant94} Herant, M., 1994 Mem. S. A. It, 65,
1013

\bibitem[Hernquist \& Katz(1989)]{hk89} Hernquist, L., Katz, N. 1989,
ApJS, 70, 419

\bibitem[Hockney \& Eastwood(1988)]{hockeast88} Hockney R. W. \&
Eastwood, J. W., 1988, Computer Simulations Using Particles,
Institute of Physics Publishing: Bristol

\bibitem[Hubber, Goodwin \& Whitworth(2006)]{HGW06} Hubber, D. A.,
Goodwin, S. P. \& Whitworth, A. P. 2006, MNRAS, 450, 881

\bibitem[Imaeda \& Inutsuka(2002)]{II02} Imaeda, Y., Inutsuka, S.,
2002, ApJ, 569

\bibitem[Koller(2004)]{Koller} Koller, J., 2004, PhD Thesis, Rice
University

\bibitem[Krumholz \etal(2004)]{KMK04} Krumholz, M. R., McKee, C. F.,
Klein, R. I., 2004, ApJ, 611, 399 

\bibitem[Leveque(2002)]{leveque02} Leveque, R. J., 2002, Finite
Volume Methods for Hyperbolic Problems, Cambridge University
Press:Cambridge

\bibitem[Lin \& Lau(1979)]{LinLau79} Lin, C. C. \& Lau, Y. Y., 1979,
Studies in Applied Mathematics, 60, 97

\bibitem[Lodato \& Rice(2004)]{LR04} Lodato, G., Rice, W. K. M., 2004,
MNRAS, 351, 630

\bibitem[Lufkin \etal(2004)]{Luf04} Lufkin, G., Quinn, T. Wadsley, J.,
Stadel, J., Governato, F., 2003, MNRAS, 347 421

\bibitem[Mayer \etal(2002)]{Mayer02} Mayer, L., Quinn, T., Wadsley,
J., Stadel, J., 2002, Science, 298, 1756

\bibitem[Mayer \etal(2004)]{Mayer04} Mayer, L., Quinn, T., Wadsley,
J., Stadel, J., 2004, ApJ, 609, 1045

\bibitem[Mejia \etal(2005)]{MDPC} Mejia, A. C., Durisen, R. H., Pickett,
M. K., Cai, K., 2005, ApJ, 619, 1098

\bibitem[Monaghan \& Lattanzio(1985)]{MonLat85} Monaghan, J. J.,
Lattanzio, J. C., 1985, A\&A, 149, 135

\bibitem[Monaghan(1992)]{Mon92} Monaghan, J. J., 1992, ARAA, 30, 543

\bibitem[Nelson \etal(1998)]{DynI} Nelson, A. F.,           
Benz, W., Adams, F. C., Arnett, W. D., 1998, ApJ, 502, 342   

\bibitem[Nelson \etal(2000)]{DynII} Nelson, A. F., Benz, W.,
Ruzmaikina, T. V., 2000, ApJ, 529, 357                       

\bibitem[Nelson(2003)]{N03} Nelson, A. F., 2003, in Scientific
Frontiers in Research on Extrasolar Planets, ASP Conference Series
v294, Deming, D. and Seager, S, editors.

\bibitem[Ostriker, Shu \& Adams(1992)]{OSA92}  Ostriker, E. C., Shu,
F. H., Adams, F. C., 1992, ApJ, 399, 192

\bibitem[Pickett \etal(1998)]{Pick98} Pickett, B. K., Cassen, P.,
Durisen, R. H., Link, R., 1998, ApJ, 504, 468

\bibitem[Pickett \etal(2000a)]{Pick00a} Pickett, B. K., Cassen, P.,
Durisen, R. H., Link, R., 2000, ApJ, 529, 1034

\bibitem[Pickett \etal(2000b)]{Pick00b} Pickett, B. K., Cassen, P.,
Durisen, R. H., Mejia, A. C., 2000, ApJL, 540, 95

\bibitem[Pickett \etal(2003)]{Pick03} Pickett, B. K., Mejia, A. C.,
Durisen, R. H., Cassen, P. M., Berry, D. K., Link, R. P., 2003, ApJ,
590, 1060

\bibitem[Press \etal(1992)]{NumRec} Press, W. H., Teukolsky, S. A.,
Vetterling, W. T., Flannery, B. P., 1992 Numerical Recipes, Cambridge
University Press, Cambridge

\bibitem[Price \& Monaghan(2004)]{PM04} Price, D. J., Monaghan, J. J.,
2004, MNRAS, 348, 139

\bibitem[Rice \etal(2003)]{RABB} Rice, W. K. M., Armitage, P. J.,
Bonnell, I. A., Bate, M. R., 2003, MNRAS, 339, 1025

\bibitem[Romeo(1994)]{Romeo94} Romeo, A. B., 1994, A\&A, 286, 799

\bibitem[Romeo(1997)]{Romeo97} Romeo, A. B., 1997, A\&A, 324, 523

\bibitem[Springel \etal(2001)]{springel2001} Springel, V., 
Yoshida, N., White, S.~D.~M., 2001, New Astronomy,  6, 79

\bibitem[Steinmetz \& M\"uller(1993)]{SM93} Steinmetz, M., M\"uller,
E., 1993, A\&A, 268, 391

\bibitem[Steinmetz(1996)]{Steinmetz96} Steinmetz, M., MNRAS, 278, 1005

\bibitem[Thacker \etal(2000)]{Thacker00} Thacker, R. J., Tittley, E.
R., Pearce, F. R., Couchman, H. M. P., Thomas, P. A., 2000, MNRAS,
319, 619

\bibitem[Thomas \& Couchman(1992)]{TC92} Thomas, P. A., Couchman,
H. M. P., 1992, MNRAS, 257, 11 (TC92)

\bibitem[Tohline(1982)]{Toh82} Tohline, J., 1982,
Fund. Cos. Phys., 8,1

\bibitem[Truelove \etal(1997)]{Truelove97} Truelove, J. K.,
Klein, R. I., McKee, C. F., Holliman, J., H., Howell, L. H.,
Greenough, J. A., 1997, ApJL, 489, 179L

\bibitem[Truelove \etal(1998)]{Truelove98} Truelove, J. K.,
Klein, R. I., McKee, C. F., Holliman, J., H., Howell, L. H.,
Greenough, J. A., Woods, D. T., 1998, ApJ, 495, 821

\bibitem[Wadsley \etal(2003)]{wadsley2003} Wadsley, J. W., Stadel, J.,
Quinn, T., 2003, New Astronomy, 9, 137

\bibitem[Williams \etal(2004)]{WCN04} Williams, P. R., Churches, D.
K., Nelson, A. H., 2004, ApJ, 607, 1


\end{thebibliography}
\end{document}